\documentclass[12pt]{article}
\usepackage{amsmath}
\usepackage{graphicx}
\usepackage{enumerate}
\usepackage{enumitem}
\usepackage{natbib}
\usepackage{url} 
\usepackage{amssymb}
\usepackage{amsthm}
\usepackage{pifont}

\usepackage{wrapfig}
\usepackage{mathtools}
\usepackage{float}
\usepackage{setspace}
\usepackage[hidelinks]{hyperref}
\usepackage{mathtools}
\usepackage{caption}
\usepackage{subcaption}
\usepackage{scalerel,stackengine}
\usepackage{relsize}
\usepackage{mathrsfs}
\usepackage{algorithm}
\usepackage{algpseudocode}
\usepackage[dvipsnames]{xcolor}
\usepackage{changebar}
\newtheorem{theorem}{Theorem}
\newtheorem{lemma}{Lemma}
\newtheorem{proposition}{Proposition}
\newtheorem{remark}{Remark}

\newtheorem{corollary}{Corollary}
\newtheorem{assumptions}{Assumptions}

\newcommand\normx[1]{\left\Vert#1\right\Vert}
\newcommand\bmat[1]{\mathbf{#1}}

\newcommand{\nacf}[0]{\operatorname{nacf}}
\newcommand{\pnacf}[0]{\operatorname{pnacf}}
\newcommand{\GNAR}[0]{\operatorname{GNAR}}

\newcommand{\rstage}[0]{\mbox{$r$-stage}}

\newcommand{\comGNAR}[0]{\mbox{community-$\alpha$}}
\newcommand{\lossfunc}[1]{g_{\boldsymbol{#1}}}
\newcommand{\gradlossfunc}[1]{\overset{\mathlarger{.}}{g}_{\boldsymbol{#1}}}

\begin{document}

\def\spacingset#1{\renewcommand{\baselinestretch}
{#1}\small\normalsize}\spacingset{1}

\begin{singlespace}
\title{\bf Generalized network autoregressive
modelling of longitudinal networks with application to presidential elections in the USA}
\author{}

\author{Guy Nason\footnotemark[1] \footnotemark[2],
    Daniel Salnikov\thanks{Dept.\ Mathematics, Huxley Building, Imperial College, 180 Queen's Gate, South Kensington, London, SW7 2AZ, UK.} \footnotemark[2]\\
    Imperial College London\\
    and \\
    Mario Cortina-Borja\thanks{Great Ormond Street Institute of Child Health, 30 Guilford Street, London WC1N 1EH} \footnotemark[1] \\
    Great Ormond Street Institute of Child Health,\\
    University College London}

\maketitle

\begin{abstract}
    Longitudinal networks are becoming increasingly relevant in the study of dynamic processes characterised by known or inferred community structure. Generalised Network Autoregressive (GNAR) models provide a parsimonious framework for exploiting the underlying network and multivariate time series. We introduce the \mbox{community-$\alpha$} GNAR model with interactions that exploits prior knowledge or exogenous variables for analysing interactions within and between communities, and can describe serial correlation in longitudinal networks. We derive new explicit finite-sample error bounds that validate analysing high-dimensional longitudinal network data with GNAR models, and provide insights into their attractive properties. We further illustrate our approach by analysing the dynamics of {\em Red, Blue} and \textit{Swing} states throughout presidential elections in the USA from 1976 to 2020, that is, a time series of length twelve on 51
    time series (US states and Washington DC). Our analysis connects network autocorrelation to eight-year long terms,  highlights a possible change in the system after the 2016 election, and 
    a difference in behaviour between {\em Red} and {\em Blue} states.
\end{abstract}

\noindent%
{\it Keywords:} high-dimensional time series, network time series, finite sample bounds, longitudinal data, network interactions, community-dependent models.

\vfill
\end{singlespace}
\newpage

\section{Introduction}
\label{sec: introduction}
Network time series models are currently of great interest. Such network-based models are useful for analysing a constant flux of temporal data characterised by large numbers of interacting variables. The interacting variables are associated with a network structure (or analysed with respect to one), for example, networks in climate science, cyber-security, genetics and political science; see, e.g., ~\cite{partialcorrnetworks, mills_introduction_2020}. Traditional models for multivariate time series, such as vector autoregressive processes, become increasingly difficult to estimate and interpret as the number of parameters grows quadratically with the number of variables, i.e., the well-known curse of dimensionality. Recently, the generalised network autoregressive (GNAR) model has been developed~\cite{gnar_org, Zhu17, gnar_paper}, which provides parsimonious models that often have simpler interpretability {\em and} superior performance in many settings, especially high-dimensional ones; see \cite{gnar_paper, NasonWei22, corbit_paper, forecastcpi}. Recent developments in this area include~\cite{Zhu19} for quantiles, \cite{Zhou20} for Network GARCH models,
\cite{ArmillottaFokianos24} for Poisson/count data processes, \cite{NasonWei22} to admit time-changing covariate variables, \cite{Mantziou23} for GNAR processes on
the edges of networks and
\cite{corbit_paper}, which introduces the Corbit
plot to aid model selection. We extend this framework by introducing the \mbox{{\em community}-$\alpha$} GNAR specification for modelling dynamic clusters in network time series. 
\begin{figure}[H]
    \centering
    \includegraphics[scale=0.45]{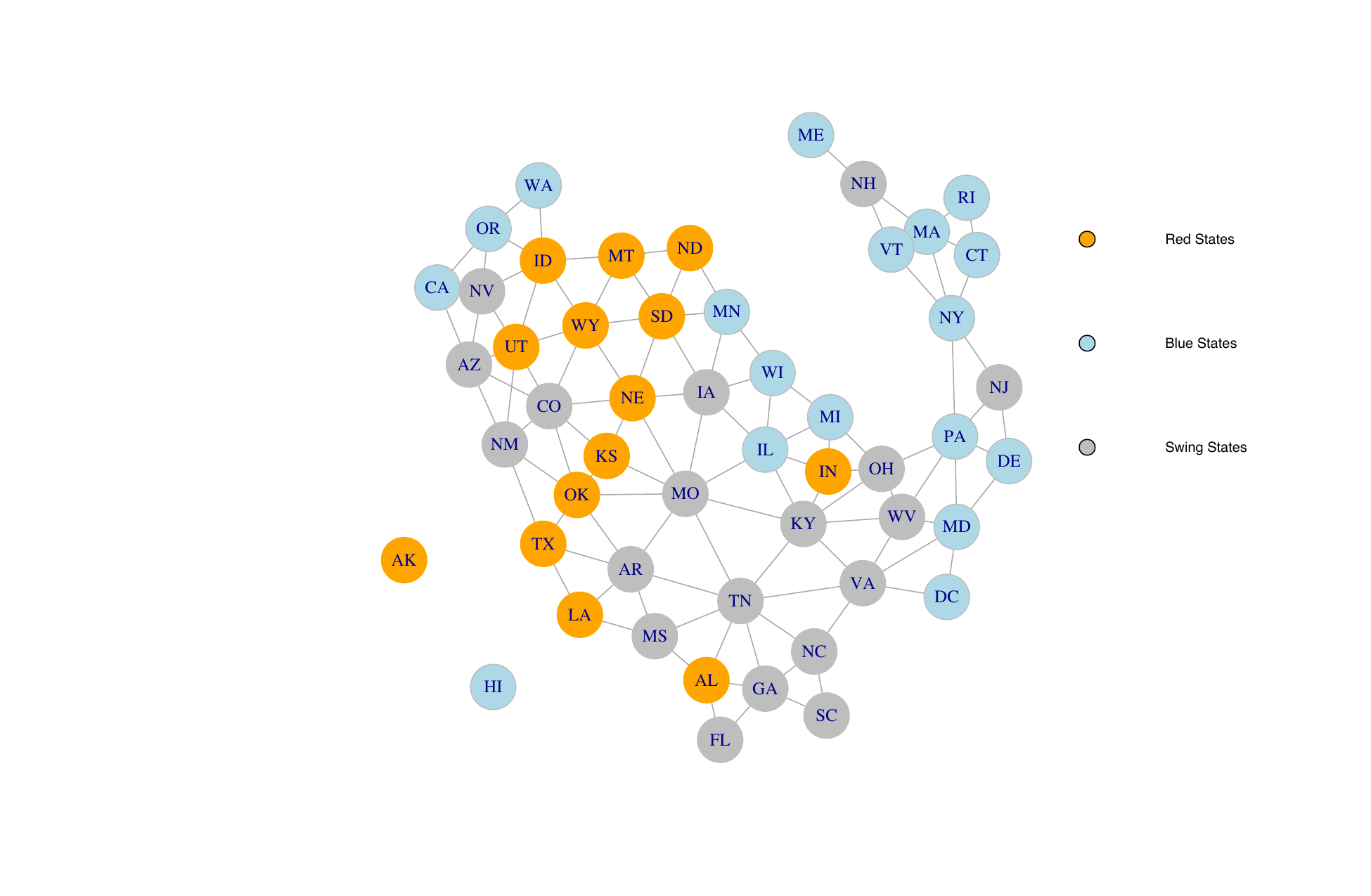}
    \caption{
    USA state-wise network, blue nodes are {\em Blue} states (Democrat nominee won at least 75$\%$ of elections), orange nodes are {\em Red} states (Republican nominee won at least $75\%$ of elections), and grey nodes are {\em Swing} states (neither party won at least $75\%$ of elections).
    }
    \label{fig: usa net}
\end{figure}
\newpage
Our \mbox{{\em community}-$\alpha$} GNAR assumes prior knowledge of the network structure, hence, it is useful when the data can be effectively described by an underlying network in which community structure is easily identifiable. Thus, it can be combined with methods that estimate network structures and/or clusters in dynamic settings. For example, \cite{clust_functional} propose methods for identifying clusters in temporal settings or \cite{parsimonious_conf} for network-linked data. These, and other clustering methods, e.g., spectral clustering; see~\cite{spectral_clust}, can aid researchers identify communities in a network time series that can be modelled as a \mbox{community-$\alpha$} GNAR model. 
\par
Alternatively, the group structure could be known before parameter estimation, for instance, longitudinal data are often associated with a known community structure, where the objective is to analyse any differences between communities. For example,~\cite{10.1093/jrsssa/qnad008} analyse the interactions within and between Democratic and Republican members of congress from 2010 to 2020. Figure~\ref{fig: usa net} shows the state-wise network for the USA, where nodes are connected if there is a land border between two states coloured by temporal party domination. The \mbox{community-$\alpha$} GNAR model can be used for studying the correlation structure of the time series linked to states and the interactions between them.

\subsection{Contribution and paper structure}
\label{subsec: comparison to existing methods}

\begin{figure}[H]
   \centering
   \includegraphics[scale=0.30]{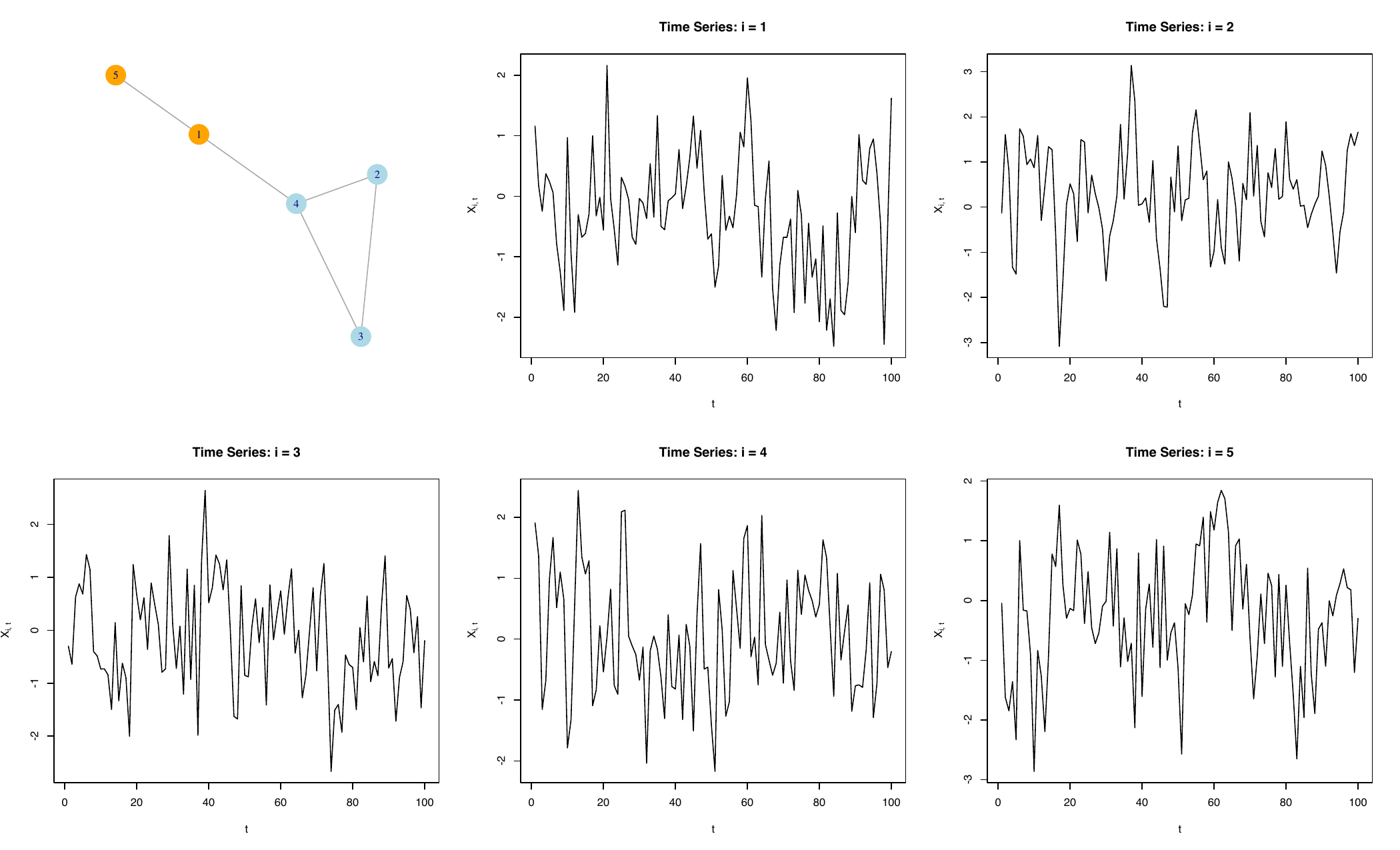}
   \caption{
       Illustration of a longitudinal network. The first plot is the \textbf{fiveNet} network (included in the \texttt{GNAR} package); see \cite{gnar_package}. The latter are nodal time series realisations of length 100 generated by a $\comGNAR$ $\GNAR \left ( [1, 2], \{ [1], [1, 1] \}, 2 \right )$ model. Blue nodes belong to community $K_1 = \{2, 3, 4 \}$ and orange ones to community $K_2 = \{ 1, 5 \}$.
   }
   \label{fig: 2-communal fiveNet}
\end{figure}

Our paper is organised as follows: Section \ref{sec: comm alpha GNAR model} introduces the $\comGNAR$ GNAR model, extends it to different model orders for different communities, and generalises it to include non-symmetric interactions between communities, i.e., when a community can affect another one without being affected by it. These extensions allow us to model interesting network-informed correlation structures that can highlight second-order behaviour in longitudinal data (grouped multivariate time series). We compare the $\comGNAR$ GNAR model to alternatives at the end of Section 2.  Figure~\ref{fig: 2-communal fiveNet} plots a longitudinal network realisation generated by a simple $\comGNAR$ GNAR model over five nodes on two communities.

Section~\ref{sec: model estimation} studies the parsimonious GNAR family, proposes a conditional least-squares estimator based on a linear model representation,  presents idealised results for our estimator and shows attractive finite sample properties of GNAR models. These form the basis of our asymptotic results and allow us to avoid using arguments from invertibility theory, as is usually done in time series. Section~\ref{sec: model selection} deals with model order choice and reviews the network autocorrelation function (NACF) and partial NACF (PNACF). 
\par
Finally, Section \ref{sec: application} exploits GNAR for studying the dynamics between {\em Red, Blue} and \textit{Swing} states throughout presidential elections in the USA by analysing the (network) time series of state-wise vote percentages for the Republican nominee in election cycles starting in 1976 and ending in 2020. This novel analysis is possible because GNAR models can handle high-dimensional time series data, in this case the realisation length is twelve and the number of time series per observation is fifty-one. Interestingly, our analysis suggests that there might be growing polarisation and a change in the governing dynamics instantiated in the 2016 election. We begin by reviewing network time series and GNAR models.

\subsection{Review of GNAR models}
\label{subsec: GNAR model review}
A network time series $\mathcal{X} := (\boldsymbol{X}_t, \mathcal{G})$ is a stochastic process that manages interactions between nodal time series $X_{i, t} \in \mathbb{R}$ based on the underlying network $\mathcal{G}$. It is composed of a multivariate time series $\boldsymbol{X}_t \in \mathbb{R}^d$ and an underlying network $\mathcal{G} = (\mathcal{K}, \mathcal{E})$, where $\mathcal{K} = \{1, \dots, d\}$ is the node set, $\mathcal{E} \subseteq \mathcal{K} \times \mathcal{K}$ is the edge set, and $\mathcal{G}$ is an undirected graph with $d \in \mathbb{Z}^+$ nodes. Each nodal time series $X_{i, t}$ is linked to node $i \in \mathcal{K}$. Throughout this work we assume that the network is static, however, $\GNAR$ processes can handle time-varying networks; see \cite{gnar_org}. GNAR models provide a parsimonious framework by exploiting the network structure. This is done by sharing information across nodes in the network, which allows us to estimate fewer parameters in a more efficient manner. A key notion is that of $\rstage$ neighbours, we say that nodes $i$ and $j$ are $\rstage$ neighbours if and only if the shortest path between them in $\mathcal{G}$ has a distance of $r$ (i.e., $d(i, j) = r$). We use $\rstage$ adjacency to define the $\rstage$ adjacency matrices $\bmat{S}_r \in \{0, 1\}^{d \times d}$, where $[\bmat{S}_r]_{ij} = \mathbb{I} \{ d(i, j) = r \}$, $\mathbb{I}$ is the indicator function, $r \in \{1, \dots, r_{\text{max}} \}$, and $r_{\text{max}} \in \mathbb{Z}^+$ is the longest shortest path in $\mathcal{G}$. The $\bmat{S}_r$ extend the notion of adjacency from an edge between nodes to the length of shortest paths (i.e., smallest number of edges between nodes). Note that $\bmat{S}_1$ is the regular adjacency matrix and that all the $\bmat{S}_r$ are symmetric. Further, assume that unique association time-varying weights $w_{ij} (t) \in [0, 1]$ between nodes are available at all times $t$. These weights measure the relevance node $j$ has for forecasting $i$, and can be interpreted as the proportion of the neighbourhood effect attributable to node $j$. We define the weights matrix $\bmat{W}_t \in \mathbb{R}^{d \times d}$ as the matrix $[\bmat{W}_t]_{ij} = w_{ij}(t)$. Note that since there are no self-loops in $\mathcal{G}$ all diagonal entries in $\bmat{W}_t$ are equal to zero, and that since $w_{ij}(t) \neq w_{ji}(t)$ is valid $\bmat{W}_t$ is not necessarily symmetric (i.e., nodes can have different degrees of relevance).
\par
If the weights are constant with respect to time, then we drop the $t$ subscript and write $\bmat{W}$. In the absence of prior weights, GNAR assigns equal importance to each $\rstage$ neighbour in a neighbourhood regression at all times $t$, i.e., $w_{ij} = \{| \mathcal{N}_r (i) |\}^{-1}$, where $|\mathcal{N}_r (i)| \leq d - 1$ is the number of $\rstage$ neighbours of node $i$ and $\mathcal{N}_r(i) \subset \mathcal{K}$ is the set of $\rstage$ neighbours of node $i$. A GNAR model assumes that effects are shared among $\rstage$ neighbours, rather than considering pairwise regressions and it focuses on the joint effect $\rstage$ neighbours have on $X_{i, t}$. 
\par
We can express the entire
autoregressive model in terms of individual $\rstage$ neighbourhood regressions. These are given by $\boldsymbol{Z}_{t}^r := \left ( \bmat{W}_t \odot \bmat{S}_r \right ) \boldsymbol{X}_t$, where $\odot$ denotes the Hadamard (component-wise) product. Each entry $Z_{i,t}^r$ in $\boldsymbol{Z}_t^r$ is the $\rstage$ neighbourhood regression corresponding to node $i$. For example, the vector-wise representation of a \mbox{\textit{global}-$\alpha$} $\GNAR \left (p, [s_k] \right )$ model is given by
\begin{equation} \label{eq: global alpha}
    \boldsymbol{X}_{t} = \sum_{k = 1}^{p} \left ( \alpha_k \boldsymbol{X}_{t - k} + \sum_{r = 1}^{s_k} \beta_{kr} \boldsymbol{Z}_{t - k}^{r} \right ) 
    + \boldsymbol{u}_{t}, 
\end{equation}
where $\alpha_k \in \mathbb{R}$ and $\beta_{kr} \in \mathbb{R}$ are the autoregressive
and network autoregressive coefficients, $p \in \mathbb{Z}^+$ is the maximum lag, $s_k \in \{1, \dots, r^*\}$ is the maximum $\rstage$ depth at lag $k = 1, \dots, p$, $r^*  \leq r_{\text{max}}$ is the maximum $\rstage$ depth across all lags, and $\boldsymbol{u}_t$ are independent and identically distributed zero-mean white noise innovations with covariance matrix $\sigma^2 \bmat{I_d}$ and $\sigma^2 > 0$. This representation is equivalent to the node-wise one in \cite{gnar_paper} and highlights the \textit{parsimonious} structure of a \text{global}-$\alpha$ GNAR model. The construction above follows the one in \cite{corbit_paper} which includes more details, interpretation and further results. Further, GNAR models can perform estimation with missing data by setting to zero the weights that involve a missing observation at time time $t$, i.e., if $X_{i, t}$ is missing, then $w_{ji} (t) = 0$ for $j \in [d]$ and the new normalising weights are computed. This is a special case of time-varying weights; see \cite{gnar_paper}.

\section{The \texorpdfstring{\mbox{community-$\alpha$}}{} GNAR model}
\label{sec: comm alpha GNAR model}
The $\comGNAR$ $\GNAR$ model is a new and extended
specification of the model introduced by \cite{gnar_org, gnar_paper}. It incorporates a cluster structure into the model by further exploiting the underlying network, and extends the weight covariate specification. This section defines the model and introduces concepts for handling cluster structure in the network. Further, we extend the model by allowing different model orders (i.e., number of lags and $\rstage$ neighbours) for each community, and incorporating interactions between communities.
\subsection{Introducing community structure into GNAR}
\label{subsec: community model}
Suppose that there is a collection of covariates $c \in \{1, \dots, C \} = [C]$ such that
the $i$th time series, $X_{i, t}$, is linked to only one covariate at all times $t \in \mathbb{Z}^+_0$, where $C \leq d$ is the number of covariates. We define  clusters $K_c := \left \{ i \in \mathcal{K} : X_{i, t} \text{ is associated with covariate } c \right \} $. By definition the $K_c$ are disjoint subsets of the node set (i.e., $K_c \subseteq \mathcal{K}$, and $K_c \cap K_{c'} = \varnothing$ if $c \neq c'$), such that $\cup_{c =1}^{C} K_c = \mathcal{K}$, each community has $d_c$ members (i.e., $|K_c| = d_c$) and $\sum_{c = 1}^C d_c = d$. Thus, the $K_c$ form a partition of $\mathcal{K}$ and define non-overlapping clusters in $\mathcal{G}$. Intuitively, each covariate is a label that indicates the cluster to which $X_{i, t}$ belongs. For example, if $\mathcal{G}$ consists of population centres, then each $X_{i, t}$ could be characterised as either urban, rural or a hub-town. Each cluster is a collection of nodes that defines a community in $\mathcal{G}$. We select the $c$-community defined by $K_c$ using the vectors $\boldsymbol{\xi}_c \in \mathbb{R}^d$. These are given by $\boldsymbol{\xi}_c := (\xi_{1, c}, \dots, \xi_{d, c} )$, where $\xi_{i, c} := \mathbb{I} (i \in K_c)$. Each entry in $\boldsymbol{\xi}_c$ is non-zero if and only if $i \in K_c$, thus, we can select the community linked to covariate $c$. Let $\boldsymbol{X}_t^c := \boldsymbol{\xi}_c \odot \boldsymbol{X}_t$. 
\par
Then, each entry in $\boldsymbol{X}_t^c$ is not constantly zero if and only if $i \in K_c$ (i.e., $X_{i, t}$ is characterised by $c$). The \mbox{\textit{community}-$\alpha$} GNAR model is given by
\begin{equation} \label{eq: com alpha gnar vector wise}
    \boldsymbol{X}_t = \sum_{k = 1}^p \sum_{c = 1}^C \left ( \alpha_{kc} \boldsymbol{X}_{t - k}^c + \sum_{r = 1}^{s_k} \beta_{krc} \boldsymbol{Z}_{t - k}^{r, c} \right ) + \boldsymbol{u}_t,
\end{equation}
where $\alpha_{kc} \in \mathbb{R}$ and $\beta_{krc} \in \mathbb{R}$ are the $c$-community autoregressive and neighbourhood regression coefficients at lag $k$, 
and $\boldsymbol{Z}_{t - k}^{r, c} := \left ( \bmat{W}_{t, c} \odot \bmat{S}_r \right) \boldsymbol{X}_{t - k}^c$
are the  $c$-community $\rstage$ neighbourhood regressions at lag $k$, and $\boldsymbol{u}_t$ are zero-mean independent and identically distributed white noise such that $\mathrm{cov} ( \boldsymbol{u}_t) = \sigma^2 \bmat{I_d}$ and $\sigma^2 > 0$. Above, $\bmat{W}_{t, c}$ is the $c$-community weights matrix, which is $\bmat{W}_t$ constrained to weights between nodes in the same community, i.e., $[\bmat{W}_{t, c}]_{ij} = w_{ij} (t) \mathbb{I} ( i \in K_c \text{ and } j \in K_c)$, and $\bmat{S}_r$ is the $\rstage$ adjacency matrix. We denote the model order of \eqref{eq: com alpha gnar vector wise} by $\GNAR \left (p, [s_k], [C] \right )$, where $k = 1, \dots, p$ is the current lag, $p \in \mathbb{Z}^+$ is the maximum lag, $C$ is the number of communities, and $c \in [C]$ is the covariate that characterises community $K_c$. 

\subsubsection{Different model order between communities}
\label{subsec: different order}
Assume that the lag and $\rstage$ order varies across communities, i.e., maximum lag and $\rstage$ are not necessarily equal for two different communities. The community-wise additive decomposition in \eqref{eq: com alpha gnar vector wise} allows expressing this extension by
\begin{equation} \label{eq: communal alpha diff orders}
    \boldsymbol{X}_t = \sum_{c = 1}^C \left \{ \sum_{k = 1}^{p_c} \left ( \alpha_{k, c} \boldsymbol{X}_{t - k}^c + \sum_{r = 1}^{s_k (c)} \beta_{k, r, c} \boldsymbol{Z}_{t - k}^{r, c} \right ) \right \} + \boldsymbol{u}_{t},
\end{equation}
where $\alpha_{k, c}$, $\beta_{k, r, c}$, $\boldsymbol{Z}_{t - k}^{r, c}$, and $\boldsymbol{u}_{t}$ are the same as in \eqref{eq: com alpha gnar vector wise}. However, instead of equal lag and $\rstage$ depth orders the model now is specified by the following: $p_c \in \mathbb{Z}^+$ and $s_{k} (c) \in \{1, \dots, r^*_c \}$, where the former is the maximum lag for community $K_c$ and the latter is the maximum $\rstage$ depth at lag $k = 1, \dots, p_c$ for community $K_c$. Also, $r^*_c$ is the maximum $\rstage$ depth for community $K_c$, and as before, $C$ is the number of non-overlapping communities that are characterised by the covariates $c \in [C]$. We denote this model order by $\GNAR \left ( [p_c], \{ [s_{k} (c)] \}, [C] \right )$.

\subsubsection{Introducing interactions between communities}
\label{subsec: interactions}
We introduce interactions between communities into the model by defining the interaction set $\mathcal{I}_c := \{ \tilde{c} \in [C] : c \text{ interacts with } \tilde{c}, \text{where }\tilde{c} \neq c \}$ for community $K_c$. Thus, $\mathcal{I}_c$ is the set that indicates which communities have an effect on $K_c$ (i.e., the interactions set for $K_c$). Note that $\tilde{c}  \in \mathcal{I}_c$ does not necessarily require that $c \in \mathcal{I}_{\tilde{c}}$, thus, it is possible to model the effect of one community on another one without imposing a symmetric effect structure. This might be useful when one community is known to ``listen'' whereas other ones tend to isolate themselves. Further, note that $\mathcal{I}_c \subset  [C]$ and $|\mathcal{I}_c | \leq C - 1$. 
\par
The linear effect community $K_{\tilde{c}}$ has on community $K_c$ at the $k$th lag and $r$th $\rstage$ is modelled by the $(c:\tilde{c})$-community interaction $\rstage$ neighbourhood regression given by $\boldsymbol{Z}_{t - k}^{r,c:\tilde{c}} = \xi_c \odot \left ( \bmat{W}_t \odot \bmat{S}_r \right) \boldsymbol{X}_{t - k}^{\tilde{c}},$ 
where each entry $Z_{i, t - k}^{r, c:\tilde{c}}$ is equal to the $\rstage$ neighbourhood regression for node $i$ restricted to nodes that belong to community $K_{\tilde{c}}$ (i.e., non-zero entries correspond to nodes in community $K_c$), and by the parameter $\gamma_{k, r, c:\tilde{c}}\in \mathbb{R}$, which is the interaction coefficient. The interaction term for the effect community $K_{\tilde{c}}$ has on community $K_{c}$ at the $k$th lag and $r$th $\rstage$ is $\gamma_{k, r, c:\tilde{c}} \boldsymbol{Z}_{t - k}^{r,c:\tilde{c}}$. The $\comGNAR$ GNAR model is composed of the following additive terms: autoregressive components $\boldsymbol{\alpha}_{c} (\boldsymbol{X}_t) :=  \sum_{k = 1}^{p_c} \alpha_{k, c} \boldsymbol{X}_{t - k}^{c},$ within community $\boldsymbol{\beta}_{c} (\boldsymbol{X}_t) := \sum_{k = 1}^{p_c} \sum_{r = 1}^{s_k (c)} \beta_{k, r, c} \boldsymbol{Z}_{t - k,}^{r, c},$ and between community $\boldsymbol{\gamma}_{c}(\boldsymbol{X}_t) := \sum_{k = 1}^{p_c} \sum_{r = 1}^{s_k (c)} \sum_{\tilde{c} \in \mathcal{I}_c} \gamma_{k, r, c:\tilde{c}} \boldsymbol{Z}_{t - k}^{r, c:\tilde{c}},$ where $\alpha_{k, c}$, $\beta_{k, r, c}$, $\boldsymbol{Z}_{t - k}^{r, c}$, and $\boldsymbol{u}_{t}$ are the same as in \eqref{eq: com alpha gnar vector wise}, and the different community orders are given as in \eqref{eq: communal alpha diff orders}. However, the model now includes interaction terms for each community $K_c$, which increase the number of parameters for each lag and $\rstage$ pair $(k, r)$ by $|\mathcal{I}_c|$. Thus, we can express the $\comGNAR$ GNAR with interactions model as
\begin{equation} \label{eq: community GNAR structural representation}
    \boldsymbol{X}_t = \sum_{c = 1}^{C} \left \{ \boldsymbol{\alpha}_{c} (\boldsymbol{X}_t) + \boldsymbol{\beta}_{c} (\boldsymbol{X}_t) + \boldsymbol{\gamma}_{c} (\boldsymbol{X}_t) \right \} + \boldsymbol{u}_t,
\end{equation}
and further, note that the model is a sum of $C$ global-$\alpha$ GNAR processes. Thus, by adding up the autoregressive components: $\boldsymbol{\alpha} (\boldsymbol{X}_t; p_c, [C]) = \sum_{c = 1}^{C} \boldsymbol{\alpha}_{c} (\boldsymbol{X}_t)$, within community components: $ \boldsymbol{\beta} \{\boldsymbol{X}_t; s_k (c), p_c, [C] \} = \sum_{c = 1}^{C}  \boldsymbol{\beta}_{c} (\boldsymbol{X}_t)$, and between community components: 
$ \boldsymbol{\gamma} \{\boldsymbol{X}_t; \mathcal{I} ([C]), s_k(c), p_c, [C] \} = \sum_{c = 1}^{C}  \boldsymbol{\gamma}_{c} (\boldsymbol{X}_t)$.
We express the model in a compact form as
\begin{singlespace}
    \begin{equation} \label{eq: GNAR compact structural representation}
    \boldsymbol{X}_t =  \boldsymbol{\alpha} \left ( \boldsymbol{X}_t; [p_c], [C] \right ) +  \boldsymbol{\beta} \{\boldsymbol{X}_t; \{ [s_k (c)] \}, [C] \} +  \boldsymbol{\gamma} \{ \boldsymbol{X}_t; \{ \mathcal{I}_c \}, [C] \} + \boldsymbol{u}_t,
\end{equation}
\end{singlespace}
\begin{flushleft}
    where each community component is given by \eqref{eq: community GNAR structural representation} and $ \mathcal{I} [C] := \{ \mathcal{I}_c \}$ is the set of interaction sets, i.e., each entry is the $c$--th interaction set for $c \in [C]$. We denote this model order as $\comGNAR$ GNAR$([p_c], \{ [s_k (c)] \}, [C], \mathcal{I}[C] )$, where $p_c$, $s_k (c)$ and $C$ are as in \eqref{eq: communal alpha diff orders} and $\mathcal{I}_c$ indicates the communities that interact with $K_c$.
\end{flushleft}
\par
Note that the model given by \eqref{eq: GNAR compact structural representation} can be extended by specifying different model orders for the interaction terms (i.e., $\rstage$ length order for interactions can be different from $\rstage$ length order for community effects). We leave these extensions for future work. If there are no interaction terms, i.e., $\mathcal{I} [C] = \varnothing$, then $\boldsymbol{\gamma} \{ \boldsymbol{X}_t; \mathcal{I} [C], s_k(c), p_c, [C] \}  \equiv 0$, so we can write the \mbox{global-$\alpha$} GNAR model as 
$$\boldsymbol{X}_t =  \boldsymbol{\alpha} \left (\boldsymbol{X}_t; p, [1] \right ) + \boldsymbol{\beta} \left ( \boldsymbol{X}_t; s_k, p, [1] \right ) + \boldsymbol{u}_t,$$ and the \mbox{local-$\alpha$} GNAR model as 
$$\boldsymbol{X}_t =  \boldsymbol{\alpha} (\boldsymbol{X}_t; p, [d]) +  \boldsymbol{\beta} \{ \boldsymbol{X}_t; s_k, p, [1] \} + \boldsymbol{u}_t. $$

\subsection{Comparison with other models}
\label{subsec: comparison with other models}
Our model has similarities with the community network autoregressive (CNAR) model introduced by \cite{CHEN20231239}, which is motivated by social and financial networks. In our notation, the CNAR model introduced by \cite{CHEN20231239} is given by
\begin{equation} \label{eq: cnar}
   \boldsymbol{X}_t =  \sum_{c = 1}^{C} \left ( \boldsymbol{\mu}_c^T \boldsymbol{v}_{t - 1} + \alpha_c \boldsymbol{X}_{t - 1}^c + \beta_{1, 1, c}  \boldsymbol{Z}_{t - 1}^{1, c} \right ) + \boldsymbol{u}_t,
\end{equation}
where $ \boldsymbol{\mu}_c \in \mathbb{R}^{d_c}$ are fixed covariate coefficients and $\boldsymbol{v}_{t - 1} \in \mathbb{R}^{p_{\mu}}$ are $p_{\mu}$-dimensional time-varying covariates, $ \boldsymbol{u}_t = \mathbf{A} \zeta_t + \boldsymbol{\epsilon}_t$, $\zeta_t \in \mathbb{R}^m$ are unknown factors, $\mathbf{A} \in \mathbb{R}^{d \times m}$ is the factor loading matrix that induces correlation into the residuals, and $\boldsymbol{\epsilon}_t$ is multivariate normal with mean zero and covariance $\sigma^2 \mathbf{I_d}$. Further, all connection weights are fixed at all times as $w_{ij} = (\sum_{j = 1}^{d} [\mathbf{S}_1]_{ij})^{-1}$, i.e., out-degree of the $i$--th node.
\par
Our model also has similarities with the batch network autoregressive (BNAR) model proposed by \cite{zhu2023simultaneous}, which is also motivated by social and financial networks. In our notation, the BNAR model is given by
\begin{equation} \label{eq: BNAR}
    \boldsymbol{X}_t = \sum_{c = 1}^{C} \{ \mathbf{V}_c \boldsymbol{\mu}_c + \alpha_c \boldsymbol{X}_{t - 1}^c + ( \sum_{\tilde{c} = 1}^C \beta_{1, 1, \tilde{c}}  \boldsymbol{Z}_{t - 1}^{1, \tilde{c}} ) \} + \boldsymbol{u}_t,
\end{equation}
where $ \boldsymbol{\mu}_c \in \mathbb{R}^{m_c}$ are fixed $m_c$-dimensional covariate coefficients and $\mathbf{V}_c \in \mathbb{R}^{d_c \times m_c}$ are static covariates, $\boldsymbol{u}_t$ is multivariate normal with mean zero and covariance $\sigma^2 \mathbf{I_d}$, where connection weights are fixed at all times as $w_{ij} = (\sum_{j = 1}^{d} [\mathbf{S}_1]_{ij})^{-1}$, i.e., out-degree of the $i$--th node.
\par
We remark that the CNAR and BNAR models assume unknown cluster memberships and incorporate clustering algorithms into model estimation. However, the CNAR model estimates its parameters by plugging in an estimated cluster membership matrix resulting from the eigendecompostion of the Laplacian for the observed adjacency matrix, i.e., $\mathrm{diag}\{ (\sum_{j = 1}^{d} [\mathbf{S}_1]_{ij}) \} - \bmat{S}_1$. Thus, said clustering algorithm ignores the temporal correlation in the data, and imposes constraints on the underlying network by assuming that it is generated by a stochastic block model, which often fail to generate networks commonly observed in practice. Further, parameter estimation ignores clustering, thus, it is equivalent to estimation of a VAR model with certain factor and residual correlation assumptions.
\par
The BNAR model proposes a clustering algorithm that depends on the magnitude differences between community-$\alpha$ coefficients, i.e., $|\alpha_c - \alpha_{\tilde{c}}| + || \boldsymbol{\mu}_c - \boldsymbol{\mu}_{\tilde{c}}||_2,$ and is constrained to one-lag model order specifications. It is computationally expensive, requires coefficients between communities to have large differences in magnitude, which is challenging given the stationarity assumption, and ignores the community interaction coefficients, i.e., $\beta_c$ and $\gamma_c$, both within and between communities. Thus, the clustering algorithm also requires particular network structures for reasonable performance. Once cluster membership has been fixed both the CNAR and BNAR models are GNAR models constrained to one-lag and one-stage model order, where either communities do not interact or all communities interact in a symmetric manner, i.e., $\gamma_{1, 1 c:\tilde{c}} = \gamma_{1, 1, \tilde{c}:c}$ for all $c, \tilde{c} \in [C]$, which cannot perform estimation with missing data, and do not exploit the network-induced correlation structure for model selection and exploring departures from stationarity, which is common in practice. 
\par
We remark that we do not include static covariates in the GNAR framework because it is likely that the time series require differencing and/or standardising before any formal modelling, said procedure usually removes trend effects. Hence, it is unclear what the interpretation of static covariates and latent factors that do not relate to the network structure is, also we note that incorporating covariates and latent factors should be done with some specific question in mind given that these saturate the model with more parameters, which reduces its parsimonious nature and makes interpretation harder. Further, our modelling focus is on longitudinal and/or time series data where knowledge of cluster membership is known, thus, interest is placed on identifying the differences in correlation structure and if interaction terms have a significant effect between communities, whereas, CNAR and BNAR are intended for settings in which the network is assumed to have some structure and identification of latent clusters is the main interest. However, we note that including static covariates, factors and other linear model extensions into GNAR is straightforward by properly adjusting~\eqref{eq: GNAR compact structural representation}. 
In Table~\ref{tab: mod comparison}, tv weights stands for time-varying weights, missing data for feasible estimation with missing data, diff. or. for different model order between communities, int. spec. for specifying interaction terms in the model that are {\em different} between different communities, $\rstage$ indicates the possibility to include higher-order interactions, i.e., $r \geq 1$, net. const. indicates that the model imposes constraints on the underlying network, and latent indicates if cluster membership is assumed to be unknown.
\begin{table}[H]
    \centering
    \caption{Summary of the main differences between the following models: community-$\alpha$ GNAR in~\eqref{eq: GNAR compact structural representation}, CNAR in~\eqref{eq: cnar} and BNAR in~\eqref{eq: BNAR}. Each column indicates if the model can incorporate said specification, as well as prior assumptions imposed on the model and/or underlying network. See text for a precise description of each column. }
    \begin{tabular}{cccccccc}
         Model & tv weights & missing data & diff. or. & int. spec. & $\rstage$ & net. const. & latent \\
         \hline
         GNAR & \ding{51} & \ding{51} & \ding{51}& \ding{51} & \ding{51} & \ding{55} & \ding{55} \\
         CNAR  & \ding{55} & \ding{55} & \ding{55}& \ding{55} & \ding{55} & \ding{51} & \ding{51} \\
         BNAR & \ding{55} & \ding{55} & \ding{55}& \ding{55} & \ding{55} & \ding{51} & \ding{51}
    \end{tabular}
    \label{tab: mod comparison}
\end{table}

\section{The parsimonious GNAR family}
\label{sec: model estimation}
GNAR processes impose network-informed constraints on the parameters of a vector
autoregressive (VAR) model. The number of parameters in GNAR models grows linearly with dimension, i.e., $O(d)$, whereas, for VAR models the number of parameters grows in polynomial fashion, i.e., $O(d^2)$. Essentially, network-informed constraints perform variable selection, reduce estimator variance and induce a correlation structure that our GNAR modelling framework exploits (via the network autocorrelation function for quick identification of departures from stationarity and model order selection); see~\cite{corbit_paper}. Thus, our GNAR modelling framework produces parsimonious models that result from network-informed reparameterizations of VAR models. Also, note that the GNAR framework is model-agnostic with respect to the underlying network, i.e., we do not impose any constraints on the process that generates the network. The $\comGNAR$ GNAR model can be thought of as being `between' a local-$\alpha$
and a global-$\alpha$ GNAR specification.
We make this precise next.
\begin{singlespace}
\begin{remark} \label{rem: GNAR from global to local and var}
    Assume that $\boldsymbol{X}_t \in \mathbb{R}^d$ is a
    particular $\comGNAR$ $\GNAR$ process with the same lag order for all communities, that the underlying network is
    \underline{fully} connected and static, and that each node $i$ is a community (i.e., $K_c = \{ i \}$ for $i \in \{1, \dots, d \}$, where $C = d$). Then, this $\comGNAR$ $\GNAR$ model can satisfy one of the following:
    \begin{itemize}
        \item[(a)] If we include all community interactions for all nodes and assume that $\rstage$ order includes all possible $\rstage$ neighbours at all lags for all nodes (i.e., $\gamma_{k, r, c:\tilde{c}} \neq 0$ for all $r \leq r_{\max}$ at all $k = 1, \dots, p$), then the model given by \eqref{eq: GNAR compact structural representation} is identical to an unconstrained $\mathrm{VAR}(p)$.

        \item[(b)] If we constrain community interactions to a single effect for all lag and $\rstage$ pairs $(k, r)$, i.e., $\gamma_{k, r, c:\tilde{c}} = \beta_{kr}$ for all $c$ and $\tilde{c}$ at all lags and 
        $\rstage$s, then the model given by \eqref{eq: GNAR compact structural representation} is identical to a local-$\alpha$ $\GNAR(p, [s_k])$.
        (Note, where each single node $i$ is a community, then there are no
        other nodes in their community, so there are only interaction
        effects, $\gamma$, but for single effects these become $\beta$s.)

        \item[(c)] If we constrain community interactions and local effects for all nodes at all lags (i.e., $\alpha_{k, c} = \alpha_k$ and $\gamma_{k, r, c:\tilde{c}} = \beta_{kr}$ for all $c$ and $\tilde{c}$ at all lags and $\rstage$s), then the model given by \eqref{eq: GNAR compact structural representation} is identical to a global-$\alpha$ $\GNAR(p, [s_k])$.
    \end{itemize}
\end{remark}
\end{singlespace}
The insight Remark~\ref{rem: GNAR from global to local and var} provides is that not enforcing network-informed constraints (i.e., case $(a)$ in Remark \ref{rem: GNAR from global to local and var}) results in an unconstrained, and likely overparameterized, VAR$(p)$ model. We interpret this as not including nor benefiting from knowledge of the network, nor accounting for network-effects.
This unconstrained model requires more data to estimate efficiently and has $p d^2$ unknown parameters. Accounting for network-effects at $\rstage$ neighbourhood levels, i.e., case $(b)$ in Remark \ref{rem: GNAR from global to local and var}, results in the more parsimonious \mbox{local-$\alpha$} $\GNAR(p, [s_k])$ model. We interpret it as local autoregressive effects that interact with sets of $\rstage$ neighbours, thus, we use knowledge of the network to study the effect that $\rstage$ neighbours have, rather than the node-wise effects on the local processes, i.e., the $\rstage$ neighbourhood regressions are interaction terms (linear combinations of $\rstage$ neighbours). 
This model has $p(d + \sum_{k = 1}^p s_k) \leq p d^2$ (usually $p(d + \sum_{k = 1}^p s_k) \ll p d^2$) unknown parameters, thus, it is feasible even in high-dimensional settings (i.e., $d > n = T-n$).
\par
Further, accounting for network-effects at the global autoregressive level and $\rstage$ neighbourhood levels, i.e., case $(c)$ in Remark \ref{rem: GNAR from global to local and var}, results in an even more parsimonious global-$\alpha$ $\GNAR(p, [s_k])$ model. We interpret this model as strong network-wise effects that are shared across all nodes and $\rstage$ neighbourhood regressions, i.e., the \textit{entire} network is \textit{one} community.
For example, in modelling the spread of a particular disease,
which does not change its characteristics depending on its
geographical location. This model has $p(1 + \sum_{k = 1}^p s_k) \ll p d^2$ unknown parameters. If $d$ is large, then $p(1 + \sum_{k = 1}^p s_k) \ll p(d + \sum_{k = 1}^p s_k)$, thus, if a GNAR model is valid, then it is possible to model an ultra high-dimensional time series (i.e., $d \gg n$). 

\begin{singlespace}
\begin{remark} \label{rem: VAR representation}
    Let $\mathbf{\Phi}_{k} (t)$ be the $d \times d$ matrix given by
    \begin{equation} \label{eq: GNAR VAR representation autoregressive matrices}
        \mathbf{\Phi}_{k} (t) = \sum_{c = 1}^C \left [ \mathrm{diag} ( \alpha_{k, c} \, \boldsymbol{\xi}_c ) + \sum_{r = 1}^{s_k (c)} \left \{ \beta_{k, r, c} (\mathbf{W}_{t, c} \odot \mathbf{S}_r ) + \sum_{\tilde{c} \in \mathcal{I}_c} \gamma_{k, r, c:\tilde{c}} 
        (\mathbf{W}_{t, c:\tilde{c}} \odot \mathbf{S}_r ) \right \} \right ],
    \end{equation}
    where terms for larger model order are set to zero, e.g., if $p_c < p_{\tilde{c}}$, then $\alpha_{k, c} \equiv 0$ for $k > p_c$, $[\mathbf{W}_{t, c}]_{ij} = w_{ij} (t)\mathbb{I} ( i \, \text{ and } \, j \in K_c)$, and $[\mathbf{W}_{t, c:\tilde{c}}]_{ij} = w_{ij}(t) \mathbb{I} ( i \in K_c \text{ and } j \in K_{\tilde{c}})$. Then, the {\em VAR}$(p)$ model given by
    \begin{equation} \label{eq: GNAR VAR representation}
         \boldsymbol{X}_t = \sum_{k = 1}^p \mathbf{\Phi}_{k} (t - k) \boldsymbol{X}_{t - k} + \boldsymbol{u}_t,
    \end{equation}
    where $\boldsymbol{u}_t$ are {\em i.i.d.} white noise, is identical to the {\em GNAR} model given by \eqref{eq: GNAR compact structural representation}.
\end{remark}
\end{singlespace}
\newpage
For all models introduced in Section~\ref{sec: comm alpha GNAR model} we
establish conditions for stationarity, derive conditional least-squares
estimators and compare the statistical efficiency of our estimators depending on idealised assumptions about the GNAR data-generating process.
We note that $\comGNAR$ GNAR models balance between cases $(b)$ and $(c)$ in Remark \ref{rem: GNAR from global to local and var}, \textit{and} incorporate community structure into network time series models, i.e., \eqref{eq: GNAR compact structural representation} can be expressed as a VAR by imposing networked-informed constraints on the autoregressive matrices, Remark~\ref{rem: VAR representation} makes this statement precise.

\subsection{Conditions for stationarity}
\label{subsec: conditions for stationarity}

Stationarity conditions can be immediately derived from the following result of~\cite{gnar_paper}. Our model can be thought of as a sum of $C$ global-$\alpha$ GNAR models, so we use Theorem~\ref{th: stationary cond} and obtain Corollary~\ref{cor: conditions for stationary processes}.
\begin{singlespace}
\begin{theorem}{[\textbf{\cite{gnar_paper}}]} \label{th: stationary cond}
Let $\boldsymbol{X}_t$ be a \mbox{local-$\alpha$} $\GNAR (p, [s_k] )$ process linked to a static network. If the parameters in $ X_{i, t} = \sum_{k = 1}^{p} ( \alpha_{i, k} X_{i, t - k} + \sum_{r = 1}^{s_k} \sum_{c = 1}^C \beta_{k r c} Z_{i, t - k}^{r, c} ) + u_{i, t},$ where $u_{i, t}$ are independent white noise, satisfy $\sum_{k = 1}^{p}  ( |\alpha_{i, k}| + \sum_{r = 1}^{s_k} \sum_{c = 1}^C |\beta_{k r c} | ) < 1$ for all $X_{i, t}$, and $i \in \mathcal{K}$. Then $\boldsymbol{X}_t$ is stationary.
\end{theorem}

\begin{corollary} \label{cor: conditions for stationary processes}
Let $\boldsymbol{X}_t$ be a {\em community-$\alpha$} $\GNAR \left ( [p_c], \{[s_{k} (c)] \}, [C], \mathcal{I}[C] \right )$ process such that all of the coefficients in \eqref{eq: GNAR compact structural representation} satisfy 
\begin{equation} \label{eq: simple stationarity condition}
    \sum_{k = 1}^{p_c} \left \{ |\alpha_{k, c} | + \sum_{r = 1}^{s_k (c)} \left ( |\beta_{k, r, c} | +  \sum_{\tilde{c} \in \mathcal{I}_c}  |\gamma_{k, r, c:\tilde{c}}| \right )  \right \} < 1,
\end{equation}
for all covariates $c \in [C]$. Then $\boldsymbol{X}_t$ is stationary.
\end{corollary}
\end{singlespace}

See \cite{gnar_paper} for a proof of Theorem \ref{th: stationary cond} and the supplementary material Section~\ref{sec: sty cond proof}
for a proof of Corollary~\ref{cor: conditions for stationary processes}. In view of Remark~\ref{rem: VAR representation}, we immediately derive a more general condition by exploiting the VAR representation in~\eqref{eq: GNAR VAR representation} and well--known conditions for stationarity of general VAR processes; see Chapter 2.1.1 in~\cite{luth}.

\begin{singlespace}
\begin{corollary} \label{cor: VAR conditions for stationarity}
    Let $\boldsymbol{X}_t$ be a {\em community-$\alpha$} $\GNAR \left ( [p_c], \{[s_{k} (c)] \}, [C], \mathcal{I}[C] \right )$ process such that the autoregressive matrices in \eqref{eq: GNAR VAR representation autoregressive matrices} satisfy 
    \begin{equation} \label{eq: VAR stationairty condition}
        \operatorname{det} \left \{ \mathbf{I_d} - \sum_{k = 1}^p z^p \, \mathbf{\Phi}_{k} (t - k) \right \} \neq 0,
    \end{equation}
    for all $|z| \leq 1$. Then, $\boldsymbol{X}_t$ given by the {\em VAR} representation in~\eqref{eq: GNAR VAR representation} is stationary.
\end{corollary}
\end{singlespace}
We remark that condition~\eqref{eq: simple stationarity condition} implies condition~\eqref{eq: VAR stationairty condition} and is easier to verify, and that these are sufficient but not necessary conditions, i.e., a GNAR process given by~\eqref{eq: GNAR compact structural representation} could still be stationary, if these are not satisfied.

\subsection{Conditional least-squares estimation for GNAR models}
\label{subsec: ls estimator}
Estimation of GNAR models is straightforward by noting that these are VAR models. However, we present a conditional linear model that exhibits the parsimonious nature of GNAR processes and, under idealised assumptions, allows us to obtain nonasymptotic bounds for our conditional least-squares estimators; see Section \ref{subsec: ls estimator error bounds}. 
\par
Assume that we observe $T \in \mathbb{Z}^+$ time-steps of a stationary $\comGNAR$ GNAR process with known order as defined in Section~\ref{subsec: interactions}.
The data $\bmat{X} := [\boldsymbol{X}_1, \dots, \boldsymbol{X}_T ]$ are realisations coming from $\boldsymbol{X}_t \sim \GNAR([p_c], \{[s_k (c)] \}, [C], \mathcal{I}[C])$. Let $n_c = T - p_c$ be the number of realisations with $p_c$ lags between them.
Then, we can expand \eqref{eq: GNAR compact structural representation} as an additive model by concatenating the following blocks. We begin by writing \eqref{eq: GNAR compact structural representation} as
\begin{singlespace}
    \begin{equation} \label{eq: sum expression}
    \sum_{c = 1}^C \boldsymbol{X}_t^c =\sum_{c = 1}^{C} \left \{ \boldsymbol{\alpha}_{c} (\boldsymbol{X}_t) + \boldsymbol{\beta}_{c} (\boldsymbol{X}_t) + \boldsymbol{\gamma}_{c} (\boldsymbol{X}_t) \right \} + \boldsymbol{u}_{t}^c,
\end{equation}
\end{singlespace}
where the terms are given by \eqref{eq: community GNAR structural representation}, and note that $\boldsymbol{u}_{t}^c := \boldsymbol{\xi}_c \odot \boldsymbol{u}_{t}$ are orthogonal in $\mathbb{R}^d$. By comparing the expression within curly brackets in \eqref{eq: sum expression} with the right-hand side in \eqref{eq: global alpha} we see that it is a vector-wise representation of a \mbox{global-$\alpha$} GNAR process restricted to community $K_c$. Hence,
$\boldsymbol{X}_t = \sum_{c = 1}^C \boldsymbol{X}_t^c$ is a sum of $C$ \mbox{global-$\alpha$} GNAR processes, where $\boldsymbol{X}_t^c$ is
the multivariate time series associated with cluster $c$, which was defined immediately
prior to~\eqref{eq: com alpha gnar vector wise}. This allows us to express the data by considering covariate blocks. Define the $c$-community `response vector' $\boldsymbol{y}_{c} := (\boldsymbol{X}_{p + 1}^c, \dots, \boldsymbol{X}_{T}^c ) \in \mathbb{R}^{n_c d}$. We proceed to build the design matrix: let $\bmat{R}_{k, c}$ be the matrix obtained from concatenating by rows the predictor columns for $t = p + 1, \dots, T$, i.e., $ \bmat{R}_{k, c} = [\boldsymbol{X}_{t - k}^c |  \boldsymbol{Z}_{t - k}^{1, c} | \dots |  \boldsymbol{Z}_{t - k}^{s_k (c), c} |  \boldsymbol{Z}_{t - k}^{1, c:\tilde{c}} | \dots |  \boldsymbol{Z}_{t - k}^{s_k (c), c:\tilde{c}} ].$ Then, each $\bmat{R}_{k, c}$ builds the block of the design matrix at lag $k$.  Concatenating the $\bmat{R}_{k, c}$ column wise for $k = 1, \dots, p_c$, i.e., $\bmat{R}_{c} = \left [\bmat{R}_{1, c} | \dots | \bmat{R}_{p_c, c} \right ], \label{eq:ccommunitydesign}$ results in the complete $c$-community design matrix $\bmat{R}_c \in \mathbb{R}^{n_c d \times q_c}$, where $q_c := p_c + \sum_{k = 1}^p s_k (c) (1 + |\mathcal{I}_c| )$ is the number of parameters in each block. Let $\boldsymbol{\theta}_c := (\alpha_{k, c}, \beta_{k, c}, \gamma_{k, c}) \in \mathbb{R}^{q_c},$ for $k = 1, \dots, p_c$, where $\beta_{k, c} = (\beta_{k, r, c})$ for $r = 1, \dots, s_k (c)$, and $\gamma_{k, c}$ concatenates $\gamma_{k, r, c:\tilde{c}}$ in ascending order with respect to  $\tilde{c} \in [C]$ for each $r = 1, \dots, s_k (c)$. Finally, concatenating the coefficients by lag, in ascending order, gives the complete parameter vector for $K_c$.
Thus, $\boldsymbol{\theta}_c$ is the $c$-community vector of unknown autoregressive coefficients in \eqref{eq: communal alpha diff orders} ordered by lag (i.e., it concatenates by rows the parameters for each lag starting at lag one). 
\par
Next we expand the block and write it as the linear model 
$$\boldsymbol{y}_c = \bmat{R}_c \boldsymbol{\theta}_c + \boldsymbol{u}_c,$$ 
where $\boldsymbol{u}_c := (\boldsymbol{u}_{p + 1}^c, \dots, \boldsymbol{u}_{T}^c )  \in \mathbb{R}^{n_c d}$. We define $\boldsymbol{\theta} := (\boldsymbol{\theta}_1, \dots, \boldsymbol{\theta}_C ) \in \mathbb{R}^{q}$, where $q = \sum_{c = 1}^C q_c$. Hence, $\boldsymbol{\theta}$ is the vector of unknown parameters for all communities. Further, note that $\sum_{c = 1}^C \boldsymbol{y}_c = \boldsymbol{y}$, where $\boldsymbol{y} := \left (\boldsymbol{X}_{p + 1}, \dots, \boldsymbol{X}_{T} \right ) \in \mathbb{R}^{nd}$ is the `response vector' for the whole model, and we set $n = \min (n_c)$. Let $\bmat{R}$ be the matrix that results from concatenating by columns each $\bmat{R}_c$ for $c = 1, \dots, C$ and further expand the linear model to obtain

\begin{equation} \label {eq: linear model}
    \boldsymbol{y} = \bmat{R} \boldsymbol{\theta} + \boldsymbol{u},
\end{equation}
where $\boldsymbol{u} = \sum_{c = 1}^C \boldsymbol{u}_c$ is a vector of independent and identically distributed zero-mean random variables.
\newpage
Therefore, we can estimate $\boldsymbol{\theta}$ by ordinary least-squares. Following the above and exploiting~\eqref{eq: linear model}, we use the estimator
\begin{singlespace}
    \begin{equation} \label{eq: ls estimator}
    \hat{\boldsymbol{\theta}} = \left ( \bmat{R}^T \bmat{R} \right )^{-1} \bmat{R}^T \boldsymbol{y},
\end{equation}
throughout this work.
    \begin{remark} \label{rem: estimators}
    If $\boldsymbol{X}_t$ is a stationary {\em $\comGNAR$ GNAR} model as in~\eqref{eq: community GNAR structural representation}, then the $\bmat{R}_c$ have zeros in different rows (non-overlapping communities). Hence, $\bmat{R}$ is orthogonal by blocks, and since $\mathrm{cov} (\hat{\boldsymbol{\theta}} ) = (\bmat{R}^T \Sigma_{\boldsymbol{u}} \bmat{R})^{-1}$ and $\Sigma_{\boldsymbol{u}}$ is diagonal, the $\hat{\boldsymbol{\theta}}_c$ are uncorrelated and non-zero entries in the precision matrix belong to the same community (i.e., $\mathrm{cov} (\hat{\alpha}_{kc}, \hat{\alpha}_{k \tilde{c}}) = 0$ if $c \neq \tilde{c}$). Further, if we assume that $\boldsymbol{u}_t \sim \mathrm{N}_{\boldsymbol{d}} (\boldsymbol{0}, \sigma^2_{\boldsymbol{u}} \bmat{I_d} )$, then $\hat{\boldsymbol{\theta}}$ is the conditional maximum likelihood estimator and $\hat{\boldsymbol{\theta}}_c = \left ( \bmat{R}_c^T \bmat{R}_c \right )^{-1} \bmat{R}_c^T \boldsymbol{y}$ are community-wise independent.
\end{remark}
\end{singlespace}
The estimator given by \eqref{eq: ls estimator} is an M-estimator (see Proposition~\ref{prop: m-estimator} in Section~\ref{apend: proof of asymp normality summary} of the supplementary material). Note that by Remark~\ref{rem: estimators}, it is possible to estimate model parameters separately and simultaneously for $\comGNAR$ GNAR models. This allows us to use more observations for communities with a smaller maximum lag, remove unnecessary predictors from each $c$-community linear model in~\eqref{eq: linear model}, and perform estimation in parallel and/or by different strategies, e.g., some communities could be regularised and/or estimated using more robust estimators. We leave these extensions for future work. Further, adapting $\hat{\boldsymbol{\theta}}$ to a generalised least-squares setting is straightforward. Suppose that $\mathrm{cov} (\boldsymbol{u}) = \bmat{\Sigma}_T$, then we estimate $\boldsymbol{\theta}$ by generalised least-squares, i.e., $\hat{\boldsymbol{\theta}}_{\mathrm{gls}} = \left ( \bmat{R}^T \bmat{\Sigma}_T^{-1} \bmat{R} \right )^{-1} \bmat{R}^T  \bmat{\Sigma}_T^{-1} \boldsymbol{y},$ where $\bmat{\Sigma}_T \in \mathbb{R}^{n d \times n d}$ is a valid covariance matrix, e.g., $ \bmat{\Sigma}_T = \bmat{I_d} \otimes \bmat{\Sigma}_{\boldsymbol{u}}$ can be block-diagonal with entries $\mathrm{cov}(\boldsymbol{u}_t) = \bmat{\Sigma}_{\boldsymbol{u}}$ at all times $t$, where each block corresponds to one community. 

\subsection{Statistical efficiency for GNAR model specifications}
\label{subsec: ls estimator error bounds}
Asymptotic results for GNAR models can be obtained by properly reformulating the model into a VAR setting and directly translating results from VAR theory. If estimation is more elaborate, then consistency can be directly derived from VAR theory, e.g., by controlling the conditional sum of squares at each iteration of the estimation algorithm if it is composed of least-squares type loss functions.
However, by Remark~\ref{rem: GNAR from global to local and var} (b, c) we see that GNAR models are far more parsimonious than VAR models. Thus,  this subsection presents nonasymptotic $\ell_2$-norm upper bounds for our conditional least-squares estimator given by \eqref{eq: ls estimator}, which further evidences the benefits of GNAR models if their use is valid.
\begin{singlespace}
    \begin{assumptions} \label{assum: theorem assumptions}
    Assume that $\boldsymbol{X}_t \in \mathbb{R}^d$ is a stationary $\GNAR$ process given by \eqref{eq: GNAR compact structural representation} from which we observe a
    realisation of length $T$  and perform estimation by least-squares, i.e., $\boldsymbol{\hat{\theta}}$ as given by \eqref{eq: ls estimator}. By conditioning, we obtain a fixed design setting for \eqref{eq: linear model}, i.e., we can express the data as $\boldsymbol{y} - \bmat{R} \boldsymbol{\theta_0} = \boldsymbol{u}$, where $\boldsymbol{\theta_0} \in \mathbb{R}^q$ is the {\em ``true''} vector of parameters, and we can assume that
    \begin{enumerate}
        \item \label{asum: A1} Recall the
        $c$-community matrices $\bmat{R}_c$,
        which were defined just after
        \eqref{eq:ccommunitydesign}.
        The matrices $(\bmat{R}_c^T \bmat{R}_c)$,
         are such that their smallest eigenvalues are lower bounded by a positive constant, i.e., there exists a $\tau_c > 0$ s.t. $\lambda_{\min} (\bmat{R}_c^T \bmat{R}_c) \geq \{ \tau_c |K_c| (T - p_c) \}^{1/2}$ for all $c \in [C]$.
        \item \label{asum: A2}There exists a positive constant $\Gamma$ such that the $\ell_2$-norm of all columns in $ \bmat{R}$ are upper bounded by $ n^{-1/2} \Gamma$, i.e., there exists $\Gamma > 0$ s.t. $\max_{j \in [q]} \{ || [\bmat{R}]_{\cdot j} ||_2 \} \leq n^{-1/2} \Gamma$, where $n := \min_{c \in [C]} \{ |K_c| (T - p_c) \}$ {\em (i.e., the number of rows in $\bmat{R}$)}.
        \item \label{asum: A3} The residuals $\boldsymbol{u}$ are independent sub-Gaussian white noise, i.e., for each $u_{i, t}$ there exists a $\sigma_i^2 > 0$ such that $E \{ \exp(z u_{i, t}) \} \leq \exp(z^2 \sigma_i^2 / 2)$ for all $z \in \mathbb{R}$, and $u_{i, t}$ are uncorrelated at all times $t = p + 1, \dots, T$ for all $i \in \mathcal{K}$. 
    \end{enumerate}
    \end{assumptions}
\end{singlespace}
Assumption {\em A.\ref{asum: A1}} is standard in (conditional) least-squares settings, it is equivalent to assuming that $\mathrm{rank}(\bmat{R}) = d(T - p)$, if satisfied, then $\boldsymbol{\hat{\theta}}$ is the unique least-squares solution for \eqref{eq: linear model}, i.e., $\boldsymbol{\hat{\theta}} = \underset{\boldsymbol{\theta}}{\mathrm{argmin}} \{ || \boldsymbol{y} - \bmat{R} \boldsymbol{\theta}||_2^2 \}$. 
Note that {\em A.\ref{asum: A1}} exhibits the challenges that collinearity creates, if the columns of $\bmat{R}$ are correlated, then $\tau$ should be closer to zero and equal to zero if one column is a linear combination of the other ones, i.e., $\mathrm{rank}(\bmat{R}) < q$. 
\par
Assumption  {\em A.\ref{asum: A2}} states that no column in $\bmat{R}$ presents a large variance, i.e., the $\ell_2$-norm can be bounded uniformly across columns. Interestingly, this analysis can be made more precise by considering that the $\ell_2$-norm of each column is the estimated variance of a stationary process. Thus, it is related to the spectral distribution of the process, and that $(\bmat{R}^T \bmat{R})$ is related to the autocovariance matrix of the process. We leave these investigations for future work.
\par
Assumption  {\em A.\ref{asum: A3}} is the most idealised, or at least as idealised as assuming that we observe a stationary process. It, loosely speaking, states that the residuals are stochastically dominated by a normal distribution, hence, outliers (i.e., large deviations from zero) are as unlikely as when sampling from a Gaussian distribution. This condition can be relaxed and studied under sub-exponential and/or sub-Weibull assumptions. However, the resulting bounds will not be as tight, thus, it is likely that estimation requires a more robust method than least-squares. Moreover, a common assumption is that $E(u_{i, t}^4) < +\infty$, i.e., a finite fourth moment; see, e.g.,~\cite{luth, brockwell_davis}, which does not admit residuals with infinite variance, e.g., Cauchy distributed residuals, and could strongly thin-out the assumed heavier tail. We state our results below.

\begin{singlespace}
    \begin{theorem} \label{th: l2 bounds for ls estimator}
    Assume that A.\ref{asum: A1} and A.\ref{asum: A2} in Assumptions \ref{assum: theorem assumptions} hold and define the following
    \begin{itemize}
        \item $q_c :=  \left \{ p_c + ( 1 + | \mathcal{I}_c |) \sum_{k = 1}^{p_c} s_{k} (c) \right \}$, i.e., the number of unknown parameters in $K_c$, 
        \item $q_{[C]} := \max_{c \in [C]} \left \{ q_c \right \}$, i.e., the maximum number of unknown parameters,
        \item $p := \max_{c \in [C]} (p_c)$, i.e., the largest active lag across all communities,
        \item $\mathcal{K}_{[C]} := \min_{c \in [C]} \{ |K_c| \}$, i.e., the size of the smallest community,
        \item  $\tau := \min_{c \in [C]} (\tau_c) $, i.e., the largest constant for uniformly lower bounding the smallest eigenvalues of all $(\bmat{R}_c^{T} \bmat{R}_c)$ {\em (i.e., $\lambda_{\min} (\bmat{R}_c^T \bmat{R}_c) \geq \{ \tau \mathcal{K}_{[C]} (T - p) \}^{1/2}$ for all $c \in [C]$)}.
    \end{itemize}
    Then, we have that
    \begin{equation} \label{eq: deterministic bound}
        || \boldsymbol{\hat{\theta}} - \boldsymbol{\theta_0} ||_2
        \leq 2 \{\tau \mathcal{K}_{[C]} (T - p) \}^{-1/2} \left ( C q_{[C]} \right )^{1/2} || \bmat{R}^T \boldsymbol{u} ||_{\infty}, 
    \end{equation}
    further, if we assume that A.\ref{asum: A3} holds, then for any $\delta > 0$ we have that 
    \begin{equation} \label{eq: sub-G noise bound}
         || \boldsymbol{\hat{\theta}} - \boldsymbol{\theta_0} ||_2 \leq  2 ( C q_{[C]} )^{1/2} \sigma_{\boldsymbol{u}} \Gamma \sqrt{2} \left [ \left \{ \log \left ( \sum_{c = 1}^C q_c \right ) \{ \tau \mathcal{K}_{[C]} (T - p)\}^{-1} \right \}^{1/2} + \delta \right ],
    \end{equation}
    holds with probability at least 
    \begin{equation} \label{eq: sub-G noise bound two}
        1 - 2 \exp \left [ - \delta^2 \left \{ \tau \mathcal{K}_{[C]} (T - p) \right \}  \right],
    \end{equation}
   where $\sigma_{\boldsymbol{u}}^2 = \underset{i \in \mathcal{K}}{\max} (\sigma_i^2)$ is the largest variance among nodal white noise processes $u_{i, t}$.
\end{theorem}
\end{singlespace}
Theorem \ref{th: l2 bounds for ls estimator} highlights the strengths and weaknesses of GNAR models. For the bound in \eqref{eq: deterministic bound} to be meaningful, it is necessary that no community is empty, and that the linear model obtained by conditioning the realisation is not underdetermined (i.e., $\mathcal{K}_{[C]} > 0$ and $\tau >0$). Also, as the number of communities and/or interactions increases the bound becomes looser, i.e., larger $ \sum_{c = 1}^C q_c $ result in a looser bound for a given realisation $\{ \boldsymbol{X}_t\}$ of length $T$. A small $\tau$ also results in a looser bound, i.e., the more correlated the columns of $\bmat{R}$ are, the looser the bound is. Thus, the autocorrelation structure that the network induces, i.e., $\bmat{W}_t$ and $\bmat{S}_r$,  affect how small $\tau$ is and how tight \eqref{eq: deterministic bound} is for a given network time-series, and how quickly the probability given by \eqref{eq: sub-G noise bound two} approaches one. The scaling in Theorem \ref{th: l2 bounds for ls estimator} is expected for linear models; see \cite{portnoy_consistency_m_estimator}. Further, note that \eqref{eq: deterministic bound} suggests that care must be taken with respect to residuals, i.e., in the presence of large deviations from zero the bound is likely extremely loose. Remark~\ref{rem: GNAR statistical efficiency} highlights the attractive parsimony properties of global-$\alpha$ GNAR models, if these are valid.
\begin{singlespace}
\begin{remark} \label{rem: GNAR statistical efficiency}
        For a stationary {\em global-$\alpha$} $\GNAR(p, [s_k])$ model, \eqref{eq: deterministic bound} simplifies to 
        \begin{equation*}
            || \boldsymbol{\hat{\theta}} - \boldsymbol{\theta_0} ||_2
        \leq 2 \sqrt{q} \left \{ \tau |\mathcal{K}| (T - p) \right \}^{-1/2} || \bmat{R}^T \boldsymbol{u} ||_{\infty},
        \end{equation*}
        which under $A.\ref{asum: A3}$, for any $\delta > 0$, gives that
        \begin{equation*}
            || \boldsymbol{\hat{\theta}} - \boldsymbol{\theta_0} ||_2 \leq 2  \sigma_{\boldsymbol{u}} \Gamma \sqrt{2} \left [ \left \{ q \log \left ( q \right ) \left \{ \tau |\mathcal{K} | (T - p) \right \}^{-1} \right \}^{1/2} + \delta \right ],
        \end{equation*}
        where $q = p + \sum_{k = 1}^p s_k$, holds with probability at least $1 - 2 \exp [ - \delta^2 \{ \tau |\mathcal{K}| (T - p)  \} / q ]$.
    \end{remark}
\end{singlespace}
Interestingly, the ratio, 
$ \left \{ C q_{[C]}  \log(C q_{[C]}) \right \} / \left \{ \tau \mathcal{K}_{[C]} (T - p) \right \},$
 between problem and sample size becomes smaller as GNAR models become more parsimonious, which by Theorem \ref{th: l2 bounds for ls estimator} makes estimation more efficient and precise. Thus, Theorem \ref{th: l2 bounds for ls estimator} shows the increase in statistical efficiency as the model becomes more parsimonious, whereas, by Remark \ref{rem: GNAR from global to local and var}, we see that as $\GNAR$ models become less parsimonious, then the bounds become looser. Hence, if a GNAR model is valid, then increasing dimension {\em \underline{can be an asset}} when performing estimation, which enables modelling of very {\em high-dimensional} (network) time series. Further, by Theorem \ref{th: l2 bounds for ls estimator} and considering that $\hat{\boldsymbol{\theta}}$ is an M-estimator we obtain the following result.
\begin{singlespace}
    \begin{corollary} \label{cor: conditional ls estimarot const.}
        Assume that A.\ref{asum: A1}, A.\ref{asum: A2} and A.\ref{asum: A3} in Assumptions \ref{assum: theorem assumptions} hold and that
        $$\lim_{T \to \infty} \left \{ C q_{[C]}  \log(C q_{[C]}) \right \} \left \{ \tau \mathcal{K}_{[C]} (T - p) \right \}^{-1} = 0, \enspace a.s., $$
        i.e., the ratio between problem size and observations goes to zero. Let $\bmat{Z}_{\boldsymbol{t}} \in \mathbb{R}^{d \times q}$ be the matrix with columns equal to the autoregresssive, between community and within community terms in \eqref{eq: community GNAR structural representation}, i.e., 
            $$\bmat{Z}_{\boldsymbol{t}} \boldsymbol{\theta} = \sum_{c = 1}^{C} \left \{ \boldsymbol{\alpha}_{c} (\boldsymbol{X}_t) + \boldsymbol{\beta}_{c} (\boldsymbol{X}_t) + \boldsymbol{\gamma}_{c} (\boldsymbol{X}_t) \right \},$$ 
            for all $t \geq p + 1$. Then, for all $\delta > 0$ we have that
        \begin{equation} \label{eq: consistency}
            \lim_{T \to \infty} \mathbf{Pr} \left ( || \boldsymbol{\hat{\theta}} - \boldsymbol{\theta_0} ||_2 > \delta \right ) = 0,
        \end{equation}
            i.e., the conditional least-squares estimator given by \eqref{eq: ls estimator} is consistent. Further, since $\boldsymbol{\hat{\theta}}$ is a consistent {\em M-estimator}, we also have that
        \begin{equation} \label{eq: asym normality}
            \sqrt{|\mathcal{K}| (T - p)} \left ( \boldsymbol{\hat{\theta}} - \boldsymbol{\theta_0} \right ) \overset{\boldsymbol{\mathrm{d}}}{\longrightarrow} \mathrm{N}_{\boldsymbol{q}} \left ( 0, \bmat{\Sigma}_{\boldsymbol{\hat{\theta}}} \right ),
        \end{equation}
        i.e., $( \boldsymbol{\hat{\theta}} - \boldsymbol{\theta_0} )$ is asymptotically normal with zero-mean and covariance $\{ |\mathcal{K}| (T - p)\}^{-1} \bmat{\Sigma}_{\boldsymbol{\hat{\theta}}}$, 
        \begin{equation*}
            \bmat{\Sigma}_{\boldsymbol{\hat{\theta}}} = |\mathcal{K}|^{-1} \left \{ E (\bmat{Z}_{\boldsymbol{t}}^T \bmat{Z}_{\boldsymbol{t}}) \right \}^{-1} \{ E (\bmat{Z}_{\boldsymbol{t}}^T \bmat{\Sigma}_{\boldsymbol{u}} \bmat{Z}_{\boldsymbol{t}}) \}  \{E (\bmat{Z}_{\boldsymbol{t}}^T \bmat{Z}_{\boldsymbol{t}}) \}^{-1},
        \end{equation*} 
        where $\bmat{\Sigma}_{\boldsymbol{u}}$ is a diagonal white noise covariance. If we further assume that $\bmat{\Sigma}_{\boldsymbol{u}} = \sigma_{\boldsymbol{u}}^2 \bmat{I_d} $, then 
        $$\bmat{\Sigma}_{\boldsymbol{\hat{\theta}}} = \sigma^2_{\boldsymbol{u}} / |\mathcal{K}| \, \{E (\bmat{Z}_{\boldsymbol{t}}^T \bmat{Z}_{\boldsymbol{t}}) \}^{-1},$$ 
        and $\hat{\sigma}_{\boldsymbol{u}}^2 := \{ |\mathcal{K}| (T - p)\}^{-1} || \boldsymbol{y} - \bmat{R} \boldsymbol{\hat{\theta}}||_2^2$
        is a consistent estimator of $\sigma^2_{\boldsymbol{u}}$.
        \end{corollary}
\end{singlespace}
See Sections~\ref{apen: error bounds proof} and~\ref{apen: proof of Cor asym norm} in the supplementary material for a proof of Theorem~\ref{th: l2 bounds for ls estimator} and Corollary \ref{cor: conditional ls estimarot const.}. We note that known consistency and asymptotic results from univariate time series analysis can be derived immediately from Theorem~\ref{th: l2 bounds for ls estimator} and Corollary~\ref{cor: conditional ls estimarot const.}, albeit obtained without using invertibility theory arguments; see Chapter 7 in~\cite{brockwell_davis}. Remark~\ref{rem: yule walker an} connects GNAR processes with autoregressive processes.

\begin{singlespace}
\begin{remark} \label{rem: yule walker an}
    Let $X_t \in \mathbb{R}$ be a stationary zero-mean $\mathrm{AR}(p)$ process, which we analyse as a one-node network $\GNAR(p)$ process. Then, by Corollary \ref{cor: conditional ls estimarot const.} we see that $\hat{\boldsymbol{\theta}}$ given by \eqref{eq: ls estimator} is consistent and asymptotically normal, i.e., 
    \begin{equation*}
         \sqrt{(T - p)} \left ( \boldsymbol{\hat{\theta}} - \boldsymbol{\theta_0} \right ) \overset{\boldsymbol{\mathrm{d}}}{\longrightarrow} \mathrm{N}_p \left ( 0,\sigma^2_{\boldsymbol{u}} \bmat{\Gamma}^{-1}_p  \right ),
    \end{equation*} 
    where $\bmat{\Gamma}_p \in \mathbb{R}^{p \times p}$ is the autocovariance matrix of $X_t$ up to lag $p - 1$. This recovers a known results for conditional least-squares estimators for scalar autoregressive processes.
\end{remark}
\end{singlespace}
\newpage
Theorem \ref{th: l2 bounds for ls estimator} and Corollary \ref{cor: conditional ls estimarot const.} highlight that our GNAR framework provides an easily generalisable family of linear models (i.e., network-informed models are easier to generalise and estimate). For example, robust estimators based on~\eqref{eq: ls estimator} could be computed by a different choice of loss function with respect to linear model~\eqref{eq: linear model}, e.g., Huber or least--absolute deviation. This will be useful in settings where realised network time-series contain outliers,
for example, in econometrics. We close by simply mentioning that these are typical results for linear models; see \cite{portnoy_asymp_normality}, albeit differently expressed within the highly {\em parsimonious} GNAR framework, which enables us to study the unknown parameters as {\em vectors} rather than as {\em matrices} and seeing `dimension as an asset' (as mentioned above).


\section{Model selection}
\label{sec: model selection}
We perform model selection by examining Corbit (correlation-orbit) and R-Corbit plots as suggested by \cite{corbit_paper}. These plots show how the network autocorrelation (NACF) and partial NACF (PNACF) functions decay or cut off with respect to lag and $r$-stage order. \cite{corbit_paper} introduced NACF and PNACF, which are network-enabled extensions of the autocorrelation and partial autocorrelation functions from univariate time series analysis, which are useful tools for model selection; see~\cite{brockwell_davis}.
\par
The R-Corbit plot allows us to quickly visualise and compare the autocorrelation structure of {\em different communities}. Points in a R-Corbit plot correspond to $\mathrm{(p)nacf}_c (h, r)$, where $h$ is the $h$th lag, $r$ is $\rstage$ depth and $c \in [C]$ denotes community $K_c$. These values are placed in circular rings, where the mean value, i.e., $C^{-1} \sum_{c = 1}^{C} \pnacf_c (h, r)$, is shown at the centre of each subring of points for a given lag and $\rstage$. The numbers outside
the outermost ring indicate lag and each inner ring of circle rings indicates $\rstage$ depth starting from one. Both plots show (P)NACF equal to zero (i.e., $\nacf = 0$ or $\pnacf = 0$) at the centre of the entire plot to aid comparison.
\par
\begin{figure}[H]
   \centering
   \includegraphics[scale=0.45]{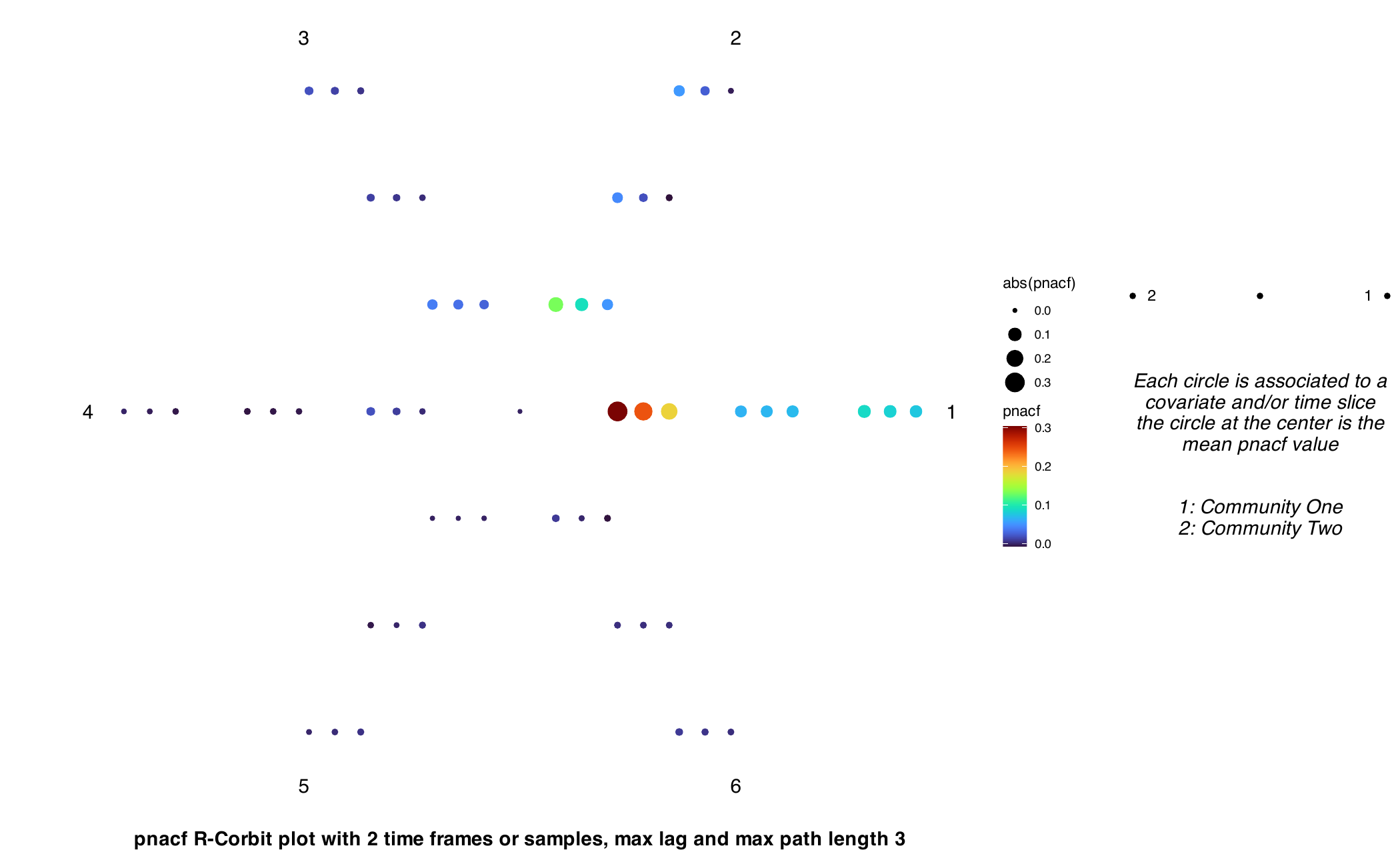}
   \caption{
       R-Corbit plot for 1000 long realisation generated by a stationary $\comGNAR$ $\GNAR \left ( [1, 2], \{ [1], [1, 1] \}, 2 \right )$, where the underlying network is \textbf{fiveNet}, $K_1 = \{2, 3, 4\}$ and $K_2 = \{1, 5 \}$; see Figure \ref{fig: 2-communal fiveNet}. The maximum lag is equal to six and maximum $\rstage$ depth is equal to three. The PNACF cut-offs are $(1, [1])$ for $K_1$ and $(2, [1, 1])$ for $K_2$.
        }
   \label{fig: R-Corbit pnacf}
\end{figure}
\begin{figure}[H]
    \centering
    \begin{subfigure}{0.48\textwidth}
            \includegraphics[width=\textwidth]{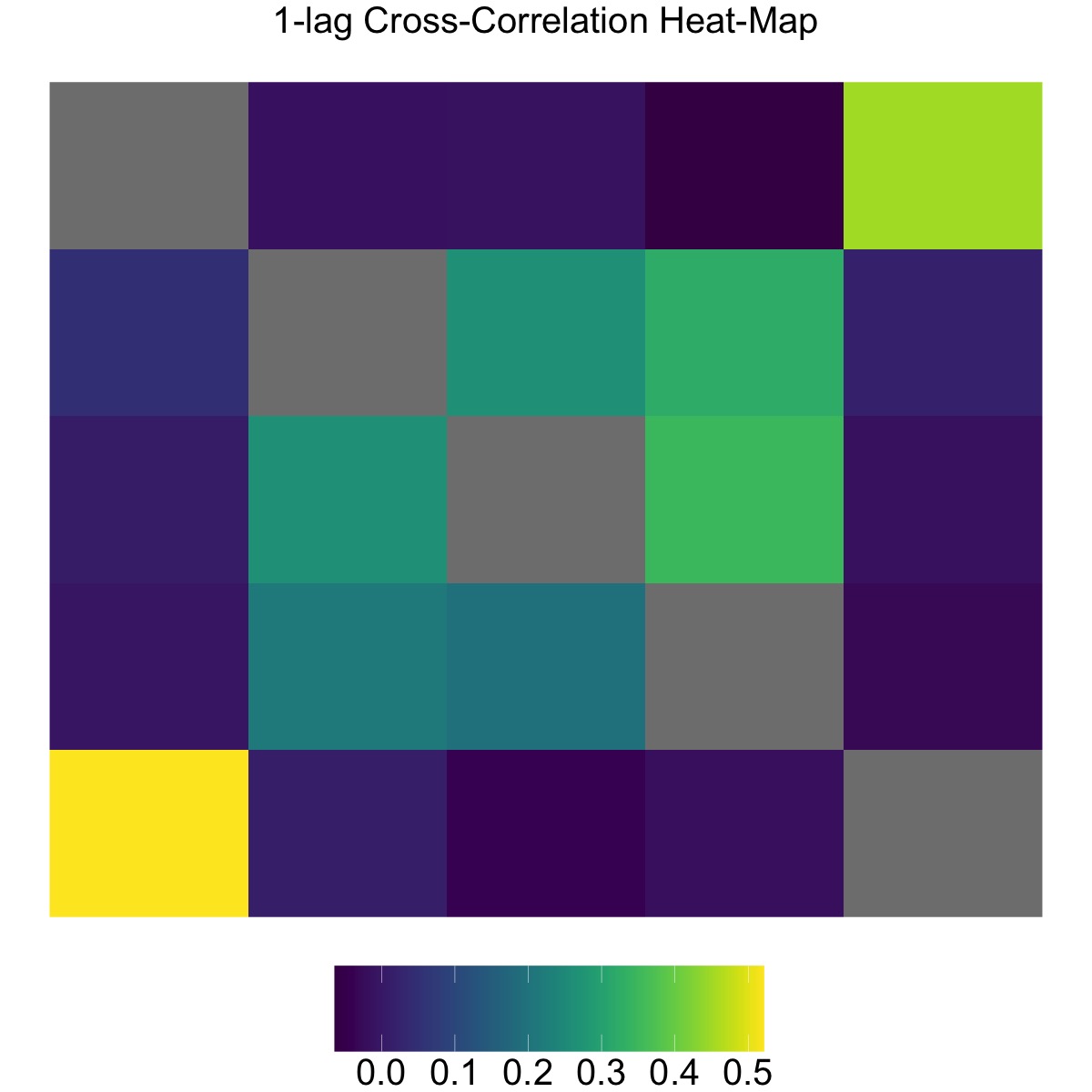}
            \caption{}
            \label{fig: one-lag cross-corr}
        \end{subfigure}
        \hfill
        \begin{subfigure}{0.48\textwidth}
            \includegraphics[width=\textwidth]{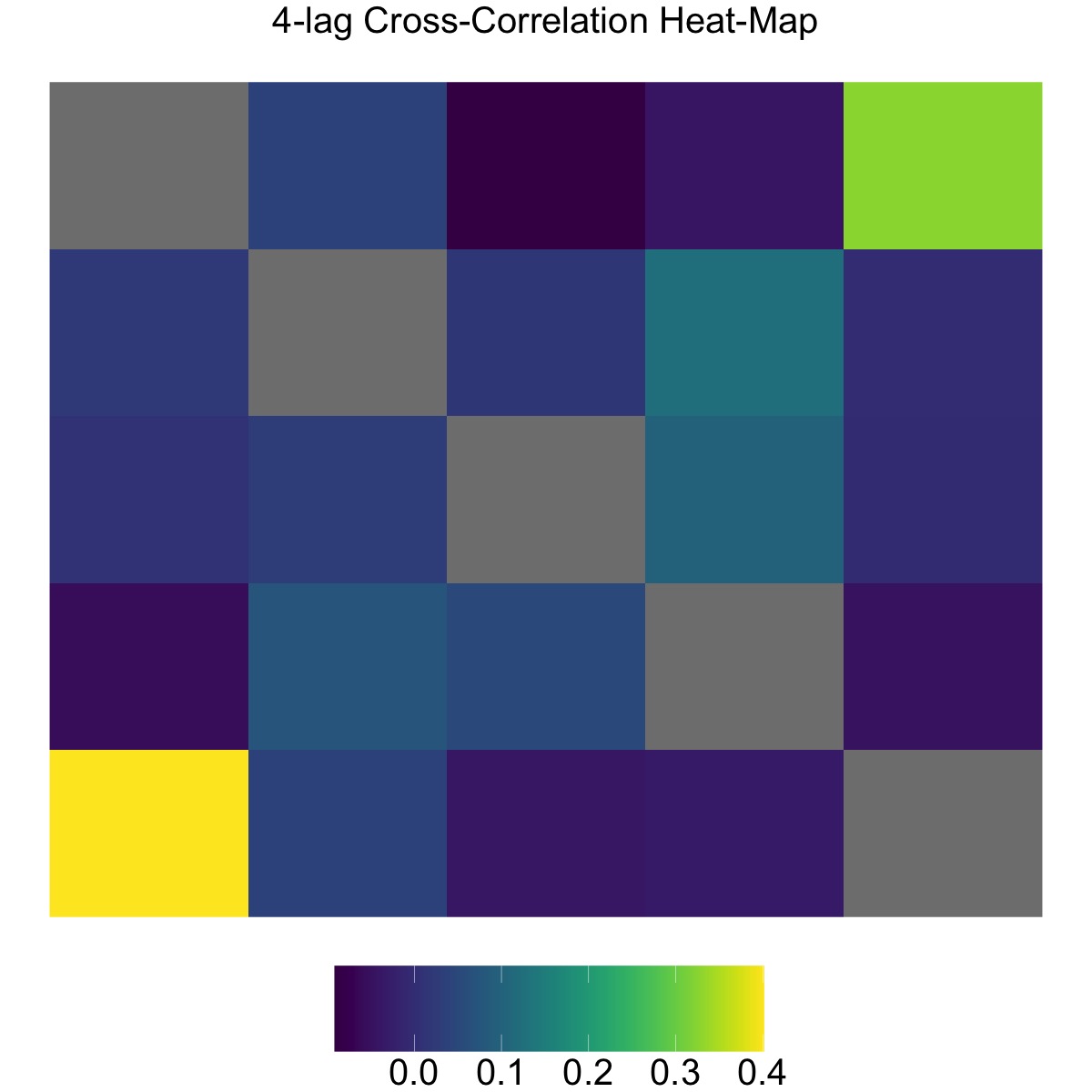}
            \caption{}
            \label{fig: four-lag cross-corr}
        \end{subfigure}
        \caption{Cross-correlation heat-map plots for a 1000 long realisation generated by a stationary $\comGNAR$ $\GNAR \left ( [1, 2], \{ [1], [1, 1] \}, 2 \right )$, where the underlying network is \textbf{fiveNet}, $K_1 = \{2, 3, 4\}$ and $K_2 = \{1, 5 \}$; see Figure \ref{fig: 2-communal fiveNet}. Figure~\ref{fig: one-lag cross-corr} shows the one-lag cross-correlation matrix, and \ref{fig: four-lag cross-corr} the four-lag one. The plots highlights the two communities and that the sample cross-correlation between nodes in different communities is close to zero. The diagonal corresponds to
        node-wise autocorrelations, which are coloured grey so as not to
        detract from the main object of study: the cross-correlations.
        }
        \label{fig: cross-corrs}
\end{figure}

As an example, we simulate a one-thousand-long realisation generated by a stationary $\comGNAR$ GNAR model given by \eqref{eq: communal alpha diff orders}. The \textbf{fiveNet} network (included in the \texttt{GNAR} package) is the underlying network; see \cite{gnar_package} and Figure \ref{fig: 2-communal fiveNet}. The simulated data come from a stationary $\GNAR \left ( [1, 2], \{ [1], [1, 1] \}, 2 \right )$, where $K_1 = \{2, 3, 4\}$ and $K_2 = \{1, 5 \}$ are the two communities, and we fix parameters to be $\alpha_{1, 1} = 0.23, \, \beta_{1, 1, 1} = 0.47$, and $\alpha_{1, 2} = 0.20, \, \beta_{1, 1, 2} = 0.30, \, \alpha_{2, 2} = 0.18, \, \beta_{2, 1, 2} = 0.27$. Suppose we no longer know the model order but still know the communities. We study the lag and $\rstage$ order, i.e., the pair $(p_c, [s_{k_c}] )$, by comparing the community correlation structure via the R-Corbit plot in Figure \ref{fig: R-Corbit pnacf}. The R-Corbit plot in Figure \ref{fig: R-Corbit pnacf} shows that the PNACF cuts-off after the second lag for both communities. Further, it shows that the PNACF cuts-off after the first lag and $\rstage$ one for $K_1$, i.e., it suggests the order $(1, [1])$ for $K_1$, and after the second lag at the first $\rstage$ for lags one and two for $K_2$, i.e., it suggests the pair $(2, [1, 1])$ for $K_2$. In this case, the R-Corbit in Figure~\ref{fig: R-Corbit pnacf} reflects the underlying model order.
Figure~\ref{fig: cross-corrs} compares the cross-correlations between nodes in $K_1$ and $K_2$, it reflects the known clusters and highlights the different lag orders
(in that the centre $3\times 3$ block values have decayed compared to the lag-one
ones, but the cross-1-5 lag correlation has not).
\begin{singlespace}
\begin{remark}
We restrict our analysis of model selection to the {\em PNACF} using Corbit plots, which are more interpretable and computationally efficient than {\em AIC} or {\em BIC}. See~\cite{corbit_paper} for a detailed discussion on the use of Corbit plots for network time series model selection and their advantages over {\em AIC} and {\em BIC}.
Nevertheless, one could use {\em AIC} or {\em BIC}, of course.
\end{remark}
\end{singlespace}

We end this section by noting that Corbit and R-Corbit plots enable quick identification of model order, different correlation structures between communities, and other behaviours such as seasonality and trend for network time series.

\section{Modelling presidential elections in the USA}
\label{sec: application}
We study the twelve presidential elections in the USA from 1976 to 2020. The data are obtained from the MIT Election Data and Science Lab ({\tt doi.org/10.7910/DVN/42MVDX}). We model the network time series $(\boldsymbol{X}_t, \mathcal{G})$, where $X_{i, t}$ is the percentage of votes for the Republican nominee in the $i$th state (ordered alphabetically) for election year $t \in \{1976, 1980, \dots, 2020 \}$ as a $\comGNAR$ GNAR model. The underlying network, $\mathcal{G}$, where $i \in \{1, \dots, 51 \}$ and $d = 51$, is built by connecting states that share a land border.  Alaska and Hawaii are not connected to the network, however, each one is related to other states by sharing the community-wise $\alpha$ coefficients. Following \cite{state_classification_science_journal, state_classification_econ_enquiry}, we classify states into communities based on the percentage of elections won by either party, i.e., each node (state) is classified as either \textit{Red}, \textit{Blue} or \textit{Swing}, which are identified by the covariates $c \in \{1, 2, 3 \}$ and communities: $i \in K_1$ if the Republican nominee won at least $75\%$ of elections, $i \in K_2$ if the Democrat nominee won at least $75\%$ of elections, and $i \in K_3$ if neither one won at least $75\%$ of elections; see Figure~\ref{fig: usa net}.
\subsection{Analysis of network autocorrelation and model selection}
\label{subsec: network autocorrelation analysis}
\begin{figure}[H]
    \centering
    \includegraphics[scale=0.45]{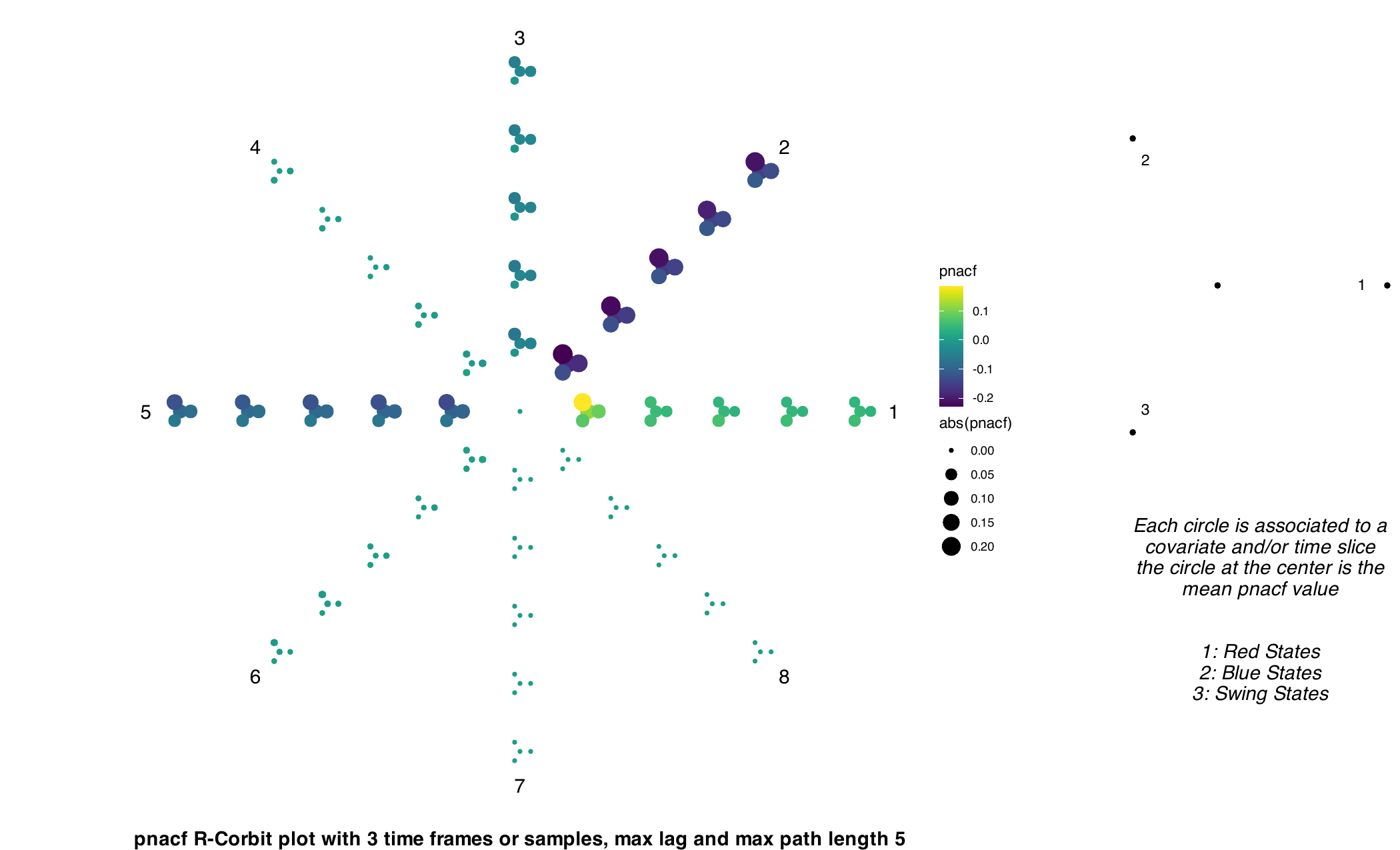}
    \caption{PNACF R-Corbit plot of the series $\boldsymbol{X}_{t}$.
    The underlying network is the USA state-wise network; see Figure \ref{fig: usa net}, and Section \ref{sec: model selection} for plot description. 
    }
    \label{fig: R-Corbit pnacf votes}
\end{figure}
The R-Corbit plot in Figure \ref{fig: R-Corbit pnacf votes} shows that the PNACF is positive at the first lag, negative and strongest at the second lag, considerably smaller at lags three and four, and interestingly, appears to be strong at the fifth lag across all $r$-stages, and
cuts off after the fifth. At the first lag, the PNACF is visually smaller after the first $r$-stage. At both the second and fifth lags, the PNACF actually
decays as $r$-stage grows, but is not easy to see from
the plot. It  does not cut-off at any $r$-stage. The positive correlation at lag one corresponds
to  elections in which a president is running for reelection,
and that network affects influence said election. Remarkably, the strong negative
correlation at the second lag across all $r$-stages suggests a change in the system, which we interpret as alternating between Republican and Democrat nominees once the incumbent president has completed their eight-year term. This has been the case with the exception of Jimmy Carter (1976--1980), George Bush (1988--1992) and Donald J. Trump (2016--2020). Interestingly, the exception cases correspond to elections where an incumbent president was running for reelection (i.e., half-terms). We believe that the fifth lag might be identifying with these exception cases. 
\par
Based on the PNACF analysis in Figure~\ref{fig: R-Corbit pnacf votes}, and that the lag five PNACF values are
smaller than the lag two ones, and given that we only
have data from twelve elections,
we initially compare models of order two.
Hence, we  trial
a~\mbox{community-$\alpha$} $\mathrm{GNAR} (2, \{[1, 0]\}, 3)$ model. The parsimonious nature of GNAR models allows us to estimate multiple models and analyse autocorrelation up to lag eight, in spite of having twelve observations for a vector of dimension fifty-one.
\par
\newpage
Our chosen model requires at least nine observations for estimation, in contrast, a VAR(1) model for these data requires at least 51 observations. Thus, it is not possible to fit a VAR($p$) to these data without performing regularisation
and limiting lag order to $p = 1$. Table \ref{tab: usa vote estimate}
shows results for our chosen GNAR model.

\begin{table}[H]
    \caption{Estimated coefficients for the standardised network time series of vote percentages for the Republican nominee in presidential elections in the USA from 1976 to 2020. The fit is a \mbox{community-$\alpha$} $\mathrm{GNAR} (2, \{[1, 0]\}, 3)$.
    [Recall  from~\eqref{eq: com alpha gnar vector wise} that the coefficients are $\alpha_{kc}$ and
    $\beta_{krc}$, where $k$ is lag,
    $r$ is network $r$-stage and $c$ is
    community.]}
    \centering
    \begin{tabular}{lrrrrrrrrr}
        & $\hat{\alpha}_{11}$ & $\hat{\beta}_{111}$ & $\hat{\alpha}_{21}$ & $\hat{\alpha}_{12}$ & $\hat{\beta}_{112}$ & $\hat{\alpha}_{22}$ & $\hat{\alpha}_{13}$ & $\hat{\beta}_{113}$ & $\hat{\alpha}_{23}$ \\
        \hline
        Estimate & 0.393 & 0.183 & -0.593 & 0.558 & 0.069 & -0.351 & 0.905 & -0.747 & -0.591 \\
        Std. Err. & 0.104 & 0.200 & 0.066 & 0.108 & 0.178 & 0.063 & 0.089 & 0.162 & 0.061 \\
    \end{tabular}
\label{tab: usa vote estimate}
\end{table}

Table~\ref{tab: usa vote estimate} shows that if~\eqref{eq: asym normality} is valid, then a
{\em Swing} states network effect
null hypothesis, $H_0: \beta_{113}=0$, is
rejected and
statistically significant at the 5\%
level (the $t$-value is $-5.05$, giving
a \mbox{$p$-value} $< 0.001$), which suggests strong inter-state communication
within the {\em Swing} state community.
The network effect for {\em Red} and {\em Blue}
states do not reject the null hypothesis of
zero for $\beta_{111}, \, \beta_{112}$, even 
though the  estimate of the former, $\hat{\beta}_{111}$, is of a reasonable size.

\subsection{Analysing interactions between communities}
We are also interested in analysing whether there are differences in magnitude for interaction coefficients, and how these states vote after a particular result, e.g., do {\em Blue} states increase their vote for a Democrat nominee more than {\em Swing} states do after a Republican win? Modelling the presidential election time series as a community-$\alpha$ GNAR model with interactions enables us to shed light on these questions. Based on the network autocorrelation in Section~\ref{subsec: network autocorrelation analysis}, we select the following model orders: $(2, [1, 1], \{2 , 3\} )$ for {\em Red} states, $(2, [1, 1], \{1, 3\} )$ for {\em Blue} and $(2, [1, 1], \{1, 2\})$ for {\em Swing}. Table~\ref{tab: interaction coefficients} shows 
the lag two estimated interaction coefficients.
\begin{table}[H]
\centering
    \caption{
        Estimated lag-two interaction coefficients $\hat{\gamma}$ for the standardised network time series of vote percentages for the Republican nominee in presidential elections in the USA from 1980 to 2020. The fit is a community-$\alpha$ GNAR$(\boldsymbol{2}; \{ \boldsymbol{1} \}; [3]; \{2 , 3\}, \{1, 3\}, \{1, 2\} )$ with interactions among {\em Blue}, {\em Red} and {\em Swing} states. [Recall  from~\eqref{eq: community GNAR structural representation} that the interaction coefficients are $\gamma_{k,r,c:\tilde{c}}$, where $k$ is lag, $r$ is network $r$-stage, $c, \tilde{c}$ are two different communities and $K_{\tilde{c}}$ is affecting $K_c$.]
    }
    \begin{tabular}{lrrrrrr}
         &  $\hat{\gamma}_{2, 1, 1:2}$ & $\hat{\gamma}_{2, 1, 1:3}$ & $\hat{\gamma}_{2, 1, 2:1}$ & $\hat{\gamma}_{2, 1, 2:3}$ & $\hat{\gamma}_{2, 1, 3:1}$ & $\hat{\gamma}_{2, 1, 3:2}$ \\
         \hline
         Estimate & -0.480 & -0.490 & -1.140 & -0.785 & -0.300 & -0.169 \\
         Std. Err. & 0.396 & 0.214 & 0.332 &  0.219 & 0.246 & 0.211
    \end{tabular}
    \label{tab: interaction coefficients}
\end{table}
Interestingly, the interaction coefficient estimates for the influence
of {\em Blue} and {\em Red} states on {\em Swing} states ($\gamma_{2, 1, 3:1},
\gamma_{2, 1, 3:2}$) are smaller in magnitude than those for the
influence of {\em Swing} states on {\em Blue} and {\em Red} states. Further, $\hat{\gamma}_{2, 1, 2:1} = -1.14$ (effect of {\em Red} states on {\em Blue} ones), suggests that {\em Blue} states vote as strongly for the Democrat nominee as {\em Red} states voted for the Republican nominee eight-years before. Whereas, {\em Red} states react half as strongly ($\hat{\gamma}_{2, 1, 1:2} = -0.48$) to {\em Blue} states. This suggests that voters in {\em Blue} states tend to flock (vote strongly for a Democrat nominee), whilst, {\em Red} states do not react as strongly. Moreover, {\em Swing} states are influenced less strongly by votes from {\em Red} and {\em Blue}, which suggests these states might be less polarised. Remarkably, {\em Blue} states have the largest polarisation as measured by the estimated interaction coefficient magnitude (i.e., $|\hat{\gamma}_{2, 1, 2:1}|$).

\subsection{Comparison with alternative models}
\label{subsec: model comparisons}
Table \ref{tab: model comparisons} compares our choice with alternative models. Before model fitting we centre and standardise the data to bring it closer to our modelling assumptions ({\em homoscedastic} noise and zero-mean). That is, we compare the model's forecasting performance using the standardised time series $Y_{i, t} := \sqrt{11} \{ \sum_{t = 1}^{11} (X_{i, t} - \overline{X}_{i} )^2  \}^{-1/2} \, (X_{i, t} - \overline{X}_i)$, where $\overline{X}_i = (1/ 11) \, \sum_{t = 1}^{11} X_{i, t}$. Each model produces a one-step ahead standardised forecast $\Hat{Y}_{i, 12}$, which we use to compute the one-step ahead forecast as $ \Hat{X}_{y_i, 12} = \{ \sum_{t = 1}^{11} (X_{i, t} - \overline{X}_{i} )^2  \}^{1/2} \, \Hat{Y}_{i, 12} / \sqrt{11} + \overline{X}_{i}$. For completeness, we also fit models to the observed time series $\{ \boldsymbol{X}_t \}$ (not standardising) and analyse the root mean-squared prediction error (RMSPE) for model output $\Hat{\boldsymbol{X}}_t$, i.e., for the $\boldsymbol{Y}_t$, RMSPE$(\Hat{\boldsymbol{Y}}_{12}) = \{ \sum_{i = 1}^{51} ( X_{i, 12} -  \Hat{X}_{y_i, 12} )^2 / 51 \}^{1/2}$, and for $\Hat{\boldsymbol{X}}_{t}$, RMSPE$(\Hat{\boldsymbol{X}}_{12}) = \{ \sum_{i = 1}^{51} ( X_{i, 12} -  \Hat{X}_{i, 12} )^2 / 51 \}^{1/2}$. For reproducibility, we fix the random seed, i.e., \texttt{set.seed(2024)}, before model fitting; see supplementary material for details. We fit sparse VAR(2) using \texttt{sparsevar}; see \cite{sparsevar}, and CARar(2), where forecasts are computed as a global-$\alpha$ GNAR as suggested by~\cite{corbit_paper}, using \texttt{CARBayesST}; see \cite{carbayes}. 
We remark that CARar forecasts are equivalent to global-$\alpha$ GNAR ones if estimation is done by generalised least-squares with a spatially-informed covariance matrix; see~\cite{corbit_paper}. 

\begin{table}[H]
    \centering
    \caption{Model comparison of root mean-squared prediction error (RMSPE). $\mathrm{GNAR}_{\dots}$ models increase in complexity from global (glo), to community (com), to interactions (int), to local (loc). Sp. VAR: sparse VAR(2). CARar: Conditional temporal autoregressive (spatially-informed noise covariance). $\# ||\boldsymbol{\theta}||_0$ is number of non-zero parameters. 
    }
     \begin{tabular}{lrrrrrr}
        & $\mathrm{GNAR}_{\mathrm{glo}}$ & $\mathrm{GNAR}_{\mathrm{com}}$ & $\mathrm{GNAR}_{\mathrm{int}}$ & $\mathrm{GNAR}_{\mathrm{loc}}$ & Sp. VAR & CARar \\
        \hline
         RMSPE$(\Hat{\boldsymbol{X}}_{12})$ & {\bf 2.34} & 2.94 & 3.53 & 24.95 & 72.57 & 3.60 \\
        RMSPE$(\Hat{\boldsymbol{Y}}_{12})$ & 3.32 & 4.44 & 3.67 & 5.22 & 4.07 & {\bf 2.75} \\
        $\# ||\boldsymbol{\theta}||_0$ & {\bf 6} & 12 & 42 & 106 & 176 & 6(+51)
    \end{tabular}
    \label{tab: model comparisons}
\end{table}

Remarkably, the global-$\alpha$ GNAR$(2, [1, 0])$ model produces the smallest RMSPE, and is the only model that {\em outperforms} the Naive (previous observation) forecast, which has a RMSPE equal to {\bf 2.45}. This could be related to the small sample error bounds in Theorem~\ref{th: l2 bounds for ls estimator}, since the highly parsimonious global-$\alpha$ GNAR model can be estimated precisely, even in (ultra) high-dimensional settings. We see that parsimonious GNAR models produce accurate forecasts with a limited amount of data. Further, all GNAR models perform comparatively well with standardised and non-standardised data, whereas, not standardising the data severely impacts the performance of sparse VAR.

\subsection{Discussion}
\label{subsec: application discussion}

We use the community-$\alpha$ with interactions GNAR model for analysing the effects within and between state-wise communities in the USA during presidential elections from 1976 to 2020. Our main interest is identifying noteworthy differences in the voting patterns among each community.  However, it is worth mentioning that, as expected, {\em Red} states vote mostly for the Republican nominee, {\em Blue} states for the Democrat nominee, and that {\em Swing} states play the deciding role, given that the community-wise mean percentage of votes for the Republican nominee is $50.80\%$. However, somewhat surprisingly, based on the global-$\alpha$ GNAR model forecasting results, our analysis suggests that communities behave more similarly than expected, i.e., there appear to be (spatio-temporal) global and communal network effects. Further, {\em Blue} states vote in a block-wise manner, and have the strongest reaction (interaction coefficient magnitude) by increasing the number of votes for a Democrat nominee with respect to previous elections. {\em Red} and {\em Swing} states react at $50\%$ and at $25\%$ the strength of {\em Blue} states.
Our autocorrelation analysis suggests that alternating between a Democrat and Republican president every eight-years might not be the only governing dynamic. Following the 2016 election, the new pattern could be that parties take turns at the presidency every four rather than eight years. Interestingly, our results corroborate the analysis of political polarisation in Twitter and Reddit longitudinal networks in~\cite{10.1093/jrsssa/qnad008}. We highlight that the observed data is overly sparse, thus, inferences and forecasts should be treated with caution, and that these claims require further investigation.
\par
Put simply, the parsimonious GNAR model allows us to investigate the dynamics of presidential election cycles in the USA. Our analysis highlights the eight-year long presidential cycle, a possible change in the system after the 2016 election, the deciding role {\em Swing} states have, and behaviour differences
between {\em Blue} and {\em Red} states.

\section{Conclusion}
\label{sec: conclusion}
We presented a \mbox{community}-$\alpha$ GNAR model
with
different lag and \mbox{$r$-stage} orders for different communities and inter-community
interactions, which are not constrained to be symmetric. Such models can detect differences in the dynamics of communities present in a network time series. Further, their parsimonious framework allows us to analyse high-dimensional data
with reasonable length time series,
such as the presidential election data studied in
Section~\ref{sec: application}. However, it is important to be clear about its limitations: our
models require knowledge of the number and membership of communities and assume that the network time series is stationary. Future work will focus on relaxing these limitations and developing a clustering algorithm aimed at exploiting the community-wise correlation structure in the network time series by assigning each node to the community that minimises some goodness-of-fit criterion based on the network autocorrelation function (NACF).
\par
Another area for further investigation might be
to connect GNAR with mixed effects models for longitudinal data, since incorporating {\em serial correlation} structure into the model is an important consideration. This correlation structure can be modelled as a community-$\alpha$ GNAR process, which is capable of describing different model orders, interactions and network effects in each community. 
\newpage
Essentially, extending GNAR to nonstationary biological network linked and spatio-temporal settings could enable modelling of complex (tensor-valued) longitudinal mixed type data, and provide insights into the dynamics of complex processes that are relevant in engineering, public policy and science.
\\
\newline
The code for fitting \mbox{community-$\alpha$} GNAR models is available at \href{https://github.com/dansal182/community_GNAR}{GitHub link}.

\begin{singlespace}
\section*{Acknowledgments}
We gratefully acknowledge the following support: Nason from EPSRC NeST Programme grant EP/X002195/1; 
Salnikov from the UCL Great Ormond Street Institute of Child Health, NeST, Imperial College London, 
the Great Ormond Street Hospital DRIVE Informatics Programme and the Bank of Mexico.
Cortina-Borja supported by the NIHR Great Ormond Street Hospital Biomedical Research Centre. The views expressed are those of the authors and not necessarily those of the EPSRC.

\begin{center}
    {\Large\bf Supplementary Material}
\end{center}
\begin{itemize}
    \item {\bf Appendix~\ref{sec: simulation study}:} GNAR simulations and plot of finite-sample error bounds.
    \item {\bf Appendix~\ref{apen: one-lag diff data}:} Analysis of one-lag differenced presidential election data.
    \item {\bf Appendix~\ref{sec: sty cond proof}:} Proof of stationarity conditions for GNAR models.
    \item {\bf Appendix~\ref{apen: error bounds proof}:} Proof of Theorem ~\ref{th: l2 bounds for ls estimator} (finite-sample error bounds).
    \item {\bf Appendix~\ref{apen: proof of Cor asym norm}:} Proof of Corollary~\ref{cor: conditional ls estimarot const.} (asymptotic normality and consistency).
    \item {\bf Appendix~\ref{apen: notation}:} Notation summary.
    \item {\bf GitHub link:} \texttt{\href{https://github.com/dansal182/community_GNAR}{https://github.com/dansal182/community\_GNAR}}.
\end{itemize}

\bibliographystyle{agsm}
\bibliography{references}
\end{singlespace}

\appendix
\setcounter{equation}{0}
\counterwithin{equation}{section}

\newpage

\section{Simulating multiple GNAR processes}
\label{sec: simulation study}
This section explores our methods via multiple simulation studies. For illustration purposes we employ the USA state-wise network; see Figure \ref{fig: usa net}, as the underlying structure, and analyse multiple realisations from different network time series models. Subsequently, we use the model selection and diagnostic tools suggested by \cite{corbit_paper} for selecting model order for a given network time-series and we use Akaike's information criteria (AIC) to assess goodness of fit.
\par
We simulate one thousand instances of different community-$\alpha$ GNAR processes using the USA state-wise network
with time-varying weights of period four
(i.e., each weight function is seasonal with a four year period). We illustrate Theorem~\ref{th: l2 bounds for ls estimator} by looking at the $\ell_2$-error-norm of the estimator~\eqref{eq: ls estimator} and a
true parameter vector,
which we change at every iteration. All simulated GNAR processes in this subsection are given by
\begin{multline} \label{eq: seed GNAR model for simulations}
    \boldsymbol{X}_t =  \boldsymbol{\alpha} \left ( \boldsymbol{X}_t; [3, 2, 1], [3] \right ) +  \boldsymbol{\beta} \{\boldsymbol{X}_t; \{[2, 2, 1], [3, 2], [3] \}, [3] \} + \cdots \\
    \boldsymbol{\gamma} \{ \boldsymbol{X}_t; \{(3), (3), (1, 2)\}, [3] \} + \boldsymbol{u}_t,
\end{multline} 
where $\boldsymbol{u}_t \sim \mathrm{N}_d (0, \bmat{I_d})$ are independent at all times $t$, and $d = 51$.
Model~\eqref{eq: seed GNAR model for simulations} is a community-$\alpha$ GNAR with interactions for three communities and fifty-one nodes. The model order for community one is $(3, [2, 2, 1], \{ 3 \})$ and interactions with community three, i.e., $\gamma_{\cdot, \cdot, 1:2} \equiv 0$, for community two it is $(2, [3, 2], \{ 3 \})$ and interactions with community three, i.e., $\gamma_{\cdot, \cdot, 2:1} \equiv 0$, and for community three it is $(1, [3], \{1, 2 \})$ and interactions with communities one and three (all possible $r$-stage neighbours). The time-varying weights are given by
\begin{align} \label{eq: time-varying weights}
    w_{ij} (t) &= \{1 + \cos(t \pi / 2) + 0.1 \} 2^{-d_{ij} / 2}\, \mathbb{I} (j \in K_1), \nonumber \\
    w_{ij} (t) &= \{1 + \sin(t \pi / 2) + 0.1 \} 2^{-d_{ij} } \,\mathbb{I} (j \in K_2), \\
     w_{i j} (t) &= \{1 + \cos(t \pi / 4) \sin(t \pi / 4)  + 0.1 \} 2^{-d_{ij} } \,\mathbb{I} (j \in K_3), \nonumber
\end{align}
where $d_{ij} \in \{1, \dots, 11 \}$ is the shortest finite path length between nodes $i$ and $j$ in the USA state-wise network. Each simulated realisation is produced from
a valid $\boldsymbol{\theta_0}$, chosen to ensure
a stationary simulated process. The experiment is summarised in psuedo-code in Algorithm \ref{alg: simulation study} below.

\begin{figure}
    \centering
    \includegraphics[scale=0.40]{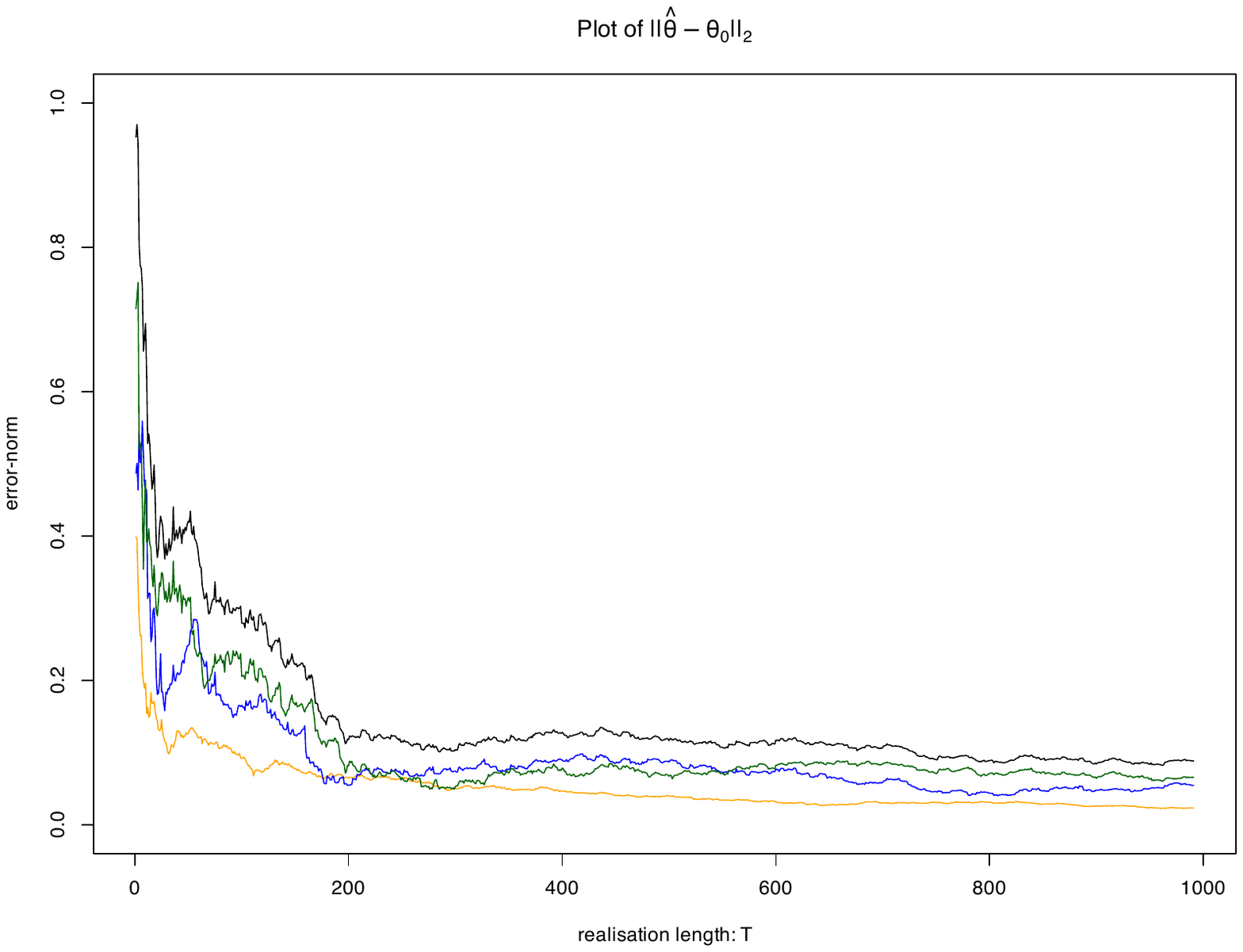}
    \caption{Multiple simulations of stationary GNAR processes given by \eqref{eq: seed GNAR model for simulations}. Each simulated realisation has a different length $T = 1, \dots, 1000$, and {\em ``true''} parameter vector $\boldsymbol{\theta_0}$; see Algorithm \ref{alg: simulation study}. Each curve is $\ell_2$-error-norm, {\color{orange} orange} is for $K_1$, set of nodes identified
    as {\em Red states}, {\color{blue} blue} is for $K_2$,
    ({\em Blue states}), {\color{ForestGreen} green} is for $K_3$, ({\em Swing states}), and {\color{black} black} is for the whole model. }
    \label{fig: l2 error sims}
\end{figure}

\begin{algorithm}[H]
\caption{Simulation study}
\label{alg: simulation study}
\begin{algorithmic}[1]
    \State \texttt{set.seed(2024)}
    \For{$T \in \{1, \dots, 1000 \}$}
        \State $\boldsymbol{\theta_0} \gets \texttt{simulate}(\theta)$ \Comment{Simulate a stationary $\boldsymbol{\theta_0}$, model order given by \eqref{eq: seed GNAR model for simulations}}
        \State $\bmat{W}_t \gets w_{ij} (t)$ for $t = 1, \dots, T$ \Comment{weight functions are given by \eqref{eq: time-varying weights}}
        \State $\{ \boldsymbol{X}_t \} \gets \texttt{community\_GNARsim}(\boldsymbol{\theta_0}, T)$ \Comment{Simulate $T$ time-steps of \eqref{eq: seed GNAR model for simulations}, with $\boldsymbol{\theta_0}$}
        \State $ \hat{\boldsymbol{\theta}} \gets \left ( \bmat{R}^T \bmat{R} \right )^{-1} \bmat{R}^T \boldsymbol{y}$ \Comment{Compute least-squares estimator given by \eqref{eq: ls estimator} }
        \For{$c = 1, 2, 3$}
            \State $\delta^{(T)}_c \gets \normx{\hat{\boldsymbol{\theta}}_c -  \boldsymbol{\theta_0}_{, c} }_2$ \Comment{Compute community $\ell_2$-error-norm}
        \EndFor
        \State $\delta^{(T)} \gets \normx{\hat{\boldsymbol{\theta}} -  \boldsymbol{\theta_0} }_2 $ \Comment{Save $\ell_2$-error-norm for the whole model}
    \EndFor
    \State \textbf{return:} $\{ (\delta^{(T)}_1, \delta^{(T)}_2, \delta^{(T)}_3, \delta^{(T)} ) \}$
\end{algorithmic}
\end{algorithm}
Figure~\ref{fig: l2 error sims} shows the results. Code for replicating the study can be found in the CRAN \texttt{GNAR} package in due course.

We close this subsection by noting that all four error curves in Figure \ref{fig: l2 error sims} decrease jointly towards zero as realisation length $T$ increases. This illustrates the results in Theorem \ref{th: l2 bounds for ls estimator}, and shows the reasonable performance of $\hat{\boldsymbol{\theta}}$ for realisations shorter that one-hundred time-steps with thirty-five unknown parameters. 
Table \ref{tab: sim fitted model} shows the estimated coefficients for our chosen model $\GNAR  \left ( [1, 2], \{ [1], [1, 1] \}, 2 \right )$.

\begin{table}[H]
   \centering
   \caption{
    Mean estimator values for only 10 simulations, each of length 100 coming from a $\GNAR  \left ( [1, 2], \{ [1], [1, 1] \}, 2 \right )$. {\bf Mean Est.} is the mean estimator value, {\bf Sim. Sd.} is the standard deviation of the 10 simulations (standard error), and {\bf True Value} is the known parameter. The simulations are computed using random seeds 1983 to 1993 in \texttt{R}.
   }
   \begin{tabular}{lrrrrrr}
         & $\hat{\alpha}_{11}$ & $\hat{\beta}_{111}$ & $\hat{\alpha}_{12}$ & $\hat{\beta}_{112}$ & $\hat{\alpha}_{22}$ & $\hat{\beta}_{212}$ \\
        \hline
        {\bf Mean Est.} & 0.280 & 0.177 & 0.222 & 0.302 & 0.115 & 0.180 \\
        {\bf Sim. Sd.} & 0.062 & 0.109 & 0.061 & 0.098 &  0.083 & 0.127 \\
        {\bf True Value} & 0.270 & 0.180 & 0.250 & 0.300 & 0.120 & 0.200
   \end{tabular}
   \label{tab: sim fitted model}
\end{table}
Table \ref{tab: sim fitted model} shows the reasonable performance of our estimator $\hat{\boldsymbol{\theta}}$.

\section{Analysis of one-lag differenced data}
\label{apen: one-lag diff data}
The estimated coefficients in Table~\ref{tab: usa vote estimate} do not satisfy the stationarity conditions in Corollary~\ref{cor: conditions for stationary processes}, however, these are sufficient, but not necessary.

\begin{figure}[H]
    \centering
    \includegraphics[scale=0.45]{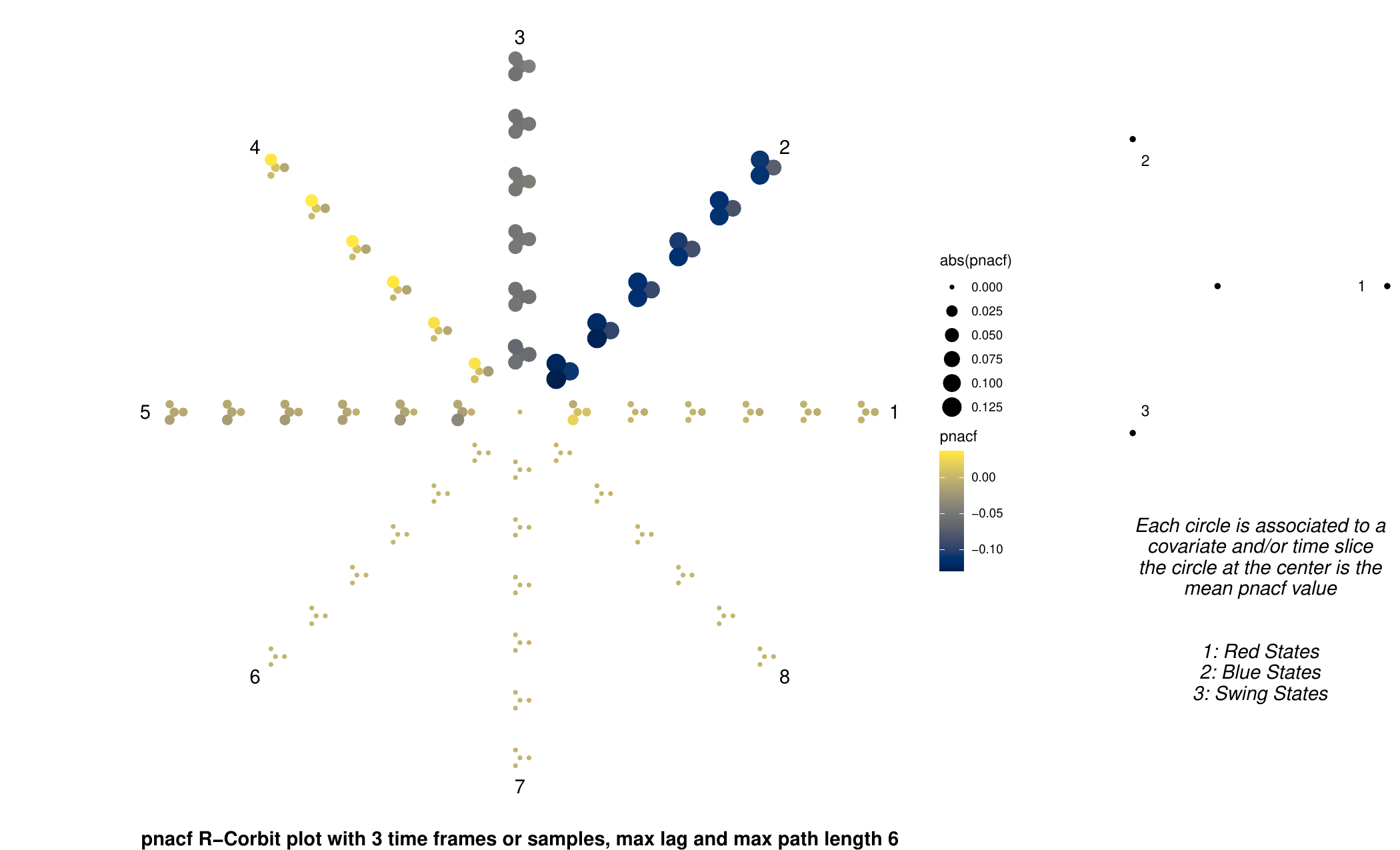}
    \caption{R-Corbit plot for the one-lag differenced network time series $\boldsymbol{X}_t - \boldsymbol{X}_{t - 1}$. Each $X_{i, t}$ is the difference in the percentage of votes for the Republican nominee in the current election minus the previous one from 1980 to 2020. The communities are {\em Red, Blue} and {\em Swing} states; see Figure \ref{fig: usa net}.}
    \label{fig: R-Corbit pnacf one-lag diff}
\end{figure}

 Hence, it is prudent to additionally study the
  one-lag differenced series. The R-Corbit plot in Figure~\ref{fig: R-Corbit pnacf one-lag diff} suggests that one-lag differencing removes correlation across all $r$-stages for the first lag, and that the PNACF
 seems to cut off after lag three, and that it does not cut-off at particular $r$-stages. We hence fit a community-$\alpha$ $\mathrm{GNAR}(3, \{[1, 1, 1]\}, 3)$ to the one-lag differenced data with estimates shown in Table~\ref{tab: usa vote one-lag diff estimate}.
\begin{table}[H]
\caption{Approx. statistically significant coefficients at the $5\%$ level estimated with the standardised one-lag differenced network time series $(\tilde{\boldsymbol{X}}_{t}, \mathcal{G})$ of vote percentages for the Republican nominee in presidential elections in the USA from 1980 to 2020. The fit is a \mbox{community-$\alpha$} $\mathrm{GNAR} (3, \{[1, 1, 1]\}, 3)$.
[Recall  from~\eqref{eq: com alpha gnar vector wise} that the coefficients are $\alpha_{kc}$ and
    $\beta_{krc}$, where $k$ is lag,
    $r$ is network $r$-stage and $c$ is
    community.]}
    \centering
    \begin{tabular}{lrrrrrrrrrr}
        & $\hat{\alpha}_{11}$ & $\hat{\beta}_{111}$ & $\hat{\alpha}_{21}$ & $\hat{\alpha}_{31}$ & $\hat{\alpha}_{12}$ & $\hat{\beta}_{112}$ & $\hat{\alpha}_{22}$ & $\hat{\alpha}_{32}$ & $\hat{\alpha}_{23}$ & $\hat{\alpha}_{33}$ \\ 
        \hline
        Estimate & -0.27 & 0.59 & -0.48 & -0.37 & -0.40 & 0.57 & -0.57 & -0.36 & -0.45 & -0.24 \\
        Std. Err. & 0.12 & 0.25 & 0.10 & 0.11 & 0.13 & 0.23 & 0.11 & 0.12 & 0.09 & 0.11
    \end{tabular}
\label{tab: usa vote one-lag diff estimate}
\end{table}


    Interestingly, {\em Blue} and {\em Red} network effects remain after one-lag differencing, whereas, {\em Swing} state network effects are not approx statistically significant at the $5\%$ level. This suggests that state-wise communities might have different second-order effects.

\newpage
\section{Stationary \texorpdfstring{\mbox{community-$\alpha$}}{} GNAR models}

\label{sec: sty cond proof}

This is an immediate consequence of stationarity conditions for local-$\alpha$ GNAR models. We proceed to show Corollary \ref{cor: conditions for stationary processes}.
\begin{proof}

We express the sum of stationary community processes as a \mbox{local-$\alpha$} GNAR process as follows.
\begin{enumerate}
    \item Let $\phi_{i, k} = \alpha_{k, c} \mathbb{I} (i \in K_c)$, where we set $\alpha_{k, c} = 0$ if $k > p_c$,
    \item $\zeta_{i, k, r} = \beta_{k, r, c} \mathbb{I} (i \in K_c)$, where we set $\beta_{k, r, c} = 0$ if $k > p_c$ or $ s_{k} (c) > r^*_c$,
    \item $\eta_{i, k, c:\tilde{c}} = \gamma_{k, r, c:\tilde{c}} \mathbb{I} (i \in K_c)$, where we set $\gamma_{k, r, c:\tilde{c}} = 0$ if $k > p_c$, or $ s_{k} (c) > r^*_c$, or $\tilde{c} \notin \mathcal{I}_{c}$.
\end{enumerate}
Hence, the community with the largest lag will have the biggest number of non-zero autoregressive coefficients, and might have the most non-zero $\rstage$ neighbourhood coefficients. Set $p = \max_{c \in  [C]} ( [p_c])$ and $s_k = \max_{c \in  [C]} \{ [s_{k} (c)] \} $ for each $k = 1, \dots, p$. Then, the node-wise representation of the $\comGNAR$ GNAR model is given by
\begin{equation} \label{eq: node wise com ganr}
    X_{i, t} = \sum_{k = 1}^{p} \left \{ \phi_{i, k} X_{i, t - k} + \sum_{r = 1}^{s_k} \left ( \zeta_{i, k, r} Z_{i, t - k}^{r, c} + \sum_{\tilde{c} \in \mathcal{I}_c} \eta_{i, k, r} Z_{i, t - k}^{r, c:\tilde{c}} \right ) \right \}
    + u_{i, t},
\end{equation}
where $u_{i, t}$ are independent white noise and $Z_{i, t - k}^{r, c}$, $Z_{i, t - k}^{r, c:\tilde{c}}$ are the $c$-community $\rstage$ neighbourhood regressions. For all lags $k \in \{1, \dots, p\}$ we have that for each $r \in \{1, \dots, s_k\}$ the node-wise within community $\rstage$ neighbourhood regression coefficients for $X_{i, t}$ satisfy
$$
    \sum_{r = 1}^{s_k} |\zeta_{i, k, r} | = \sum_{r = 1}^{s_k} |\beta_{k, r, c}| \mathbb{I}(i \in K_c),
$$
and that the between community $\rstage$ neighbourhood regression coefficients for $X_{i, t}$ satisfy
$$ 
    \sum_{r = 1}^{s_k} \sum_{\tilde{c} \in \mathcal{I}_c} |\eta_{i, k, r} | =  \sum_{r = 1}^{s_k} \sum_{\tilde{c} \in \mathcal{I}_c}  |\gamma_{k, r, c:\tilde{c}}| \mathbb{I}(i \in K_c),
$$
and that the autoregressive coefficients satisfy
$$
    \sum_{k = 1}^p |\phi_{i, k}| = \sum_{k = 1}^p |\alpha_{k, c}| \mathbb{I} (i \in K_c).
$$
By construction we have that for all $X_{i, t}$ such that $i \in K_c$ the within community $\rstage$ neighbourhood coefficients are equal for all lags and at all $\rstage$s, i.e., if $i \in K_c$, then $\zeta_{i, k, r} = \beta_{k, r, c}$ otherwise $\zeta_{i, k, r} = 0$. Note that the same holds for between community coefficients, i.e., if $i \in K_c$, then $\eta_{i, k, r} = \gamma_{k, r, c:\tilde{c}}$ otherwise $\eta_{i, k, r} = 0$. Thus, the coefficients for each nodal time series $X_{i, t}$ satisfy
\begin{align} \label{eq: stat proof coeff expression}
    &\sum_{k = 1}^{p} \left \{ |\phi_{i, k} | + \sum_{r = 1}^{s_k} \sum_{c = 1}^C \left ( |\zeta_{i, k, r} | + |\eta_{i, k, r} | \right )  \right \}  \nonumber \\
    &= \mathbb{I} (i \in K_c) \sum_{k = 1}^{p} \left \{ |\alpha_{k, c} | + \sum_{r = 1}^{s_k} \left ( |\beta_{k, r, c} | +  \sum_{\tilde{c} \in \mathcal{I}_c}  |\gamma_{k, r, c:\tilde{c}}| \right )  \right \}.
\end{align}
Assume that 
\begin{equation} \label{eq: stat cond proof coeff conditions}
    \sum_{k = 1}^{p_c} \left \{ |\alpha_{k, c} | + \sum_{r = 1}^{s_k (c)} \left ( |\beta_{k, r, c} | +  \sum_{\tilde{c} \in \mathcal{I}_c}  |\gamma_{k, r, c:\tilde{c}}| \right )  \right \} < 1,
\end{equation}
holds for all covariates $c \in [C]$. Combining this with \eqref{eq: node wise com ganr} and \eqref{eq: stat proof coeff expression} we have that each local-$\alpha$ (i.e., $\phi_{i, k}$) nodal process satisfies
$$
     \sum_{k = 1}^{p} \left \{ |\phi_{i, k} | + \sum_{r = 1}^{s_k} \sum_{c = 1}^C \left ( |\zeta_{i, k, r} | + |\eta_{i, k, r} | \right )  \right \} < 1,
$$
for all $i \in \mathcal{K}$. Therefore, by Theorem \ref{th: stationary cond} the process given by \eqref{eq: node wise com ganr} is stationary. Recalling that \eqref{eq: node wise com ganr} is identical to the process given by \eqref{eq: community GNAR structural representation}, and that the process given by \eqref{eq: communal alpha diff orders} is a special case of \eqref{eq: community GNAR structural representation} finishes the proof.
\end{proof}
\newpage
\section{Nonasymptotic bounds}
\label{apen: error bounds proof}
We present a proof of Theorem \ref{th: l2 bounds for ls estimator}. We begin by recalling the fixed design we obtain by conditioning the observed realisation, recall from Section \ref{subsec: ls estimator} that a realisation of length $T \in \mathbb{Z}^+$ coming from a stationary $\comGNAR$ GNAR process can be expressed as the linear model \eqref{eq: linear model}, i.e., 
\begin{equation*}
    \boldsymbol{y} = \bmat{ R} \boldsymbol{\theta}_{0} + \boldsymbol{u},
\end{equation*}
where $\boldsymbol{u} = \sum_{c = 1}^C \boldsymbol{u}_c$ is a vector of independent zero-mean random variables, $\boldsymbol{y} \in \mathbb{R}^{d ( T - p)}$ is the response, $\bmat{R} \in \mathbb{R}^{d (T - p) \times q}$ is the design matrix, $\boldsymbol{\theta}_0 \in \mathbb{R}^q$, where $q = \sum_{c = 1}^C q_c$, and that we estimate the {\em ``true''} vector of parameters $\boldsymbol{\theta}_0 = (\boldsymbol{\theta}_{0, 1}, \dots, \boldsymbol{\theta}_{0, C})$ by ordinary least-squares.

\subsection{Lower bound for the community norm}
We begin by inspecting {\em A.\ref{asum: A1}}, recall that
\begin{equation*}
    \bmat{R} = \left [\bmat{R}_1 | \dots | \bmat{R}_C \right ],
\end{equation*}
and notice that each $c$-community design matrix has zeros in different row entries in $\bmat{R}_c$, hence, without loss of generality we can write 
\begin{equation*}
    \bmat{R}^T \bmat{R} = \begin{pmatrix}
                            \bmat{R}_1^T \bmat{R}_1 & \dots & \bmat{0} \\
                             \overset{\boldsymbol{\cdot}}{\underset{\cdot}{\cdot}} & _{^{\bullet} \bullet_{ \bullet}} &  \overset{\cdot}{\underset{\cdot}{\cdot}} \\
                            \bmat{0} & \dots  & \bmat{R}_C^T \bmat{R}_C
                        \end{pmatrix},
\end{equation*}
i.e., $\bmat{R}^T \bmat{R}$ is diagonal by blocks. 
\par
Thus, if $\mathrm{rank}(\bmat{R}_c ) = q_c$ for all $c \in [C]$ (i.e., each community linear model is not underdetermined), then a singular value decomposition of $\bmat{R} =  \bmat{U} \bmat{D} \bmat{V}^T$, where 
$\bmat{U} \in \mathbb{R}^{n \times n}$ is an orthonormal matrix, $\bmat{D} \in \mathbb{R}^{n \times q}$ is a rectangular diagonal matrix with $q > 0$ non-zero entries (positive singular values) in the diagonal, and $\bmat{V} \in \mathbb{R}^{q \times q}$ is an orthonormal matrix, can be expressed by blocks, i.e., $\bmat{R}_c = \bmat{U_c D_c V^T_c}$, and satisfies for all $\boldsymbol{\theta} \in  \mathbb{R}^{q}$ the following
\begin{align*}
    || \bmat{R} \boldsymbol{\theta} ||_2^2 &= \boldsymbol{\theta}^T (\bmat{R}^T \bmat{R}) \boldsymbol{\theta} \\
    &= \sum_{c = 1}^C \boldsymbol{\theta}^T_c (\bmat{R}^T_c \bmat{R}_c) \boldsymbol{\theta}_c \\
    &= \sum_{c = 1}^C \normx{\bmat{D_c V_c^T} \boldsymbol{\theta}_c}_2^2
\end{align*}
where $\bmat{D_c}$ and $\bmat{V_c^T}$ are the block matrices corresponding to community $K_c$. 
\newline
$\bullet$ By applying assumption {\em A.\ref{asum: A1}} we have that each non-zero entry in $\bmat{D_c}$ is lower bounded by $\{\tau_c |K_c| (T - p_c) \}^{1/2}$. Thus, we have that
\begin{equation*}
    \normx{\bmat{D_c V_c^T} \boldsymbol{\theta}_c}_2^2 \geq  \{\tau_c |K_c| (T - p_c) \}^{1/2} \normx{\bmat{V_c^T} \boldsymbol{\theta}_c}_2^2,
\end{equation*}
for all $c \in [C]$.
\newline
$\bullet$ Since $\bmat{V_c^T}$ is an orthonormal matrix, i.e., $\normx{\bmat{V_c^T} \boldsymbol{\theta}_c}_2 =  \normx{\boldsymbol{\theta}_c}_2$, we have that
\begin{equation*}
    \normx{\bmat{D_c V_c^T} \boldsymbol{\theta}_c}_2^2 \geq  \{\tau_c |K_c| (T - p_c) \}^{1/2} \normx{ \boldsymbol{\theta}_c}_2^2,
\end{equation*}
for all $c \in [C]$.
\newline 
$\bullet$ Define $\tau := \min_{c \in [C]} (\tau_c) $, and $\mathcal{K}_{[C]} := \min_{c \in [C]} \{ |K_c| \}$, i.e., the size of the smallest community, by the above, for all $c \in [C]$, we get the following lower bound
\begin{equation*}
      || \bmat{R}_c \boldsymbol{\theta}_c ||_2^2 = \normx{\bmat{D_c V_c^T} \boldsymbol{\theta}_c}_2^2 \geq \{\tau \mathcal{K}_{[C]} (T - p)\}^{1 / 2} \normx{ \boldsymbol{\theta}_c}_2^2 .
\end{equation*}
Thus, for all $\boldsymbol{\theta}_c \neq \boldsymbol{0}$, we have that
\begin{equation}\label{eq: upp bound proof, l2 norm lower bound svd}
     \{\tau \mathcal{K}_{[C]} (T - p)\}^{1 / 2} || \boldsymbol{\theta}_c||_2^2 \leq \{ \tau_c |K_c| (T - p_c) \}^{1 / 2} || \boldsymbol{\theta}_c||_2^2 \leq || \bmat{R}_c \boldsymbol{\theta}_c ||_2^2.
\end{equation}

\subsection{Upper bounding the community error vector norm}
Now recall that the estimator given by \eqref{eq: ls estimator} is the unique solution of 
\begin{equation*}
    \boldsymbol{\hat{\theta}} = \underset{\boldsymbol{\theta}}{\mathrm{argmin}} \{ || \boldsymbol{y} - \bmat{R} \boldsymbol{\theta}||_2^2 \}.
\end{equation*}
Hence, we have that 
\begin{equation*}
     || \boldsymbol{y} - \bmat{R} \boldsymbol{\hat{\theta}}||_2^2 \leq   || \boldsymbol{y} - \bmat{R} \boldsymbol{\theta_0}||_2^2,
\end{equation*}
which after expanding the left-hand side, recalling
that $\boldsymbol{u} = \boldsymbol{y} - \boldsymbol{R}\boldsymbol{\theta}$ and then
rearranging gives the {\em basic inequality}
\begin{equation*}
    || \bmat{R} ( \boldsymbol{\hat{\theta}} - \boldsymbol{\theta_0} ) ||_2^2 \leq 2 \langle \boldsymbol{u}, \bmat{R} ( \boldsymbol{\hat{\theta}} - \boldsymbol{\theta_0} ) \rangle.
\end{equation*}
Hence, by noting that each $ \boldsymbol{\hat{\theta}}_c$ corresponds to different entries in $ \boldsymbol{y}$ (i.e., $\boldsymbol{\xi}_c \odot \boldsymbol{\hat{\theta}}_c$ are orthogonal), we have that
\begin{equation} \label{eq: upp bound proof, optima inequality}
    \sum_{c = 1}^C || \bmat{R}_c ( \boldsymbol{\hat{\theta}}_c - \boldsymbol{\theta_0}_{, c} ) ||_2^2 \leq 2  \sum_{c = 1}^C \langle \boldsymbol{u}, \bmat{R}_c ( \boldsymbol{\hat{\theta}}_c - \boldsymbol{\theta_0}_{, c} ) \rangle,
\end{equation}
where $\boldsymbol{\theta_0}_{, c}$ are the
 values of $\boldsymbol{\theta_0}$ 
 corresponding to community $c$.

By equations~\eqref{eq: upp bound proof, l2 norm lower bound svd} and \eqref{eq: upp bound proof, optima inequality} for each community $K_c$, we have that
\begin{align} \label{eq: upp bound proof, error l2 norm bound}
     || \bmat{R}_c ( \boldsymbol{\hat{\theta}}_c - \boldsymbol{\theta_0}_{, c} ) ||_2^2 &\leq 2 \langle \boldsymbol{u}, \bmat{R}_c ( \boldsymbol{\hat{\theta}}_c - \boldsymbol{\theta_0}_{, c} ) \rangle \nonumber \\
     &= 2 \langle \bmat{R}_c^T \boldsymbol{u}, ( \boldsymbol{\hat{\theta}}_c - \boldsymbol{\theta_0}_{, c} ) \rangle \nonumber \\
     &\overset{(i)}{\leq} 2 ||  \bmat{R}_c^T \boldsymbol{u}||_{\infty} || ( \boldsymbol{\hat{\theta}}_c - \boldsymbol{\theta_0}_{, c} )||_1 \nonumber \\
     &\overset{(ii)}{\leq} 2 \left \{ p_c + \sum_{k = 1}^{p_c} s_{k} (c) ( 1 + | \mathcal{I}_c |) \right \}^{1/2} ||  \bmat{R}_c^T \boldsymbol{u}||_{\infty} || ( \boldsymbol{\hat{\theta}}_c - \boldsymbol{\theta_0}_{, c} )||_2,
\end{align}
above $(i)$ follows from Holder's inequality using the $\ell_1$- and $\ell_{\infty}$-norms as conjugates, and $(ii)$ follows from \mbox{$\ell_1-\ell_2$-norm} inequality in $\mathbb{R}^{q_c}$. By applying \eqref{eq: upp bound proof, l2 norm lower bound svd} to \eqref{eq: upp bound proof, error l2 norm bound} with respect to the error vector $( \boldsymbol{\hat{\theta}}_c - \boldsymbol{\theta_0}_{, c} )$ we have that
\begin{equation*}
     \{ \tau \mathcal{K}_{[C]} (T - p) \}^{1 / 2} ||( \boldsymbol{\hat{\theta}}_c - \boldsymbol{\theta_0}_{, c} )||_2^2 \leq 2  \left \{ p_c + \sum_{k = 1}^{p_c} s_{k} (c) ( 1 + | \mathcal{I}_{c} |) \right \}^{1/2} ||  \bmat{R}_c^T \boldsymbol{u}||_{\infty} || ( \boldsymbol{\hat{\theta}}_c - \boldsymbol{\theta_0}_{, c} )||_2,
\end{equation*}
or 
\begin{equation} \label{eq: upp bound proof, comm wise error bound}
    ||( \boldsymbol{\hat{\theta}}_c - \boldsymbol{\theta_0}_{, c} )||_2 \leq 2 \left \{ p_c + \sum_{k = 1}^{p_c} s_{k} (c) ( 1 + | \mathcal{I}_c |) \right \}^{1/2} \{ \tau \mathcal{K}_{[C]} (T - p) \}^{-1/2} ||  \bmat{R}_c^T \boldsymbol{u}||_{\infty}.
\end{equation}

\subsection{Jointly bounding the error for all communities}

\begin{proof}
Before bounding the error we note that each community-wise vector norm is the same if we augment the vector with zeros in the entries that correspond to members that do not belong to community $K_c$, i.e., $|| \boldsymbol{\theta}_c ||_2^2 = || \boldsymbol{\xi}_c \odot \boldsymbol{\theta} ||_2^2$, hence, by orthogonality of the $\boldsymbol{\xi}_c$, we have that $ || \boldsymbol{\theta} ||_2^2 = \sum_{c = 1}^C || \boldsymbol{\theta}_c ||_2^2$. The community-wise decomposition allows jointly bounding the error vector for all communities as follows.
\begin{align*}
    ||( \boldsymbol{\hat{\theta}} - \boldsymbol{\theta_0} )||_2^2 &= \sum_{c = 1}^C ||( \boldsymbol{\hat{\theta}}_c - \boldsymbol{\theta_0}_{, c} )||_2^2 \\
    &\overset{(i)}{\leq} \sum_{c = 1}^C \left [ 2 \left \{ p_c + \sum_{k = 1}^{p_c} s_{k} (c) ( 1 + | \mathcal{I}_c |) \right \}^{1/2} \{ \tau \mathcal{K}_{[C]} (T - p) \}^{-1/2} || \bmat{R}_c^T \boldsymbol{u}||_{\infty} \right ]^2 \\
    &\overset{(ii)}{\leq} \left [ 2 (C q_{[C]})^{1/2} \{ \tau \mathcal{K}_{[C]} (T - p) \}^{-1/2}  || \bmat{R}^T \boldsymbol{u}||_{\infty} \right ]^2,
\end{align*}
where $(i)$ follows from applying \eqref{eq: upp bound proof, comm wise error bound}, and $(ii)$ follows from noting that by definition $ \{ p_c + \sum_{k = 1}^{p_c} s_{k} (c) ( 1 + | \mathcal{I}_c |) \}^{1/2} \leq q_{[C]} $ and that $ \bmat{R}^T \boldsymbol{u}$ has more non-zero entries than $ \bmat{R}_c^T \boldsymbol{u}$ and contains all of its entries, i.e., $|| \bmat{R}_c^T \boldsymbol{u}||_{\infty} \leq || \bmat{R}^T \boldsymbol{u}||_{\infty}$ for all $c \in [C]$, whence the sum is upper bounded. Thus, we arrive at \eqref{eq: deterministic bound}, i.e., 
\begin{equation*}
    || \boldsymbol{\hat{\theta}} - \boldsymbol{\theta_0} ||_2
        \leq 2 \{\tau \mathcal{K}_{[C]} (T - p) \}^{-1/2} \left ( C q_{[C]} \right )^{1/2} || \bmat{R}^T \boldsymbol{u} ||_{\infty}.
\end{equation*}
\end{proof}

\subsection{Bounding the probability}
Before proving the second part we mention a well known result for sub-Gaussian variables.
\begin{lemma}\label{lemma: prob upper bound for sub-G variables}
Assume that each $u_i$ is a zero-mean sub-Gaussian variable with parameter $\sigma_i^2 < +\infty$. Let $\boldsymbol{u} = (u_1, \dots, u_n) \in \mathbb{R}^n$ be the vector whose $ith$-entry is $u_i$, and define $\sigma_u^2 = \max_{i \in [n]} ( \sigma_i^2)$. Then, for all $\delta > 0$ we have that
\begin{equation} \label{eq: upp bound proof, prob upper bound sub-G}
    \mathbf{Pr} \left \{ || \boldsymbol{u} ||_{\infty} >  \sqrt{2} \sigma_u [\{ \log(n) \}^{1/2} + \delta] \right \} \leq 2 \exp( -\delta^2) .
\end{equation}
\end{lemma}
\begin{proof} Here, $[n] = \{ 1, \ldots, n\}$.
    See~\citet[Chapter 2]{wainwright_2019}.
\end{proof}
We are ready to prove the second part of Theorem \ref{th: l2 bounds for ls estimator}, i.e., \eqref{eq: sub-G noise bound two}. 
\begin{proof}
    Notice that the bound given by \eqref{eq: deterministic bound} always holds once we observe a realisation of $\{ \boldsymbol{X}_t \}$ and fix the linear model \eqref{eq: linear model}, i.e., it is a property of the estimator $\boldsymbol{\hat{\theta}}$ under our assumptions. Further, assume that $A.\ref{asum: A3}$ holds. We upper bound the sub-Gaussian parameter (i.e., $\sigma_i^2$) of each entry in $\bmat{R}^T \boldsymbol{u}$ as follows. Define $q = \sum_{c = 1}^C \{ p_c + \sum_{k = 1}^{p_c} s_{k} (c) ( 1 + | \mathcal{I}_c |) \}$ and $\sigma_{\boldsymbol{u}}^2 = \underset{i \in \mathcal{K}}{\max} (\sigma_i^2)$. Notice that all $j$th-entries in $\bmat{R}^T \boldsymbol{u}$ (i.e., $ [\bmat{R}^T \boldsymbol{u}]_j$ for $j \in [q]$) satisfy the following
    \begin{align*}
         [\bmat{R}^T \boldsymbol{u}]_j^2 &= \langle [\bmat{R}]_{\cdot j}, \boldsymbol{u} \rangle^2 \\
         &\overset{(i)}{\leq} ||[\bmat{R}]_{\cdot j}||_2^2 \cdot || \boldsymbol{u} ||_2^2 \\
         &\overset{(ii)}{\leq} n^{-1} \Gamma^{2} || \boldsymbol{u} ||_2^2, 
    \end{align*}
    where $(i)$ follows from the Cauchy-Schwarz inequality and $(ii)$ from assumption $A.\ref{asum: A2}$. 
   
    Thus, for all $j \in [q]$ we have that
    \begin{align} \label{eq: upp bound proof, Rtu variance upper bounds}
         E \{ [\bmat{R}^T \boldsymbol{u}]_j^2 \} &\leq E \left ( n^{-1} \Gamma^{2} || \boldsymbol{u} ||_2^2 \right )\nonumber \\
         &= n^{-1} \Gamma^{2} E || \boldsymbol{u} ||_2^2 \nonumber \\
         &\leq n n^{-1} \Gamma^{2} \sigma_{\boldsymbol{u}}^2 \nonumber \\
         &= \Gamma^{2} \sigma_{\boldsymbol{u}}^2,
    \end{align}
    above follows from applying the upper bound, taking expectations and the independence of the $u_i$. Thus, we see that $[\bmat{R}^T \boldsymbol{u}]_j$ is sub-Gaussian with zero-mean and parameter at most $\Gamma^{2} \sigma_{\boldsymbol{u}}^2 < +\infty$, for all $j \in [q]$. 
    \par
    By \eqref{eq: upp bound proof, Rtu variance upper bounds} we see that the variables
    \begin{equation*}
        (C q_{[C]})^{1/2} \{ \tau \mathcal{K}_{[C]} (T - p) \}^{-1/2}  [\bmat{R}^T \boldsymbol{u}]_j,
    \end{equation*}
    are sub-Gaussian with parameter at most
    $\Gamma^{2} \sigma_{\boldsymbol{u}}^2 (C q_{[C]}) \{ \tau \mathcal{K}_{[C]} (T - p) \}^{-1}$ for all $j \in [q]$. 
    \par
    Thus, by Lemma \ref{lemma: prob upper bound for sub-G variables} and \eqref{eq: upp bound proof, prob upper bound sub-G} we have that
    \begin{multline} \label{eq: probability bound for residuals}
        (C q_{[C]})^{1/2} \{ \tau \mathcal{K}_{[C]} (T - p) \}^{-1/2}  || \bmat{R}^T \boldsymbol{u} ||_{\infty} \leq \\
        ( C q_{[C]} )^{1/2} \sigma_{\boldsymbol{u}} \Gamma \left [ \left \{ 2 \log \left ( \sum_{c = 1}^C q_c \right ) \{ \tau \mathcal{K}_{[C]} (T - p)\}^{-1} \right \}^{1/2} + \sqrt{2} \delta \right ],
    \end{multline}
    holds with probability at least
    \begin{equation} \label{eq: upp bound proof, exponential bound for residuals}
        1 - 2 \exp[ - \delta^2  \{ \tau \mathcal{K}_{[C]} (T - p)\}]. 
    \end{equation}
    \newpage
    Notice that the inequality between random variables given by \eqref{eq: deterministic bound}, i.e., 
    \begin{equation*}
        || \boldsymbol{\hat{\theta}} - \boldsymbol{\theta_0} ||_2
        \leq 2 \{\tau \mathcal{K}_{[C]} (T - p) \}^{-1/2} \left ( C q_{[C]} \right )^{1/2} || \bmat{R}^T \boldsymbol{u} ||_{\infty},
    \end{equation*}
    implies that
    \begin{equation*}
         || \boldsymbol{\hat{\theta}} - \boldsymbol{\theta_0} ||_2 \leq 2  ( C q_{[C]} )^{1/2} \sigma_{\boldsymbol{u}} \Gamma \left [ \left \{ 2 \log \left ( \sum_{c = 1}^C q_c \right ) \{ \tau \mathcal{K}_{[C]} (T - p)\}^{-1} \right \}^{1/2} + \sqrt{2} \delta \right ],
    \end{equation*}
    holds with probability lower bounded by the one for \eqref{eq: probability bound for residuals}. Therefore, we have that \eqref{eq: sub-G noise bound} holds with probability at least equal to the one in \eqref{eq: upp bound proof, exponential bound for residuals}, i.e., we have shown \eqref{eq: sub-G noise bound two}.

\end{proof}

\section{Proof of Corollary \ref{cor: conditional ls estimarot const.}}
\label{apen: proof of Cor asym norm}
We present a proof of Corollary \ref{cor: conditional ls estimarot const.}. We note that the following assumption is likely a consequence of stationarity, however, we include it as an assumption for completeness and leave the proper derivation for future work. 
\par
We assume that A.\ref{asum: A1}, A.\ref{asum: A2}, A.\ref{asum: A3} in Assumptions \ref{assum: theorem assumptions} hold. Further, assume that as $T \to +\infty$, all three conditions hold, i.e., for all $T > 0$ there exists a $\tau > 0$ such that $\lambda_{\min} (\bmat{R}_c^T \bmat{R}_c) \geq \{ \tau |K_c| (T - p_c) \}^{1/2}$ for all $c \in [C]$, so we have that 
\begin{equation} \label{eq: asymp norm proof, tau condition}
    \lim_{T \to +\infty} \lambda_{\min} (\bmat{R}_c^T \bmat{R}_c)  \geq \lim_{T \to +\infty} \{ \tau |K_c| (T - p_c) \}^{1/2} = +\infty, \enspace a.s.
\end{equation}
 for all $c \in [C]$. Further, that 
$$ \lim_{T \to +\infty} \max_{j \in [q]} \{ || [\bmat{R}]_{\cdot j} ||_2 \} \leq \lim_{T \to +\infty} n^{-1/2} \Gamma = 0,$$ 
where $n := \min_{c \in [C]} \{ |K_c| (T - p_c) \}$, i.e., the number of rows in $\bmat{R}$. 
\subsection{Proof of consistency}
We proceed to prove consistency, i.e., \eqref{eq: consistency}.
\begin{proof}
    By \eqref{eq: asymp norm proof, tau condition}, our assumptions and the above means that
    \begin{equation} \label{eq:  asymp norm proof, ratio limit}
        \lim_{T \to \infty} 2 \sigma_{\boldsymbol{u}}^2 \Gamma^2 \left \{ C q_{[C]}  \log(C q_{[C]}) \right \} \left \{ \tau \mathcal{K}_{[C]} (T - p) \right \}^{-1} = 0.
    \end{equation}
    \par
    By~\eqref{eq: sub-G noise bound two}, for all $T > 0$ and $\delta > 0$, we have that
    \begin{multline} \label{eq: asymp norm proof, prob bound}
        0 \leq \mathbf{Pr} \left ( || \boldsymbol{\hat{\theta}} - \boldsymbol{\theta_0} ||_2 > 2  ( C q_{[C]} )^{1/2} \sigma_{\boldsymbol{u}} \Gamma \left [ \left \{ 2 \log \left ( \sum_{c = 1}^C q_c \right ) \{ \tau \mathcal{K}_{[C]} (T - p)\}^{-1} \right \}^{1/2} + \delta \right ] \right ) \\
        \leq 2 \exp \left [ - \frac{\delta^2}{2} \left \{ \tau \mathcal{K}_{[C]} (T - p) \right \} \right ],
    \end{multline} 
    hence, by applying \eqref{eq: asymp norm proof, tau condition}, \eqref{eq:  asymp norm proof, ratio limit} and the squeeze theorem to \eqref{eq: asymp norm proof, prob bound} we arrive at \eqref{eq: consistency}, i.e., as $T \to \infty$ we have that $|| \boldsymbol{\hat{\theta}} - \boldsymbol{\theta_0} ||_2 \overset{\mathbf{Pr}}{\longrightarrow} 0$.
\end{proof}

\subsection{Asymptotic normality}

    We introduce a general notation for random variables and M-estimators below. This is required for presenting the general results for asymptotic normality in \ref{conditions for asymp normality}. These are adapted into our GNAR problem in \ref{apend: proof of asymp normality summary} for proving asymptotic normality (Corollary~\ref{cor: conditional ls estimarot const.}). Readers familiar with M-estimators and empirical processes may go to \ref{apend: proof of asymp normality summary}. Throughout this subsection we use the following definitions.

\subsubsection{Preliminary definitions}
Let $\boldsymbol{X} \in \mathcal{X}$ be a random variable with distribution function $F_{\boldsymbol{x}}$, and $\lossfunc{\theta}$ be a loss function indexed by $\boldsymbol{\theta} \in \bmat{\Theta} \subseteq \mathbb{R}^q$. Assume that we observe $n$ realisations of $\boldsymbol{X}$, which are independent and identically distributed. Define the following.
\begin{itemize}
    \item $F_{\boldsymbol{x}} \lossfunc{\theta} := E_{\boldsymbol{x}} \{ \lossfunc{\theta} (\boldsymbol{X}) \}$, i.e., the expected value of the loss function.
    \item $\hat{F}_{\boldsymbol{x}} \lossfunc{\theta} := n^{-1} \sum_{i = 1}^n  \lossfunc{\theta} (\boldsymbol{X}_i) $, i.e., the empirical mean (average) of the loss function for $n$ observations.
    \item $\gradlossfunc{\theta} := \nabla_{\boldsymbol{\theta}} \lossfunc{\theta} (\boldsymbol{x})$, i.e., the gradient with respect to $\boldsymbol{\theta}$ of the loss function with $\boldsymbol{X} = \boldsymbol{x}$ a fixed realisation.
    \item The estimator $\hat{\boldsymbol{\theta}}$ is asymptotically linear if there exists a function $L_{F}$ such that $F_{\boldsymbol{x}} L_{F} = 0$, $F_{\boldsymbol{x}} (L_{F} L_{F}^T) < +\infty$, i.e., the mean of $L_{F}$ is zero, and all its entries have a finite variance, and
    \begin{equation*}
        \sqrt{n} (\boldsymbol{\hat{\theta}} - \boldsymbol{\theta} ) = \sqrt{n} \Hat{F}_{\boldsymbol{x}} \, L_{F} (\boldsymbol{X}_i) + o_{\mathbf{Pr}} (1),
    \end{equation*}
    where $L_{F}$ is the {\em influence function}. Intuitively, the influence function measures the precision of our estimator and its sensitivity to a data point $\boldsymbol{x}$. It allows us to define a score-like function for a given distribution $F_{\boldsymbol{x}}$ and class of loss functions; see Chapter 10.1 in~\cite{emp_process_theory}. 
\end{itemize}

\subsubsection{Conditions for asymptotic normality} \label{conditions for asymp normality}

    Our proof of asymptotic normality relies on arguments from empirical process theory and M-estimators. The conditions below are taken from Chapter 10 in~\cite{emp_process_theory}. The proof of asymptotic normality for M-estimators relies on asymptotic continuity, consistency and exploits Vapnik-Chervonenkis theory. See Chapters 7 to 10 in~\cite{emp_process_theory}.

\begin{enumerate}
    \item \label{cond: cond one asymp norm} There exists an $\epsilon > 0$ such that for all $\boldsymbol{\theta}$ that satisfy $\normx{\boldsymbol{\theta} - \boldsymbol{\theta_0}}_2 < \epsilon $, the function $\lossfunc{\theta}$ is differentiable with gradient $\gradlossfunc{\theta}$ for all $x \in \mathcal{X}$, i.e., all partial derivatives exist in an $\ell_2$-norm neighbourhood of $\boldsymbol{\theta_0}$. See Condition (a) in Chapter 10.2 of \cite{emp_process_theory}.
    \item \label{cond: cond two asymp norm} As $\boldsymbol{\theta} \to \boldsymbol{\theta_0}$, we have that 
    $$F_{\boldsymbol{x}} (\gradlossfunc{\theta} - \gradlossfunc{\theta_0}) = \bmat{H} (\boldsymbol{\theta} - \boldsymbol{\theta_0}) + o( |\boldsymbol{\theta} - \boldsymbol{\theta_0}|),$$
    where $\bmat{H} \in \mathbb{R}^{q \times q}$ is a symmetric positive definite matrix, i.e., the second order approximation is well defined as $\boldsymbol{\theta} \to \boldsymbol{\theta_0}$. See Condition (b) in Chapter 10.2 of \cite{emp_process_theory}.
    \item \label{cond: cond three asymp norm} The $\ell_2$-norm of gradient differences is continuous, i.e., 
    $$\lim_{\boldsymbol{\theta} \to \boldsymbol{\theta_0}} \normx{\gradlossfunc{\theta} - \gradlossfunc{\theta_0}}_2 = 0. $$
    See Condition (c.2) in Chapter 10.2 of \cite{emp_process_theory}.
    \item \label{cond: cond four asymp norm} There exists an $\epsilon > 0$ such that the function class $\mathfrak{G}_{\epsilon} := \{ \gradlossfunc{\theta} : ||\boldsymbol{\theta} - \boldsymbol{\theta_0}||_{2} < \epsilon \}$ is asymptotically continuous at $\gradlossfunc{\theta_0}$, i.e., as $n \to +\infty$, if the random sequence $\{ \gradlossfunc{\theta_n} \} \subset \mathfrak{G}_{\epsilon}$ satisfies $ F_{\boldsymbol{x}} ( \normx{\gradlossfunc{\theta_n} - \gradlossfunc{\theta_0}}_2^2 ) \overset{\mathbf{Pr}}{\longrightarrow} 0$, then  
    $$|z_n (\boldsymbol{\theta_n}) - z_n (\boldsymbol{\theta_0} ) | \overset{\mathbf{Pr}}{\longrightarrow} 0, $$
    where $z_n(\boldsymbol{\theta}) := \sqrt{n} (F_{\boldsymbol{x}} - \hat{F}_{\boldsymbol{x}} ) \gradlossfunc{\theta}$ is the empirical process indexed by the class $\mathfrak{G}_{\epsilon}$ for the distribution $F_{\boldsymbol{x}}$. That is, the function class $\mathfrak{G}_{\epsilon}$ is such that approximating the expected value of the idealised (theoretical) loss by the empirical one in a neighbourhood of $\boldsymbol{\theta_0}$ is valid. See Condition (c.1) in Chapter 10.2 of \cite{emp_process_theory}.
\end{enumerate}
\subsubsection{Preliminary results for showing asymptotic normality}
\begin{theorem} \label{th: asymp. continuity for VC class}
    Assume that $\mathfrak{G}_{\theta} := \{ g_{\theta}: \boldsymbol{\theta} \in \bmat{\Theta} \}$ is a Vapnik-Chervonenkis graph class with envelope function $G_{\theta} := \underset{g \in \mathfrak{G}_{\theta}}{\sup} \{ | g_{\theta} (\boldsymbol{x}) | \},$ $\boldsymbol{x} \in \mathcal{X}$, such that $G_{\theta} \in L_2 (F_{\boldsymbol{x}})$, i.e., $F_{\boldsymbol{x}} G_{\boldsymbol{\theta}}^2 < +\infty$, and that all {\em (possibly random)} sequences $\{ g_{\boldsymbol{\theta}_n} \} \subset \mathfrak{G}_{\theta}$ satisfy $ \normx{\lossfunc{\theta_n} - \lossfunc{\theta_0}}_2 \overset{\mathbf{Pr}}{\longrightarrow} 0$ as $ \boldsymbol{\theta_n}  \overset{\mathbf{Pr}}{\longrightarrow} \boldsymbol{\theta_0}$, i.e., the function class is mean square continuous with respect to $\boldsymbol{\theta_0}$. Then, 
    \begin{equation*}
        |z_n (\boldsymbol{\theta_n}) - z_n (\boldsymbol{\theta_0} ) | \overset{\mathbf{Pr}}{\longrightarrow} 0,
    \end{equation*}
    i.e., the class $\mathfrak{G}_{\boldsymbol{\theta}}$ is asymptotically continuous at $\lossfunc{\theta_0}$.
\end{theorem}
\begin{proof}
    See Chapter 9 in \cite{emp_process_theory}, and Chapter 6 in \cite{emp_process_theory} and Chapter 4 in \cite{wainwright_2019} for the definition and properties of a Vapnik-Chervonenkis graph class and Vapnik-Chervonenkis dimension.
\end{proof}
\begin{lemma} \label{lemma> asymp norm of m-estimators}
    Assume that $\boldsymbol{\hat{\theta}}$ is a consistent {\em M-estimator} of $\boldsymbol{\theta_0} \in \mathbf{\Theta} \subseteq \mathbb{R}^q$, where $\boldsymbol{\theta_0}$ is an interior point. Further, assume that conditions {\em \ref{cond: cond one asymp norm}, \ref{cond: cond two asymp norm}, \ref{cond: cond three asymp norm} and \ref{cond: cond four asymp norm}} in Conditions \ref{conditions for asymp normality} hold. Then, we have that
    \begin{equation*}
        \sqrt{n} (\boldsymbol{\hat{\theta}} - \boldsymbol{\theta_0}) = - \sqrt{n} \hat{F} \left ( \bmat{H}^{-1} \gradlossfunc{\theta_0} \right ) + o_{\mathbf{Pr}}(1),
    \end{equation*}
    i.e., $\boldsymbol{\hat{\theta}}$ is asymptotically linear with influence function $- \bmat{H}^{-1} \gradlossfunc{\theta_0} $. Thus, we have that 
    \begin{equation*}
         \sqrt{n} (\boldsymbol{\hat{\theta}} - \boldsymbol{\theta_0}) \overset{\boldsymbol{\mathrm{d}}}{\longrightarrow} \mathrm{N}_q \left(0, \bmat{\Sigma}_{\boldsymbol{\hat{\theta}}} \right ),
    \end{equation*}
    where $\bmat{\Sigma}_{\boldsymbol{\hat{\theta}}} = \bmat{H}^{-1} \{ F (\gradlossfunc{\theta_0} \gradlossfunc{\theta_0}^T) \}  \bmat{H}^{-1} $. That is, $\boldsymbol{\hat{\theta}}$ is asymptotically normal with covariance given by the local behaviour of the loss function around $\boldsymbol{\theta_0}$.
\end{lemma}
\begin{proof}
    See~\citet[Chapter 10]{emp_process_theory}.
\end{proof}

\subsection{Proof of asymptotic normality} \label{apend: proof of asymp normality summary}

Before proving asymptotic normality we present a summary of the proof. 
\begin{enumerate}
    \item Show that the problem is equivalent to analysing $n := T - p \in \mathbb{Z}^*$ {\em i.i.d.} observations of a $d$-dimensional linear regression problem with $q < n d$ (usually $q \ll n d$) unknown parameters;
    \item Show that the $q$ leading singular values of the expected value of the design matrix are positive and finite, i.e., the {\em idealised} (expected) linear regression problem has a unique minimum and well defined Hessian;
    \item Show that $\boldsymbol{\hat{\theta}}$ is an M-estimator, and that $\boldsymbol{\theta_0}$ minimises the idealised loss and is an interior point;
    \item Apply Theorem \ref{th: l2 bounds for ls estimator} for showing that $\boldsymbol{\hat{\theta}}$ is consistent, i.e., applying \eqref{eq: consistency};
    \item Show that conditions \ref{cond: cond one asymp norm}, \ref{cond: cond two asymp norm}, \ref{cond: cond three asymp norm} and  \ref{cond: cond four asymp norm} in Conditions \ref{conditions for asymp normality} hold;
    \item Apply Lemma \ref{lemma> asymp norm of m-estimators} for finishing the proof.
\end{enumerate}

\subsubsection{Problem setting}
Assume Assumptions~\ref{assum: theorem assumptions}, \eqref{eq: asymp norm proof, tau condition} and \eqref{eq:  asymp norm proof, ratio limit} hold and that we observe a realisation of length $T$ of a stationary GNAR$(p)$ (maximum lag is $p$) process $\boldsymbol{X}_t \in \mathbb{R}^d$ with $q$ unknown parameters. Let $t_i := p + i$ for $i \in [n]$, where $n = T - p$. Also, $\boldsymbol{y}_{t_i} := \boldsymbol{X}_{p + i}$ for all $t = p + 1, \dots, T$. Then, we can write \eqref{eq: GNAR compact structural representation} as
\begin{equation} \label{eq: time-step linear model}
    \boldsymbol{y}_{t_i} = \bmat{Z}_{\boldsymbol{t_i}} \boldsymbol{\theta_0} + \boldsymbol{u}_{t_i},
\end{equation}
where $\bmat{Z}_{\boldsymbol{t_i}} \in \mathbb{R}^{d \times q}$ is the design matrix with $p$ previous lags and the corresponding within and between community components, i.e, 
\begin{align} \label{eq: time-step desing matrix}
    \bmat{R}_{k, t, c} &:= \left [\boldsymbol{X}_{t - k}^c |  \boldsymbol{Z}_{t - k}^{1, c} | \dots |  \boldsymbol{Z}_{t - k}^{s_k (c), c} |  \boldsymbol{Z}_{t - k}^{1, c:\tilde{c}} | \dots |  \boldsymbol{Z}_{t - k}^{s_k (c), c:\tilde{c}} \right], \nonumber \\
    \bmat{R}_{k, t} &:= [\bmat{R}_{k, t, c}| \cdots | \bmat{R}_{k, t, C} ], \nonumber \\
    \bmat{Z}_{\boldsymbol{t_i}} &:= \left [ \bmat{R}_{1, t}| \cdots | \bmat{R}_{p, t} \right ],
\end{align}
where each predictor column for $\tilde{c} \in \mathcal{I}_c$ is concatenated in ascending order with respect to $\tilde{c}$, i.e., if $\tilde{c}_2 > \tilde{c}_1$, then the column for $\tilde{c}_1$ precedes the one for $\tilde{c}_2$, $ \boldsymbol{u}_{t_i}$ is a sub-Gaussian multivariate white noise process with a diagonal covariance, and $ \boldsymbol{\theta_0} \in \bmat{\Theta} \subset \mathbb{R}^q$ is the ``true'' vector of parameters.
\newpage
Let $F_{\boldsymbol{y}_{t}, \bmat{Z}_{\boldsymbol{t}}}$ denote the joint distribution function, $F_{\boldsymbol{y}_{t} | \bmat{Z}_{\boldsymbol{t}}}$ denote the conditional distribution function, and $F_{\bmat{Z}_{\boldsymbol{t}}}$, $F_{\boldsymbol{u}}$ denote the marginal distribution functions. Then, for any function $\lossfunc{\theta}$ we have that
\begin{align*}
    F_{\boldsymbol{y}_{t}, \bmat{Z}_{\boldsymbol{t}}} (\lossfunc{\theta}) &= F_{\bmat{Z}_{\boldsymbol{t}}} \{ F_{\boldsymbol{y}_{t} | \bmat{Z}_{\boldsymbol{t}}} (\lossfunc{\theta} ) \}, \\
    F_{\boldsymbol{u}, \bmat{Z}_{\boldsymbol{t}}} (\lossfunc{\theta}) &= F_{\bmat{Z}_{\boldsymbol{t}}} \{ F_{\boldsymbol{u}} (\lossfunc{\theta}) \},
\end{align*}
i.e., we can compute the relevant risks by iterated expectations.
\par
Notice that by \eqref{eq: time-step linear model} we can factorise the joint density of the data by exploiting that $\boldsymbol{u}_{t_i}$ are {\em i.i.d.} and independent of $\bmat{Z}_{\boldsymbol{t_i}}$ for all $i \in [n]$. Hence, we have that 
\begin{equation*}
    f ( \boldsymbol{y}_{t_i} - \bmat{Z}_{\boldsymbol{t_i}}  \boldsymbol{\theta_0} ) = f ( \boldsymbol{u}_{t_i}),
\end{equation*}
thus, by exploiting the Markov property above for $t_i$, using stationarity of $\bmat{Z}_{\boldsymbol{t_i}}$ for each $t_i$, and letting $(\boldsymbol{y}, \bmat{Z}) \in \mathbb{R}^{nd} \times \mathbb{R}^{nd \times q}$ denote the sample, we have that
\begin{equation*}
    f(\boldsymbol{y}, \bmat{Z}) = f(\bmat{Z}_{\boldsymbol{t}}) \prod_{i = 1}^n f_{\boldsymbol{u}} (\boldsymbol{y}_{t_i} - \bmat{Z}_{\boldsymbol{t_i}} \boldsymbol{\theta_0}),
\end{equation*}
thus, we see that there are $n$ {\em i.i.d.} observations of the linear regression problem \eqref{eq: time-step linear model}, albeit we still have to account for the first $p$ time-steps, i.e., $f(\bmat{Z}_{\boldsymbol{t_1}})$. We circumvent this by performing iterated expectations with respect to the $n$ linear regression problems. We finish the problem setting with Proposition \ref{prop: bounded eigenvalues}. 
\begin{proposition} \label{prop: bounded eigenvalues}
    Assume that each stationary process $[\bmat{Z}_{\boldsymbol{t}}]_{ij} \neq 0$ in $\bmat{Z}_{\boldsymbol{t}} \in \mathbb{R}^{d \times q}$, where $q_c \leq d_c$ for all $c \in [C]$, i.e., lagged entries of the same process for each community, has a spectral density $f_k (\omega)$ such that there exist two constants $0 < m \leq M < +\infty$, which satisfy 
        $$m \leq f_{k_i k_j} (\omega) \leq M,$$
    for all $\omega \in [-\pi, \pi]$. Then, we have that
    \begin{equation} \label{eq: design eigenvalu bounds}
        0 < 2 \pi m d \leq \lambda_{\min} \left [ E \{ (\bmat{Z}_{\boldsymbol{t}}^T \bmat{Z}_{\boldsymbol{t}}) \} \right ] \leq \lambda_{\max} \left [ E \{ (\bmat{Z}_{\boldsymbol{t}}^T \bmat{Z}_{\boldsymbol{t}}) \} \right ] \leq 2 d \pi M < +\infty,  
    \end{equation}
    i.e., all eigenvalues of $E \{ (\bmat{Z}_{\boldsymbol{t}}^T \bmat{Z}_{\boldsymbol{t}}) \}$ are positive and finite. Further, the matrix $ E \{ (\bmat{Z}_{\boldsymbol{t}}^T \bmat{Z}_{\boldsymbol{t}}) \} $ is symmetric positive definite.
\end{proposition}
\begin{proof}
    Notice that all entries in $(\bmat{Z}_{\boldsymbol{t}}^T \bmat{Z}_{\boldsymbol{t}})$ correspond to the sum of products of either autocovariances between the same entries in $\bmat{Z}_{\boldsymbol{t}}$ or covariances between different entries in $\bmat{Z}_{\boldsymbol{t}}$ at lags $h = 0, \dots, p - 1$, i.e, for $(i, j) \in [q] \times [q]$ we have that
    \begin{equation*}
        [(\bmat{Z}_{\boldsymbol{t}}^T \bmat{Z}_{\boldsymbol{t}})]_{ij} = \sum_{k = 1}^d [\bmat{Z}_{\boldsymbol{t}}]_{ki} [\bmat{Z}_{\boldsymbol{t}}]_{kj},
    \end{equation*}
    where the columns $i$ and $j$ will vary between covariances and autocovariances at different lags for autoregressive, between and within community columns in $\bmat{Z}_{\boldsymbol{t}}$ for the $k$th stationary process with a maximum lag equal to $p -1$. 
    \par
    Define $\bmat{\Gamma}_k (p, [C]) \in \mathbb{R}^{q \times q}$ as the matrix with entries 
    \begin{equation*}
        [\bmat{\Gamma}_k (p, [C])]_{ij} := E \{ [\bmat{Z}_{\boldsymbol{t}}]_{ki} [\bmat{Z}_{\boldsymbol{t}}]_{kj} \},
    \end{equation*}
    Thus, we have that
    \begin{equation} \label{eq: expectation sum}
        E \{ (\bmat{Z}_{\boldsymbol{t}}^T \bmat{Z}_{\boldsymbol{t}}) \} = \sum_{k = 1}^d \bmat{\Gamma}_k (p, [C]).
    \end{equation}
    \par
    Notice that by assumption we have that the spectral density for each $k$th process and the cross-spectral density between two correlated processes for all $k \in \mathcal{K}$ and all $\omega \in [-\pi, \pi]$ satisfy
    \begin{align}
        &0 < m \leq f_{k_i} (\omega) \leq M < +\infty,  \label{eq: spec density bounds one}\\
        &|f_{k_i, k_j} (\omega)| \leq \{ f_{k_i} (\omega) f_{k_j} (\omega) \}^{1/2}, \label{eq: spec density bounds two}
    \end{align}
    where $k_i$ denotes the $i$th process linked to node $k$ (e.g., number of within community $r$-stage neighbourhood regressions). Further, if the cross spectral density is non-zero, then
    \begin{align} \label{eq: cross-spec density bounds}
        &0 < m \leq \mathrm{Re} \{f_{k_i, k_j} (\omega) \}, \\
        &0 < m \leq \mathrm{Im} \{f_{k_i, k_j} (\omega) \}.
    \end{align}
    
    Let $\gamma_{k, ij} (h) = [\bmat{\Gamma}_k (p, [C])]_{ij}$. Then, it is the (auto)covariance of the $i$th and $j$th columns for the $k$th row in $\bmat{Z}_{\boldsymbol{t}}$ at lag $h = 0, \dots, p -1$, i.e,
    \begin{equation}  \label{eq: autocovariance cases}      
    \gamma_{k, ij} (h) = \begin{cases}
            E \left( X_{k, t + h} X_{k, t} \right ), &\text{ if } i, j \text{ are } s.t. \enspace k \text{ is autoregressive}, \\
            E \left( X_{k, t + h} Z_{k, t}^{r, c} \right ), &\text{ if } i, j \text{ are } s.t. \enspace k \text{ is within}, \\
            E \left( X_{k, t + h} Z_{k, t}^{r, c:\tilde{c}} \right ), &\text{ if } i, j \text{ are } s.t. \enspace k \text{ is interaction}, \\
            E \left( Z_{k, t + h}^{r, c} Z_{k, t}^{r, c} \right ), &\text{ if }  i, j \text{ are } s.t. \enspace k \text{ is cross-within}, \\
            E \left( Z_{k, t + h}^{r, c:\tilde{c}} Z_{k, t}^{r, c:\tilde{c}} \right ), &\text{ if } i, j \text{ are } s.t. \enspace k \text{ is cross-interaction}, \\
            E \left( Z_{k, t + h}^{r, c} Z_{k, t}^{r, c:\tilde{c}} \right ), &\text{ if } i, j \enspace \text{ are } \, s.t.\, k \text{ is within-interaction}.
        \end{cases}
    \end{equation}
    \par
    Since all $k_i$-th processes are stationary, by the spectral representation of the autocovariance function; see \cite{brockwell_davis}, we have that
    \begin{equation} \label{eq: spec representation of autocovariance}
        \gamma_{k, ij} (h) = \int_{-\pi}^{\pi} f_{k_i, k_j} (\omega) \exp ( \imath \pi h \omega) d \omega,
    \end{equation}
    where $\imath$ is the imaginary unit (i.e., $\imath^2 = -1$), and $f_{k_i, k_i}  (\omega) = f_{k_i}  (\omega)$ is the (cross)spectral density of the $k_i$-th process. Let $\boldsymbol{a} \in \mathbb{R}^q \setminus \{\boldsymbol{0} \}$ be an eigenvector of $\bmat{\Gamma}_k (p, [C])$ with eigenvalue $\lambda_k \neq 0$. Then, we have that
    \begin{align*}
        \lambda_k || \boldsymbol{a}||_2^2 &= \langle \boldsymbol{a}, \bmat{\Gamma}_k (p, [C]) \boldsymbol{a} \rangle \\
        &= \sum_{i = 1}^q \sum_{j = 1}^q a_i a_j  \gamma_{k, ij} (h)
    \end{align*}
    $\bullet$ By applying \eqref{eq: spec representation of autocovariance} we have that
    \begin{align*}
        \lambda_k || \boldsymbol{a}||_2^2  &= \sum_{i = 1}^q \sum_{j = 1}^q \int_{-\pi}^{\pi} f_{k_i, k_j} (\omega) \exp ( \imath \pi h \omega) d \omega \\
        &\leq \int_{-\pi}^{\pi} \sum_{i = 1}^q \sum_{j = 1}^q a_i a_j   |f_{k_i, k_j} (\omega)| \exp ( \imath \pi h \omega) d \omega
    \end{align*}
    $\bullet$ By applying \eqref{eq: spec density bounds two} we have that
    \begin{equation*}
        \lambda_k || \boldsymbol{a}||_2^2  \leq  \int_{-\pi}^{\pi}   \sum_{i = 1}^q \sum_{j = 1}^q a_i a_j \{ f_{k_i} (\omega) f_{k_j} (\omega) \}^{1/2}  \exp ( \imath \pi h \omega) d \omega
    \end{equation*}
    $\bullet$ By applying \eqref{eq: spec density bounds one} we have that
    \begin{align*}
        \lambda_k || \boldsymbol{a}||_2^2  &\leq \int_{-\pi}^{\pi}  \sum_{i = 1}^q \sum_{j = 1}^q a_i a_j M  \exp ( \imath \pi h \omega) d \omega 
    \end{align*}
    $\bullet$ By noting that  $\int_{-\pi}^{\pi}  \exp ( \imath \pi h \omega) d \omega = \mathbb{I}(h = 0)$ we have that
    \begin{align*}
        \lambda_k || \boldsymbol{a}||_2^2  &= M \sum_{i = 1}^q \sum_{j = 1}^q \int_{-\pi}^{\pi}  a_i a_j \mathbb{I} (i = j) d \omega \\
        &= 2 \pi M  || \boldsymbol{a}||_2^2 < +\infty,
    \end{align*}
    where $\mathbb{I} (i = j)$ indicates one of the cases in \eqref{eq: autocovariance cases} with $h = 0$. Similarly, we have that
    \begin{align*}
        \lambda_k || \boldsymbol{a}||_2^2 &= \langle \boldsymbol{a}, \bmat{\Gamma}_k (p, [C]) \boldsymbol{a} \rangle \\
        &= \sum_{i = 1}^q \sum_{j = 1}^q a_i a_j  \gamma_{k, ij} (h)
    \end{align*}
    $\bullet$ By applying \eqref{eq: spec representation of autocovariance} we have that
    \begin{align*}
         \lambda_k || \boldsymbol{a}||_2^2 &= \sum_{i = 1}^q \sum_{j = 1}^q \int_{-\pi}^{\pi} a_i a_j \exp ( \imath \pi h \omega) f_{k_i, k_j} (\omega) d \omega \\
        &= \sum_{i = 1}^q \sum_{j = 1}^q \int_{-\pi}^{\pi} a_i a_j \exp ( \imath \pi h \omega) \left [\mathrm{Re} \{f_{k_i, k_j} (\omega) \} + \imath \mathrm{Im} \{f_{k_i, k_j} (\omega) \} \right ] d \omega
    \end{align*}
    $\bullet$ By noting that the left hand side is a real number, we see that
    $$\imath \sum_{i = 1}^q \sum_{j = 1}^q \int_{-\pi}^{\pi} a_i a_j \exp ( \imath \pi h \omega)  \mathrm{Im} \{f_{k_i, k_j} (\omega) \} d \omega = 0,$$ 
    and by applying \eqref{eq: cross-spec density bounds} we have that
    \begin{align*}
         \lambda_k || \boldsymbol{a}||_2^2 &= \sum_{i = 1}^q \sum_{j = 1}^q \int_{-\pi}^{\pi} a_i a_j \exp ( \imath \pi h \omega) \mathrm{Re} \{f_{k_i, k_j} (\omega) \} d \omega \\
        &\geq m \int_{-\pi}^{\pi}  \sum_{i = 1}^q \sum_{j = 1}^q a_i a_j \exp ( \imath \pi h \omega) d \omega \\
        &= m \sum_{i = 1}^q \sum_{j = 1}^q \int_{-\pi}^{\pi}  a_i a_j \mathbb{I} (i = j) d \omega \\
        &= 2 \pi m || \boldsymbol{a}||_2^2 > 0.
    \end{align*}
    \par
    Thus, we have shown that any eigenvalue $\lambda_k \neq 0$ of $\bmat{\Gamma}_k (p, [C])$ satisfies
    \begin{equation} \label{eq: kth eigenvalue bound}
        0 < 2 \pi m || \boldsymbol{a}||_2^2 \leq \lambda_k || \boldsymbol{a}||_2^2 =  \langle \boldsymbol{a}, \bmat{\Gamma}_k (p, [C]) \boldsymbol{a} \rangle \leq 2 \pi M || \boldsymbol{a}||_2^2 < +\infty.
    \end{equation}
    Let  $\boldsymbol{a} \in \mathbb{R}^q \setminus \{\boldsymbol{0} \}$ be an eigenvector of $\bmat{\Gamma} (p, [C]) := E \{ (\bmat{Z}_{\boldsymbol{t}}^T \bmat{Z}_{\boldsymbol{t}}) \}$ with eigenvalue $\lambda \neq 0$. Then, we have that
    \begin{align*}
         \lambda || \boldsymbol{a}||_2^2 &= \langle \boldsymbol{a}, \bmat{\Gamma} (p, [C]) \boldsymbol{a} \rangle \\
         &= \langle \boldsymbol{a}, \sum_{k = 1}^d \bmat{\Gamma}_k (p, [C])  \boldsymbol{a} \rangle \\
         &= \sum_{k = 1}^d \langle \boldsymbol{a}, \bmat{\Gamma}_k (p, [C]) \boldsymbol{a} \rangle,
    \end{align*}
   and notice that by \eqref{eq: kth eigenvalue bound} we have that
   \begin{equation*}
       0 < \sum_{k = 1}^d 2 \pi m || \boldsymbol{a}||_2^2 \leq \lambda || \boldsymbol{a}||_2^2 \leq  \sum_{k = 1}^d 2 \pi M  || \boldsymbol{a}||_2^2 < +\infty,
   \end{equation*}
   whence we arrive at \eqref{eq: design eigenvalu bounds}, i.e., 
   \begin{equation*}
        0 < 2 \pi m d \leq \lambda_{\min} \left [ E \{ (\bmat{Z}_{\boldsymbol{t}}^T \bmat{Z}_{\boldsymbol{t}}) \} \right ] \leq \lambda_{\max} \left [ E \{ (\bmat{Z}_{\boldsymbol{t}}^T \bmat{Z}_{\boldsymbol{t}}) \} \right ] \leq 2 d \pi M < +\infty,
   \end{equation*}
   which implies that $ E \{ (\bmat{Z}_{\boldsymbol{t}}^T \bmat{Z}_{\boldsymbol{t}}) \}$ is positive definite, and by symmetry of the autocovariance function of stationary processes and \eqref{eq: autocovariance cases} it is also symmetric.
\end{proof}
Assuming that $q_c \leq d_c$ can be relaxed by noting that by stacking $m$ time-steps coming from \eqref{eq: time-step linear model}, such that $q_c \leq m d_c$ for all $c \in [C]$. Since the $m$ white noise vectors $\boldsymbol{u}_t$ are independent, augmenting the design in this way results in a setting where \eqref{eq: time-step linear model} is not underdetermined (i.e., $q \leq m d$).

\newpage
\begin{proposition} \label{prop: m-estimator}
    Let $\boldsymbol{\hat{\theta}}$ be given by \eqref{eq: ls estimator} and $\boldsymbol{\theta_0}$ be the ``true'' vector of parameters, both in $\bmat{\Theta} \subset \mathbb{R}^q$, where $\bmat{\Theta}$ is the open set which satisfies the stationarity constraints given by Corollary \ref{cor: conditions for stationary processes}. Define the squared error loss function $\lossfunc{\theta}$ for \eqref{eq: time-step linear model} and all $\boldsymbol{\theta} \in \bmat{\Theta}$ as
    \begin{align} \label{eq: squared error loss definition}
        \lossfunc{\theta}: \enspace &\bmat{\Theta} \longrightarrow \mathbb{R}_0^+ \nonumber \\
        &\boldsymbol{\theta} \mapsto \left(2 d \right)^{-1} \normx{\boldsymbol{y}_{t_i} - \bmat{Z_{t_i}} \boldsymbol{\theta}}_2^2. 
    \end{align}
    Then, we have that
    \begin{align} 
         \boldsymbol{\theta_0} &= \underset{\boldsymbol{\theta}}{ \mathrm{argmin}} \left \{ F_{\boldsymbol{y}_{t}, \bmat{Z}_{\boldsymbol{t}}} \left ( \lossfunc{\theta} \right ) \right \}, \label{eq: m estimator equation one} \\
         \boldsymbol{\hat{\theta}} &= \underset{\boldsymbol{\theta}}{ \mathrm{argmin}} \left \{ \Hat{F}_{\boldsymbol{y}_{t} | \bmat{Z}_{\boldsymbol{t}}} \left ( \lossfunc{\theta} \right ) \right \} \label{eq: m estimator equation two},
    \end{align}
    i.e., $\boldsymbol{\hat{\theta}}$ is the {\em M-estimator} under $\lossfunc{\theta}$ with idealised value $\boldsymbol{\theta_0}$ in our problem setting, where $\boldsymbol{\theta_0}$ is an interior point of $\bmat{\Theta}$ that minimises the idealised loss. Further, we have that the {\em M-estimator} is consistent, i.e., as $T \to \infty$ we have that $|| \boldsymbol{\hat{\theta}} - \boldsymbol{\theta_0} ||_2 \overset{\mathbf{Pr}}{\longrightarrow} 0$.
\end{proposition}
\begin{proof}
    We start by analysing the idealised loss. Notice that
    \begin{align*}
        F_{\boldsymbol{y}_{t} | \bmat{Z}_{\boldsymbol{t}}} \left ( \lossfunc{\theta} \right ) &=  F_{\boldsymbol{y}_{t} | \bmat{Z}_{\boldsymbol{t}}} \left(2 d \right)^{-1} \normx{\boldsymbol{y}_{t_i} - \bmat{Z_{t_i}} \boldsymbol{\theta}}_2^2 \\
        &=  F_{\boldsymbol{y}_{t} | \bmat{Z}_{\boldsymbol{t}}} \left(2 d \right)^{-1} \normx{\boldsymbol{y}_{t_i} - \bmat{Z_{t_i}} \boldsymbol{\theta} + \bmat{Z_{t_i}} \boldsymbol{\theta_0} - \bmat{Z_{t_i}} \boldsymbol{\theta_0}}_2^2 \\
        &= \left(2 d \right)^{-1}  F_{\boldsymbol{u}_{t} } \normx{\boldsymbol{u}_{t_i} - \bmat{Z_{t_i}} (\boldsymbol{\theta} - \boldsymbol{\theta_0})}_2^2 \\
        &=\left(2 d \right)^{-1} \left \{  F_{\boldsymbol{u}_{t} } \normx{\boldsymbol{u}_{t_i}}_2^2 + F_{\boldsymbol{u}_{t} } \normx{\bmat{Z_{t_i}} (\boldsymbol{\theta} - \boldsymbol{\theta_0) }}_2^2 \right \} \\
        &= \left(2 d \right)^{-1} \left \{ \mathrm{tr} ( \bmat{\Sigma_{\boldsymbol{u}}} ) + \normx{\bmat{Z_{t_i}} (\boldsymbol{\theta} - \boldsymbol{\theta_0) }}_2^2 \right \}.
    \end{align*}
    Notice that the above is the sum of two non-negative numbers, thus, it is minimised if $\boldsymbol{\theta} = \boldsymbol{\theta_0}$, i.e., $\boldsymbol{\theta_0}$ is an idealised minimiser of $\lossfunc{\theta}$.
    \par
    By Proposition \ref{prop: bounded eigenvalues} the risk is finite, and notice that 
    $$0 \leq E \normx{\bmat{Z_{t_i}} (\boldsymbol{\theta} - \boldsymbol{\theta_0) }}_2^2 \leq  \tau_{(q)} \normx{(\boldsymbol{\theta} - \boldsymbol{\theta_0) }}_2^2,$$ 
    thus, by iterated expectation we have that
    \begin{equation} \label{eq: bounded expected loss}
         F_{\boldsymbol{y}_{t}, \bmat{Z}_{\boldsymbol{t}}} \left ( \lossfunc{\theta_0} \right ) \leq F_{\boldsymbol{y}_{t}, \bmat{Z}_{\boldsymbol{t}}} \left ( \lossfunc{\theta} \right )  \leq \sigma_{\boldsymbol{u}}^2 + \tau_{(q)} \normx{(\boldsymbol{\theta} - \boldsymbol{\theta_0) }}_2^2 < +\infty,
    \end{equation}
    where $\tau_{(q)} := \lambda_{\max} \left [ E \{ (\bmat{Z}_{\boldsymbol{t}}^T \bmat{Z}_{\boldsymbol{t}}) \} \right ]$. 
    \par
   Notice that squared error loss is differentiable for all $\boldsymbol{\theta} \in \bmat{\Theta}$ and all $(\boldsymbol{y}_{t}, \bmat{Z}_{\boldsymbol{t}}) \in \mathbb{R}^{d} \times \mathbb{R}^{d \times q}$ because it is the composition of differentiable maps. Thus, we have that
   \begin{equation} \label{eq: loss function gradient}
       \gradlossfunc{\theta} = - d^{-1} \left \{ \bmat{Z}_{t_i}^T \left ( \boldsymbol{y}_{t_i} - \bmat{Z_{t_i}} \boldsymbol{\theta} \right ) \right \},
   \end{equation}
   for all $\boldsymbol{\theta} \in \bmat{\Theta}$ and all $(\boldsymbol{y}_{t}, \bmat{Z}_{\boldsymbol{t}}) \in \mathbb{R}^{d} \times \mathbb{R}^{d \times q}$. 
   \par
   By Proposition \ref{prop: bounded eigenvalues} we have that the matrix $E \{ (\bmat{Z}_{\boldsymbol{t}}^T \bmat{Z}_{\boldsymbol{t}}) \}$ is non-singular. 
   Thus, the risk is
   \begin{align*}
        F_{\boldsymbol{y}_{t}, \bmat{Z}_{\boldsymbol{t}}} \left ( \gradlossfunc{\theta} \right ) &= - d^{-1} F_{\boldsymbol{y}_{t}, \bmat{Z}_{\boldsymbol{t}}} \left \{ \bmat{Z}_{t_i}^T \left ( \boldsymbol{y}_{t_i} - \bmat{Z}_{t_i} \boldsymbol{\theta} \right ) \right \} \\
        &= - d^{-1} \left [  F_{\boldsymbol{y}_{t}, \bmat{Z}_{\boldsymbol{t}}} \bmat{Z}_{t_i}^T \left \{ \boldsymbol{u}_{t_i} - \bmat{Z_{t_i}} (\boldsymbol{\theta} - \boldsymbol{\theta_0}) \right \} \right ],
    \end{align*}
    and by iterated expectation we have that
    \begin{equation} \label{eq: zero gradient loss function}
         F_{\boldsymbol{y}_{t}, \bmat{Z}_{\boldsymbol{t}}} \left ( \gradlossfunc{\theta} \right ) = d^{-1} \left [ -  F_{ \bmat{Z}_{\boldsymbol{t}}} \bmat{Z}_{t_i}^T F_{\boldsymbol{u}_{t} } ( \boldsymbol{u}_{t_i}) + F_{ \bmat{Z}_{\boldsymbol{t}}}  (\bmat{Z}_{\boldsymbol{t}}^T \bmat{Z}_{\boldsymbol{t}}) (\boldsymbol{\theta} - \boldsymbol{\theta_0}) \right ],
    \end{equation}
    since $E \{ (\bmat{Z}_{\boldsymbol{t}}^T \bmat{Z}_{\boldsymbol{t}}) \}$ is symmetric positive definite and non-singular, and $F_{\boldsymbol{u}_{t} } ( \boldsymbol{u}_{t_i}) = \boldsymbol{0}$, above equals zero if and only if $\boldsymbol{\theta} = \boldsymbol{\theta_0}$. Moreover, the Hessian of $\lossfunc{\theta}$ is 
    \begin{equation} \label{eq: loss function Hessian}
        \nabla^{2}_{\theta} \lossfunc{\theta} = d^{-1} (\bmat{Z}_{\boldsymbol{t}}^T \bmat{Z}_{\boldsymbol{t}}),
    \end{equation}
    for all $\boldsymbol{\theta} \in \bmat{\Theta}$ and all $(\boldsymbol{y}_{t}, \bmat{Z}_{\boldsymbol{t}}) \in \mathbb{R}^{d} \times \mathbb{R}^{d \times q}$. Hence, by noting that the expectation of $\nabla^{2}_{\theta} \lossfunc{\theta}$ given by \eqref{eq: loss function Hessian} is a positive definite Hessian matrix, we have that $\boldsymbol{\theta_0}$ is the idealised global minima of the convex loss function \eqref{eq: squared error loss definition}, i.e., we have shown \eqref{eq: m estimator equation one}. Thus, we can solve for $\boldsymbol{\theta_0}$ by integrating \eqref{eq: loss function gradient} the result to zero. Thus, we have the well known result from linear regression and least-squares for the conditional problem
   \begin{equation*}
       \boldsymbol{\theta_0} = E \{ (\bmat{Z}_{\boldsymbol{t}}^T \bmat{Z}_{\boldsymbol{t}}) \}^{-1} E \left ( \bmat{Z}_{\boldsymbol{t}}^T \boldsymbol{y}_{t} \right ),
   \end{equation*}
   which, by noticing that the expectations are autocovariances, can be written in a Yule-Walker like way
   \begin{equation*}
        E \{ (\bmat{Z}_{\boldsymbol{t}}^T \bmat{Z}_{\boldsymbol{t}}) \} \boldsymbol{\theta_0} = E \left ( \bmat{Z}_{\boldsymbol{t}}^T \boldsymbol{y}_{t} \right ).
   \end{equation*}
    We proceed with the empirical risk minimiser. Note that when we concatenate all $ \bmat{Z}_{t_i}$ for all time steps $t_i$ by rows, i.e., each matrix is placed below the preceding one according to the $t_i$. 
    
    Then, we have that the resulting matrix is equal to $\bmat{R}$ in Section \ref{subsec: ls estimator}, hence, we can write
    \begin{align*}
        \boldsymbol{\hat{\theta}} &= \underset{\boldsymbol{\theta}}{ \mathrm{argmin}} \left \{ \Hat{F}_{\boldsymbol{y}_{t} | \bmat{Z}_{\boldsymbol{t}}} \left ( \lossfunc{\theta} \right ) \right \}, \\
        &= \underset{\boldsymbol{\theta}}{ \mathrm{argmin}} \left(2 d \right)^{-1} \sum_{i = 1}^n \normx{\boldsymbol{y}_{t_i} - \bmat{Z}_{t_i} \boldsymbol{\theta}}_2^2 \\
        &= \underset{\boldsymbol{\theta}}{ \mathrm{argmin}} \left(2 d \right)^{-1} \normx{\boldsymbol{y} - \bmat{R} \boldsymbol{\theta}}_2^2, \\
        &\overset{(i)}{=} \underset{\boldsymbol{\theta}}{ \mathrm{argmin}} \normx{\boldsymbol{y} - \bmat{R} \boldsymbol{\theta}}_2^2,
    \end{align*}
    where $(i)$ follows by noting that $\left(2 d \right)^{-1}$ is a positive constant that does not affect the optimal value, and $(\boldsymbol{y}, \bmat{R})$ are the same ones as in \eqref{eq: linear model} in Section \ref{subsec: ls estimator}, thus, we have that
    \begin{equation*}
         \boldsymbol{\hat{\theta}} = (\bmat{R}^T \bmat{R})^{-1}\bmat{R}^T \boldsymbol{y},
    \end{equation*}
    i.e., the M-estimator is the (conditional) least-squares estimator defined in Section \ref{subsec: ls estimator}, i.e., we have shown \eqref{eq: m estimator equation two}. Finally, by Theorem \ref{th: l2 bounds for ls estimator} and the first part of Corollary \ref{cor: conditional ls estimarot const.} we have that the M-estimator is consistent, i.e., as $T \to \infty$ we have that $|| \boldsymbol{\hat{\theta}} - \boldsymbol{\theta_0} ||_2 \overset{\mathbf{Pr}}{\longrightarrow} 0$.
\end{proof}
\newpage
\begin{proposition}
    Let $\lossfunc{\theta}$ be given by \eqref{eq: squared error loss definition} in Proposition \ref{prop: m-estimator}. Then, there exists an $\epsilon > 0$ such that
    \begin{equation} \label{eq: condition two}
        F_{\boldsymbol{y}_{t}, \bmat{Z}_{\boldsymbol{t}}} \left ( \gradlossfunc{\theta} - \gradlossfunc{\theta_0} \right ) = d^{-1}
        E \{ (\bmat{Z}_{\boldsymbol{t}}^T \bmat{Z}_{\boldsymbol{t}}) \} \left (\boldsymbol{\theta} - \boldsymbol{\theta_0} \right ) + o (|\boldsymbol{\theta} - \boldsymbol{\theta_0}|),
    \end{equation}
    for all $\boldsymbol{\theta}$ such that $|\boldsymbol{\theta} - \boldsymbol{\theta_0}| < \epsilon$, i.e., the loss function and problem setting satisfy Condition \ref{cond: cond two asymp norm} in Conditions \ref{conditions for asymp normality}.
\end{proposition}
\begin{proof}
    Notice that by \eqref{eq: loss function Hessian} and \eqref{eq: loss function gradient} for an $\epsilon > 0$ by Taylor's Theorem for multivariate functions for all $\boldsymbol{\theta} \in \mathbf{\Theta}$ such that $\normx{\boldsymbol{\theta} - \boldsymbol{\theta_0} }_2 < \epsilon$, we have that
    \begin{equation} \label{eq: second oreder approximation loss function}
        \lossfunc{\theta} - \lossfunc{\theta_0} = \langle \gradlossfunc{\theta_0}, (\boldsymbol{\theta} - \boldsymbol{\theta_0}) \rangle + 2^{-1} (\boldsymbol{\theta} - \boldsymbol{\theta_0})^T (\bmat{Z}_{\boldsymbol{t}}^T \bmat{Z}_{\boldsymbol{t}}) ( \boldsymbol{\theta} - \boldsymbol{\theta_0}) + o(1) \normx{\boldsymbol{\theta} - \boldsymbol{\theta_0} }_2^2,
    \end{equation}
    where $o(\normx{\boldsymbol{\theta} - \boldsymbol{\theta_0} }_2^2) \to 0$ as $\epsilon \to 0$. 
    \par
    Hence, by taking iterated expectations in \eqref{eq: second oreder approximation loss function} and using \eqref{eq: zero gradient loss function}, i.e., $ F_{\boldsymbol{y}_{t}, \bmat{Z}_{\boldsymbol{t}}} (\gradlossfunc{\theta_0}) = 0$, we have that
    \begin{equation} \label{eq: second order expectation}
         F_{\boldsymbol{y}_{t}, \bmat{Z}_{\boldsymbol{t}}} ( \lossfunc{\theta} - \lossfunc{\theta_0}) = 2^{-1} (\boldsymbol{\theta} - \boldsymbol{\theta_0})^T E \{ (\bmat{Z}_{\boldsymbol{t}}^T \bmat{Z}_{\boldsymbol{t}}) \} ( \boldsymbol{\theta} - \boldsymbol{\theta_0}) + o(1) \normx{\boldsymbol{\theta} - \boldsymbol{\theta_0} }_2^2.
    \end{equation}
    Notice that $\boldsymbol{\theta_0}$ is an interior point and that by \eqref{eq: bounded expected loss} the integral is bounded for all $\boldsymbol{\theta}$, thus, we can differentiate both sides of \eqref{eq: second order expectation} to get
    \begin{equation*}
        F_{\boldsymbol{y}_{t}, \bmat{Z}_{\boldsymbol{t}}} ( \gradlossfunc{\theta} - \gradlossfunc{\theta_0}) = E \{ (\bmat{Z}_{\boldsymbol{t}}^T \bmat{Z}_{\boldsymbol{t}}) \} ( \boldsymbol{\theta} - \boldsymbol{\theta_0}) + o(1) \normx{\boldsymbol{\theta} - \boldsymbol{\theta_0} }_2, 
    \end{equation*}
    i.e., we have shown \eqref{eq: condition two}. Alternatively, we can look at \eqref{eq: zero gradient loss function} and notice that the error is accounted for by the smoothness and nice properties of squared error loss.
\end{proof}
\begin{proposition} \label{prop: condition three}
    Let $\lossfunc{\theta}$ be given by \eqref{eq: squared error loss definition} in Proposition \ref{prop: m-estimator} in the same problem setting. Then, we have that
    \begin{equation} \label{eq: condition three}
        \lim_{\boldsymbol{\theta} \to \boldsymbol{\theta_0}} \normx{\gradlossfunc{\theta} - \gradlossfunc{\theta_0}}_2 = 0,
    \end{equation}
    i.e., the loss function $\lossfunc{\theta}$ is mean squared continuous.
\end{proposition}
\begin{proof}
    Notice that
\begin{align*}
    \normx{\gradlossfunc{\theta} - \gradlossfunc{\theta_0}}_2^2 &= \langle \gradlossfunc{\theta} - \gradlossfunc{\theta_0}, \gradlossfunc{\theta} - \gradlossfunc{\theta_0} \rangle \\
    &= d^{-2} \bigg \langle \bmat{Z}_{t_i}^T \left ( \boldsymbol{y}_{t_i} -  \bmat{Z}_{t_i} \boldsymbol{\theta} \right ) -  \bmat{Z}_{t_i}^T \left ( \boldsymbol{y}_{t_i} -  \bmat{Z}_{t_i} \boldsymbol{\theta_0} \right ), \\
    &\bmat{Z}_{t_i}^T \left ( \boldsymbol{y}_{t_i} -  \bmat{Z}_{t_i} \boldsymbol{\theta} \right ) -  \bmat{Z}_{t_i}^T \left ( \boldsymbol{y}_{t_i} -  \bmat{Z}_{t_i} \boldsymbol{\theta_0} \right ) \bigg \rangle \\
    &= d^{-2} \left \langle (\bmat{Z}_{\boldsymbol{t}}^T \bmat{Z}_{\boldsymbol{t}}) (\boldsymbol{\theta} -  \boldsymbol{\theta_0}),  (\bmat{Z}_{\boldsymbol{t}}^T \bmat{Z}_{\boldsymbol{t}}) (\boldsymbol{\theta} -  \boldsymbol{\theta_0}) \right \rangle \\
    &= d^{-2} \normx{(\bmat{Z}_{\boldsymbol{t}}^T \bmat{Z}_{\boldsymbol{t}}) (\boldsymbol{\theta} -  \boldsymbol{\theta_0})}_2^2.
\end{align*}
$\bullet$ Thus, as $\boldsymbol{\theta} -  \boldsymbol{\theta_0} \to \boldsymbol{0}$, the above goes to zero, i.e., 
\begin{equation*}
     \lim_{\boldsymbol{\theta} \to \boldsymbol{\theta_0}} \normx{\gradlossfunc{\theta} - \gradlossfunc{\theta_0}}_2^2 = 0
\end{equation*}
$\bullet$ By continuity of the square root function we arrive at \eqref{eq: condition three}, i.e., 
\begin{equation*}
     \lim_{\boldsymbol{\theta} \to \boldsymbol{\theta_0}} \normx{\gradlossfunc{\theta} - \gradlossfunc{\theta_0}}_2 = 0.
\end{equation*}
$\bullet$ Further, by linearity of expectation and Proposition \ref{prop: bounded eigenvalues} we also have that
\begin{equation*}
       F_{\boldsymbol{y}_{t}, \bmat{Z}_{\boldsymbol{t}}} \normx{\gradlossfunc{\theta} - \gradlossfunc{\theta_0}}_2 \leq d^{-1} \tau_{(q)} \normx{\boldsymbol{\theta} -  \boldsymbol{\theta_0}}_2. 
\end{equation*}
\end{proof}

\begin{proposition} \label{prop: condition four}
    Let $\mathfrak{G}_{\epsilon} := \{ \gradlossfunc{\theta} : \normx{\boldsymbol{\theta} -  \boldsymbol{\theta_0}}_2 < \epsilon \}$ for some $\epsilon > 0$, and define the envelope function $G_{\epsilon} := \underset{\gradlossfunc{\theta} \in \mathfrak{G}_{\epsilon}}{\sup} \{ | [\gradlossfunc{\theta} (\boldsymbol{y}_{t_i}, \bmat{Z}_{t_i} )]_j | \}$ for $j \in [q]$, i.e., the absolute value supremum among all entries in $\gradlossfunc{\theta}$. Then, we have that
    \begin{enumerate}
        \item The class $\mathfrak{G}_{\epsilon}$ is a finite dimension Vapnik-Chervonenkis graph class.
        \item $G_{\epsilon} \in L_2 (  F_{\boldsymbol{y}_{t}, \bmat{Z}_{\boldsymbol{t}}} )$, i.e., $ F_{\boldsymbol{y}_{t}, \bmat{Z}_{\boldsymbol{t}}} [G_{\epsilon}^2]_j < +\infty$, for all $j \in [q]$.
    \end{enumerate}
    Thus, by the above and Proposition \ref{prop: condition three}, the class $\mathfrak{G}_{\epsilon}$ satisfies the assumptions in Theorem \ref{th: asymp. continuity for VC class}. Hence, $\mathfrak{G}_{\epsilon}$ is asymptotically continuous at $\gradlossfunc{\theta_0}$.
\end{proposition}
\begin{proof}
    We start by looking at the squared norm of the gradient at $\boldsymbol{\theta_0}$, by \eqref{eq: loss function gradient}
    \begin{align*}
        F_{\boldsymbol{u}} \normx{\gradlossfunc{\theta_0}}_2^2 &= F_{\boldsymbol{u}} d^{-2} \normx{\bmat{Z}_{t_i}^T \boldsymbol{u}_{t_i} }_2^2 \\
        &= d^{-2} F_{\boldsymbol{u}} \{ \boldsymbol{u}_{t_i}^T (\bmat{Z}_{\boldsymbol{t}} \bmat{Z}_{\boldsymbol{t}}^T) \boldsymbol{u}_{t_i} \} \\
        &= d^{-2} F_{\boldsymbol{u}} [ \mathrm{tr} \{ \boldsymbol{u}_{t_i}^T (\bmat{Z}_{\boldsymbol{t}} \bmat{Z}_{\boldsymbol{t}}^T) \boldsymbol{u}_{t_i} \} ] \\
        &= d^{-2} \mathrm{tr} [ F_{\boldsymbol{u}} \{ \boldsymbol{u}_{t_i}^T (\bmat{Z}_{\boldsymbol{t}} \bmat{Z}_{\boldsymbol{t}}^T) \boldsymbol{u}_{t_i} \} ] \\
        &= d^{-2} \mathrm{tr} (\bmat{Z}_{\boldsymbol{t}} \bmat{\Sigma}_{\boldsymbol{u}} \bmat{Z}_{\boldsymbol{t}}^T) \\
        &\leq d^{-1} \sigma_{\boldsymbol{u}}^2 \mathrm{tr} (\bmat{Z}_{\boldsymbol{t}}^T \bmat{Z}_{\boldsymbol{t}}) 
    \end{align*}  
    $\bullet$ By Proposition \ref{prop: bounded eigenvalues}, there exists a $0 < \tau_{(q)} < +\infty$, which upper bounds all eigenvalues of $E (\bmat{Z}_{\boldsymbol{t}}^T \bmat{Z}_{\boldsymbol{t}})$, so we have that
    \begin{equation} \label{eq: bounded sqaured norm at theta0}
        F_{\boldsymbol{y}_{t}, \bmat{Z}_{\boldsymbol{t}}} \normx{\gradlossfunc{\theta_0}}_2^2 \leq \sigma_{\boldsymbol{u}}^2 \tau_{(q)} < +\infty.
    \end{equation}
    Let $\gradlossfunc{\theta} \in \mathfrak{G}_{\epsilon}$, and notice that
    \begin{align*}
        \normx{\gradlossfunc{\theta}}_2^2 &= \normx{\gradlossfunc{\theta} - \gradlossfunc{\theta_0} + \gradlossfunc{\theta_0}}_2^2 \\
        &\leq \left ( \normx{\gradlossfunc{\theta} - \gradlossfunc{\theta_0} }_2 + \normx{\gradlossfunc{\theta_0} }_2 \right )^2 \\
        &\leq 3 \left ( \normx{\gradlossfunc{\theta} - \gradlossfunc{\theta_0} }_2^2 + \normx{\gradlossfunc{\theta_0} }_2^2 \right ).
    \end{align*}
    $\bullet$ By \eqref{eq: condition two} and the above we have that
    \begin{align*}
         \normx{\gradlossfunc{\theta}}_2^2 &\leq 3 \left ( \normx{ (\bmat{Z}_{\boldsymbol{t}}^T \bmat{Z}_{\boldsymbol{t}}) (\boldsymbol{\theta} -  \boldsymbol{\theta_0}) + o(\| \boldsymbol{\theta} -  \boldsymbol{\theta_0} \|_2)}_2^2 + \normx{\gradlossfunc{\theta_0}}_2^2 \right ).
    \end{align*}
    $\bullet$ By taking expectations, \eqref{eq: condition two} and \eqref{eq: bounded sqaured norm at theta0}, we arrive at
    \begin{equation} \label{eq: bounded squared mean}
         F_{\boldsymbol{y}_{t}, \bmat{Z}_{\boldsymbol{t}}} \normx{\gradlossfunc{\theta}}_2^2 \leq 9 \left \{ ( \epsilon^2 + \sigma_{\boldsymbol{u}}^2 ) \tau_{(q)} + \epsilon^2 \right \} < +\infty.
    \end{equation}
    By \eqref{eq: bounded squared mean} we see that $ F_{\boldsymbol{y}_{t}, \bmat{Z}_{\boldsymbol{t}}} \normx{G_{\epsilon} }_2^2 < +\infty$, and noting that the $\ell_2$-norm of a vector is bounded if and only if all of its entries are bounded, we see that 
    \begin{equation*}
         F_{\boldsymbol{y}_{t}, \bmat{Z}_{\boldsymbol{t}}} [G_{\epsilon}]_j^2 < +\infty,
    \end{equation*}
    for all entries in $ G_{\epsilon}$.
    \par
    We proceed to the second part. Notice that for all $j \in [q]$ each $j$th entry satisfies
    \begin{align*}
        [\gradlossfunc{\theta}]_j &= - [ \bmat{Z}_{t_i}^T ( \boldsymbol{y}_{t_i} - \bmat{Z}_{t_i} \boldsymbol{\theta}) ]_j \\
        &= -[  \bmat{Z}_{t_i}^T \boldsymbol{y}_{t_i} - (\bmat{Z}_{t_i}^T  \bmat{Z}_{t_i}) \boldsymbol{\theta} ]_j
    \end{align*}
    Let $\boldsymbol{\phi}_{0,j} \in \mathbb{R}^{q + 1}$ such that $\phi_{0, j} = -[  \bmat{Z}_{t_i}^T \boldsymbol{y}_{t_i}]_j$ and all other entries are $\boldsymbol{\phi}_j = [(\bmat{Z}_{t_i}^T  \bmat{Z}_{t_i})]_{j \cdot}$, i.e., the $j$th column in $(\bmat{Z}_{t_i}^T  \bmat{Z}_{t_i})$. Then, 
    \begin{equation*}
        [\gradlossfunc{\theta}]_j = \phi_{0, j} + \langle \boldsymbol{\phi}_j, \boldsymbol{\theta} \rangle,
    \end{equation*}
    for all $j \in [q]$, i.e., each entry is a hyperplane in $\mathbb{R}^{q + 1}$. Thus, it is a finite dimension Vapnik-Chervonenkis graph class because the subgraph corresponds to the subgraph of a hyperplane.
    \par
    Recall that an intersection of Vapnik-Chervonenkis graph classes is a Vapnik-Chervonenkis graph class. Thus, all $j$th entries in $\gradlossfunc{\theta}$ are finite dimension Vapnik-Chervonenkis graph classes. Hence, $\mathfrak{G}_{\epsilon}$ is a finite dimension Vapnik-Chervonenkis graph class; see \cite{emp_process_theory}. 
    \par
    Therefore, by combining the above, \eqref{eq: bounded squared mean} and \eqref{eq: condition three}, we see that $\mathfrak{G}_{\epsilon}$ satisfies the assumptions in Theorem \ref{th: asymp. continuity for VC class}. Hence, $\mathfrak{G}_{\epsilon}$ is asymptotically continuous at $\gradlossfunc{\theta_0}$.
\end{proof}
\subsubsection{Completion of asymptotic normality}
We are ready to finish our proof of asymptotic normality. 
\begin{proof}
    By Propositions \ref{prop: bounded eigenvalues}--\ref{prop: condition four} we see that $\boldsymbol{\hat{\theta}}$ satisfies the assumptions in Lemma \ref{lemma> asymp norm of m-estimators}. Define the following $\bmat{\Gamma} (p, [C]) := E \{ (\bmat{Z}_{\boldsymbol{t}}^T \bmat{Z}_{\boldsymbol{t}}) \}$. Then, by \eqref{eq: condition two} we see that the influence function is $ 
    d^{-1} \{\bmat{\Gamma} (p, [C]) \}^{-1} \gradlossfunc{\theta_0}$. Notice that
    \begin{align*}
         F_{\boldsymbol{y}_{t}, \bmat{Z}_{\boldsymbol{t}}} (\gradlossfunc{\theta_0} \gradlossfunc{\theta_0}^T ) &= F_{\bmat{Z}_{\boldsymbol{t}}} \{ F_{\boldsymbol{u}} (\gradlossfunc{\theta_0} \gradlossfunc{\theta_0}^T ) \} \\
         &= F_{\bmat{Z}_{\boldsymbol{t}}} d^{-2} \{ F_{\boldsymbol{u}} ( \bmat{Z}_{t_i}^T \boldsymbol{u}_{t_i} \boldsymbol{u}_{t_i}^T \bmat{Z}_{t_i} ) \}\\
         &= F_{\bmat{Z}_{\boldsymbol{t}}} d^{-2} \{ \bmat{Z}_{t_i}^T  \bmat{\Sigma}_{\boldsymbol{u}} \bmat{Z}_{t_i} \},
    \end{align*}
    by Lemma \ref{lemma> asymp norm of m-estimators} we see that the asymptotic covariance is
    \begin{equation*}
        \bmat{\Sigma}_{\boldsymbol{\hat{\theta}}} = d^{-1} \{ \bmat{\Gamma} (p, [C]) \}^{-1}  E \{ (\bmat{Z}_{\boldsymbol{t}}^T \bmat{\Sigma}_{\boldsymbol{u}} \bmat{Z}_{\boldsymbol{t}}) \} \{ \bmat{\Gamma} (p, [C]) \}^{-1},
    \end{equation*}
    where, we send the other $d^{-1}$ term to the sample size divisor. Finally, by recalling that $|\mathcal{K}| = d$ and applying Lemma \ref{lemma> asymp norm of m-estimators} we have that
    \begin{equation*} 
            \{ |\mathcal{K}| (T - p) \}^{1/2} ( \boldsymbol{\hat{\theta}} - \boldsymbol{\theta_0} ) \overset{\boldsymbol{\mathrm{d}}}{\longrightarrow} \mathrm{N}_q \left(0, \bmat{\Sigma}_{\boldsymbol{\hat{\theta}}} \right ).
    \end{equation*}
\end{proof}
\begin{proposition}
    Assume that A.\ref{asum: A1}, A.\ref{asum: A2}, and A.\ref{asum: A3} hold in Assumptions \ref{assum: theorem assumptions}. Further, assume that all conditions in Proposition \ref{prop: m-estimator} hold, and that $\boldsymbol{u}_t \sim \mathrm{WN} (0, \sigma_{\boldsymbol{u}}^2 \bmat{I_d})$, i.e., the white noise processes are {\em i.i.d.} at all times $t$. Then, we have that
    \begin{equation*}
        \hat{\sigma}_{\boldsymbol{u}}^2 = \{ |\mathcal{K}| (T - p)\}^{-1} || \boldsymbol{y} - \bmat{R} \boldsymbol{\hat{\theta}}||_2^2,
    \end{equation*}
    is a consistent estimator of $\sigma_{\boldsymbol{u}}^2$, i.e., $\hat{\sigma}_{\boldsymbol{u}}^2 \overset{\mathbf{Pr}}{\longrightarrow} \sigma_{\boldsymbol{u}}^2.$
\end{proposition}
\begin{proof}
    By Proposition \ref{prop: m-estimator} and the continuous mapping theorem, as $\boldsymbol{\hat{\theta}}  \overset{\mathbf{Pr}}{\longrightarrow} \boldsymbol{\theta_0}$ we have that
    \begin{align*}
         \{ |\mathcal{K}| (T - p)\}^{-1} || \boldsymbol{y} - \bmat{R} \boldsymbol{\hat{\theta}}||_2^2 &\overset{\mathbf{Pr}}{\longrightarrow} \{ |\mathcal{K}| (T - p)\}^{-1} || \boldsymbol{y} - \bmat{R} \boldsymbol{\theta_0}||_2^2 \\
         &= \{ |\mathcal{K}| (T - p)\}^{-1} || \boldsymbol{u} ||_2^2.
    \end{align*}
    Notice that for all entries $u_{i, t}$ in $\boldsymbol{u}$, we have that $E(u_{i,t}^2) =  \sigma_{\boldsymbol{u}}^2$. Since, the above is the sample mean of $ \{ |\mathcal{K}| (T - p)\}$ {\em i.i.d.} realisations of $u_{i, t}^2$, by the (weak) law of large numbers, we have that
    \begin{equation*}
        \{ |\mathcal{K}| (T - p)\}^{-1} || \boldsymbol{u} ||_2^2 \overset{\mathbf{Pr}}{\longrightarrow}  \sigma_{\boldsymbol{u}}^2,
    \end{equation*}
    whence, we arrive at
    \begin{equation*}
        \hat{\sigma}_{\boldsymbol{u}}^2 \overset{\mathbf{Pr}}{\longrightarrow} \sigma_{\boldsymbol{u}}^2.
    \end{equation*}
\end{proof}
\newpage
\section{Notation summary}\label{apen: notation}

\begin{itemize}
    \item $T$: length of realisation.
    \item  $\mathcal{X} := (\boldsymbol{X}_t, \mathcal{G})$: Network time series.
    \item $X_{i, t}$: Nodal time series.
    \item $\boldsymbol{X}_t$ : Multivariate time series at time $t$.
    \item $\mathcal{G} = (\mathcal{K}, \mathcal{E})$, where $\mathcal{K} = \{1, \dots, d\}$ is the node set, $\mathcal{E} \subseteq \mathcal{K} \times \mathcal{K}$ is the edge set.
    \item $\bmat{S}_r \in \{0, 1\}^{d \times d}$, where $[\bmat{S}_r]_{ij} = \mathbb{I} \{ d(i, j) = r \}$, $\mathbb{I}$ is the indicator function, $r \in \{1, \dots, r_{\text{max}} \}$, and $r_{\text{max}} \in \mathbb{Z}^+$ is the longest shortest path in $\mathcal{G}$.
    \item Weights matrix $\bmat{W}_t \in \mathbb{R}^{d \times d}$ as the matrix $[\bmat{W}_t]_{ij} = w_{ij}(t)$.
    \item $\boldsymbol{Z}_{t}^r := \left ( \bmat{W}_t \odot \bmat{S}_r \right ) \boldsymbol{X}_t$, where $\odot$ denotes the Hadamard (component-wise) product.
    \item $K_c := \left \{ i \in \mathcal{K} : X_{i, t} \text{ is associated with covariate } c \right \} $.
    \item $\mathcal{I}_c := \{ \tilde{c} \in [C] : c \text{ interacts with } \tilde{c}, \text{where }\tilde{c} \neq c \}$.
    \item $\boldsymbol{Z}_{t - k}^{r, c} := \left ( \bmat{W}_{t, c} \odot \bmat{S}_r \right) \boldsymbol{X}_{t - k}^c$, $\rstage$ neighbourhood regressions for community $K_c$.
    \item $[\bmat{W}_{t, c}]_{ij} = w_{ij} (t) \mathbb{I} ( i \in K_c \text{ and } j \in K_c)$.
    \item $\alpha_{k, c}$: autoregressive coefficient at lag $k$ for community $K_c$.
    \item $\beta_{k, r, c}$: Within community coefficient at lag $k$ for community $K_c$.
    \item $\gamma_{k, c, c:c'}$: Between community coefficient at lag $k$ for the effect community $K_{c'}$ has on community $K_c$.
    \item $p_c \in \mathbb{Z}^+$ and $s_{k} (c) \in \{1, \dots, r^*_c \}$, where the former is the maximum lag for community $K_c$ and the latter is the maximum $\rstage$ depth.
    \item $\boldsymbol{Z}_{t - k}^{r,c:\tilde{c}} = \xi_c \odot \left ( \bmat{W}_t \odot \bmat{S}_r \right) \boldsymbol{X}_{t - k}^{\tilde{c}}$, between community interactions.
    \item $\sigma^2_{\boldsymbol{u}}$: white noise process variance.
    \item $\bmat{R}_{k, c} = \left [\boldsymbol{X}_{t - k}^c |  \boldsymbol{Z}_{t - k}^{1, c} | \dots |  \boldsymbol{Z}_{t - k}^{s_k (c), c} |  \boldsymbol{Z}_{t - k}^{1, c:\tilde{c}} | \dots |  \boldsymbol{Z}_{t - k}^{s_k (c), c:\tilde{c}} \right ].$
    \item $\bmat{R}_{c} = \left [\bmat{R}_{1, c} | \dots | \bmat{R}_{p_c, c} \right ]$.
    \item $\boldsymbol{\theta}_c := (\alpha_{k, c}, \beta_{k, c}, \gamma_{k, c})$.
    \item $\boldsymbol{\theta} := (\boldsymbol{\theta}_1, \dots, \boldsymbol{\theta}_C )$.
    \item $\boldsymbol{y} := \left (\boldsymbol{X}_{p + 1}, \dots, \boldsymbol{X}_{T} \right )$.
    \item $\boldsymbol{y} = \bmat{R} \boldsymbol{\theta} + \boldsymbol{u}$: linear model representation.
    \item $q_c :=  \left \{ p_c + ( 1 + | \mathcal{I}_c |) \sum_{k = 1}^{p_c} s_{k} (c) \right \}$, i.e., the number of unknown parameters in $K_c$, 
    \item $q_{[C]} := \max_{c \in [C]} \left \{ q_c \right \}$, i.e., the maximum number of unknown parameters,
    \item $p := \max_{c \in [C]} (p_c)$, i.e., the largest active lag across all communities,
    \item $k_{[C]} := \min_{c \in [C]} \{ |K_c| \}$, i.e., the size of the smallest community,
    \item  $\tau := \min_{c \in [C]} (\tau_c) $, i.e., the largest constant for uniformly lower bounding the smallest eigenvalues of all $(\bmat{R}_c^{T} \bmat{R}_c)$ {\em (i.e., $\lambda_{\min} (\bmat{R}_c^T \bmat{R}_c) \geq \{ \tau k_{[C]} (T - p) \}^{1/2}$ for all $c \in [C]$)}.
\end{itemize}
The linear model representation can be exploited to express the problem in general form as follows; see Section~\ref{apend: proof of asymp normality summary} in the supplementary material.
\begin{align*}
    \bmat{R}_{k, t, c} &:= \left [\boldsymbol{X}_{t - k}^c |  \boldsymbol{Z}_{t - k}^{1, c} | \dots |  \boldsymbol{Z}_{t - k}^{s_k (c), c} |  \boldsymbol{Z}_{t - k}^{1, c:\tilde{c}} | \dots |  \boldsymbol{Z}_{t - k}^{s_k (c), c:\tilde{c}} \right], \nonumber \\
    \bmat{R}_{k, t} &:= [\bmat{R}_{k, t, c}| \cdots | \bmat{R}_{k, t, C} ], \nonumber \\
    \bmat{Z}_{\boldsymbol{t_i}} &:= \left [ \bmat{R}_{1, t}| \cdots | \bmat{R}_{p, t} \right ],
\end{align*}
where each predictor column for $\tilde{c} \in \mathcal{I}_c$ is concatenated in ascending order with respect to $\tilde{c}$, i.e., if $\tilde{c}_2 > \tilde{c}_1$, then the column for $\tilde{c}_1$ precedes the one for $\tilde{c}_2$, $ \boldsymbol{u}_{t_i}$ is a sub-Gaussian multivariate white noise process with a diagonal covariance, and $ \boldsymbol{\theta_0} \in \bmat{\Theta} \subset \mathbb{R}^q$ is the ``true'' vector of parameters.

\end{document}